\documentclass[preprint,12pt]{elsarticle}

\usepackage{amssymb}

\usepackage[T2A]{fontenc}
\usepackage[russian, english]{babel}
\usepackage[utf8]{inputenc}

\usepackage{graphicx}
\usepackage{subfigure}
\usepackage{color}
\usepackage{textgreek}
\usepackage{upgreek}
\usepackage{multirow}
\usepackage{amsmath}
\usepackage{url}
\usepackage{comment}
\usepackage[super]{nth}
\usepackage{csquotes}
\usepackage[dvipsnames]{xcolor}
\usepackage{enumitem}
\usepackage{hyperref}
\usepackage{xurl}

\usepackage[version=4]{mhchem}

\usepackage{textcomp}

\usepackage[normalem]{ulem}

\renewcommand{\numberline}[1]{#1\hspace{0.5em}}

\newcommand{\unit}[1]{\,\text{#1}}

\journal{Progress in Aerospace Sciences}

\begin{document}

\begin{frontmatter}

\title{The New Science of \\ Unidentified Aerospace-Undersea Phenomena (UAP)}

\address[UA]{Dept. of Physics, Univ. at Albany (SUNY), Albany, NY USA}
\address[Bourgogne]{Universit\'{e} de Bourgogne, Dijon, France}
\address[Chicago]{Dept. of Anthropolgy, Univ. of Chicago, Chicago, IL USA}
\address[CNES]{CNES (Centre National D'\'{e}tudes Spatiales), Toulouse, France}
\address[CUNY]{City University of New York, New York, NY USA}
\address[Geipan]{GEIPAN (Groupe D'\'{E}tudes et d'Informations sur les Ph\'{e}nom\`{e}nes A\'{e}rospatiaux Non-identifi\'{e}s)}
\address[HARV]{Harvard-Smithsonian Center for Astrophysics, Cambridge, MA USA}
\address[StJohns]{St. John's University, Queens, NY USA}
\address[JMUW]{Julius-Maximilians Universit\"{a}t W\"{u}rzburg, W\"urzburg, Germany}
\address[WMU]{Western Michigan University, Kalamazoo, MI USA, Professor Emeritus}
\address[KTH]{KTH Royal Institute of Technology, Stockholm, Sweden}
\address[Nordita]{Nordic Institute for Theoretical Physics (Nordita), Stockholm, Sweden}
\address[Stan]{Dept. of Pathology, Stanford University, Stanford CA USA}
\address[Stock]{Stockholm University, Stockholm, Sweden}
\address[WC]{Dept. of Physics and Astronomy, Wellesley College, Wellesley, MA USA}
\address[AIAA-OUT]{AIAA UAP Integration \& Outreach Committee}
\address[AIAA-STEER]{AIAA UAP Community Of Interest Steering Group}
\address[ASA]{Executive Director, Americans for Safe Aerospace (ASA)}
\address[3AF-SIGMA2]{3AF-SIGMA2, Association Aéronautique et Astronautique de France, Paris, France}
\address[Conservateur]{Conservateur en Chef des Biblioth\`{e}ques, France}
\address[DR]{Documatica Research, LLC, San Francisco, CA USA}
\address[Front]{Frontier Analysis, Ltd. (Founder)}
\address[EYE]{Eye On The Sky Project, Long Island, NY USA}
\address[HESS]{Project Hessdalen, Hessdalen, Norway \url{https://www.hessdalen.org/}}
\address[IFEX]{Interdisciplinary Research Center for Extraterrestrial Studies (IFEX), Julius-Maximilians Universit\"{a}t W\"{u}rzburg, \url{https://www.uni-wuerzburg.de/en/ifex/}}
\address[METACX]{Meta-Connexions, Toulouse, Occitanie, France}
\address[SCU]{Scientific Coalition for UAP Studies (SCU), \url{https://www.explorescu.org/}}
\address[SOL]{The SOL Foundation, \url{https://thesolfoundation.org/}}
\address[SPACEEXPLORE]{Space Exploration, Ltd., Boyle, County Roscommon, Ireland, \url{https://spaceexplorationltd.com/}}
\address[SUAPS]{Society for UAP Studies (SUAPS), \url{https://www.societyforuapstudies.org/}}
\address[UAPCHECK]{UAP Check Board Member, \url{https://www.uapcheck.com/}}
\address[UAPX]{UAPx, (\url{https://www.uapexpedition.org/}))}
\address[CUFOS]{Center for UFO Studies (CUFOS), \url{https://cufos.org/}}
\address[MUFON]{Mutual UFO Network (MUFON), \url{https://mufon.com/}}
\address[OHIO]{Ohio MUFON (Newsletter Editor)}

\author[UA,UAPX,SUAPS,SOL,SCU,IFEX]{Kevin H.~Knuth}
\author[AIAA-STEER]{Philippe Ailleris}
\author[Chicago]{Hussein Ali Agrama}
\author[SPACEEXPLORE]{Eamonn Ansbro}
\author[Front,SCU,MUFON,OHIO]{Phyllis A. Budinger}
\author[SCU]{Tejin Cai}
\author[Conservateur]{Thibaut Canuti}
\author[StJohns,SUAPS]{Michael C. Cifone}
\author{Walter Bruce Cornet, Jr.}
\author[Geipan,CNES]{Fr\'ed\'eric Courtade}
\author{Richard Dolan}
\author[HARV,SCU]{Laura Domine}
\author[3AF-SIGMA2]{Luc Dini}
\author[Bourgogne]{Baptiste Friscourt}
\author[AIAA-OUT,ASA]{Ryan Graves}
\author{Richard F. Haines}
\author[SCU]{Richard Hoffman}
\author[JMUW,IFEX]{Hakan Kayal}
\author[WC,SCU]{Sarah Little}
\author[Stan,SOL]{Garry P. Nolan}
\author[SCU]{Robert Powell}
\author[CUFOS,SUAPS]{Mark Rodeghier}
\author[UAPCHECK]{Edoardo Russo}
\author[SOL]{Peter Skafish}
\author[HESS]{Erling Strand}
\author[WMU]{Michael Swords}
\author[UA,UAPX,SUAPS,SCU,IFEX]{Matthew Szydagis}
\author[EYE]{Gerald T. Tedesco}
\author[EYE]{John J. Tedesco}
\author[SUAPS]{Massimo Teodorani}
\author[DR,SOL]{Jacques Vall\'{e}e}
\author[METACX,UAPCHECK]{Micha\"{e}l Vaillant}
\author[Nordita,KTH,Stock]{Beatriz Villarroel}
\author[WC]{Wesley A. Watters}
\cortext[]{Corresponding author. \textit{E-mail address}: \\ kknuth@albany.edu (Kevin H. Knuth).}

\begin{abstract}
After decades of dismissal and secrecy, it has become clear that a significant number of the world’s governments take Unidentified Aerospace-Undersea Phenomena (UAP), formerly known as Unidentified Flying Objects (UFOs), seriously--—yet still seem to know little about them. As a result, these phenomena are increasingly attracting the attention of scientists around the world, some of whom have recently formed research efforts to monitor and scientifically study UAP. In this paper, we review and summarize approximately 20 historical government studies dating from 1933 to the present (in Scandinavia, WWII, US, Canada, France, Russia, China), several historical private research studies (France, UK, US), and both recent and current scientific research efforts (Ireland, Germany, Norway, Sweden, US). In doing so, our objective is to clarify the existing global and historical scientific narrative around UAP. Studies range from field station development and deployment to the collection and analysis of witness reports from around the world. We dispel the common misconception that UAPs are an American phenomenon and show that UAP can be, and have been, scientifically investigated. Our aim here is to enable future studies to draw on the great depth of prior documented experience.
\end{abstract}

\begin{highlights}
\item UAP, formerly known as UFOs, are a long-standing global phenomenon, which have not only been observed by professional engineers, scientists, and astronomers, but have also been studied by them.
\item There exist today a number of serious academic and private efforts to scientifically study UAP using the equivalent of multi-messenger astronomy, which involves a diverse array of scientific instrumentation.
\end{highlights}


\end{frontmatter}

\tableofcontents

\section{Introduction} \label{sec:intro}
Unidentified Aerial Phenomena (UAP), which have been referred to as Unidentified Aerospace-Undersea Phenomena by the United States Congress to emphasize their multi-media characteristics, are currently more simply referred to as Unidentified Anomalous Phenomena.  While it is not widely appreciated that UAP have been observed and reported for centuries
\cite{Roberts:1967,Wittmann:1968,Corliss:1995,Stothers:2007,Vallee+Aubeck:2010,Dolan:2025},  investigation into reports of unusual objects and light phenomena in the sky has resulted in the identification, characterization, and scientific understanding of a great number of aerial and astronomical phenomena, such as planets, comets, asteroids and meteors, auroras, lenticular clouds, parhelia (sundogs) and halo phenomena \cite{Neuberger:1951,Greenler:1980,Whalley:1981}, and even St. Elmo's fire \cite{Wescott+etal:1996}, ball lightning \cite{Cen+etal:2014,Singer:2012}, elves and sprites \cite{Fukunishi+etal:1996,Barrington+etal:2001,Rycroft:2006}, as well as earthquake lights \cite{Freund:2019,Freund:2003,St-Laurent+etal:2006}. 
Several of the latter examples have been recently considered pseudoscientific \cite{Stenhoff:1999}, but are now recognized by the scientific community as real phenomena worthy of study.  To this day, there remains a residuum of phenomena that continue to defy explanation.
%

The study of UAP suffers from two main difficulties.  First, the phenomena are neither repeatable nor controllable.  This makes scientific data collection extremely difficult, since one must be resigned to setting up observing stations and waiting for events.  In some ways this makes collecting data on UAPs similar to collecting data on earthquakes, gravitational waves, dark matter, or rare astronomical events, such as supernovae, except that the observation of particular UAP of interest is potentially more rare.  The second difficulty is that it has often been asserted that witness reports are not of anomalous phenomena and can be readily explained in terms of well-understood natural phenomena, misidentifications, exaggerations, and/or hoaxes; and therefore they are not worthy of serious scientific study.  This latter assertion has been widely held and promoted by the authorities and the scientific community for over 70 years, thus preventing study, discovery, and understanding; consequently, leaving us in a rather disconcerting state of ignorance.

The problem and opportunity that we face today is that the situation has changed dramatically.  In December 2017, the New York Times (NYT) published an article \cite{Cooper+Blumenthal+Kean:2017}, which revealed that the US Defense Intelligence Agency (DIA) had conducted a six-year covert program, the Advanced Aerospace Threat Identification Program (AATIP), to study UAP.  AATIP, which was led by Luis Elizondo \cite{Elizondo:2024}, focused on military-only encounters and existed by operating under the more extensive Advanced Aerospace Weapon System Application Program (AAWSAP), which was funded by \$22 million secured by Senators Reid, Inouye, and Stevens.  The objective of AAWSAP was to study the many aspects of UAP, which included the psychic and paranormal correlates to UAP interactions \cite{Kelleher+Knapp:2005, Vallee+Davis:2005, Lacatski+Kelleher+Knapp:2021, Ballester-Olmos+Cayetano:2024}, while identifying ``potential breakthrough technology applications employed in future aerospace weapon systems.''\cite{Knapp:2023, Ballester-Olmos+Cayetano:2024}   AAWSAP was managed by veteran intelligence analyst and rocket expert James Lacatski \cite{Elizondo:2019, Lacatski+Kelleher+Knapp:2021, Knapp:2023, Ballester-Olmos+Cayetano:2024, Elizondo:2024}.  At its peak, the effort employed 50 full-time investigators (far more than any other program), which compiled the largest UFO data warehouse covering more than 200,000 cases.  George Knapp, in his statement to Congress \cite{Knapp:2023}, notes that the AAWSAP Program produced more than 100 research papers on UAP, some of them more than 100 pages long.  He also notes that the first case investigated by AAWSAP was the 2004 Nimitz TicTac event and that the report on the TicTac and its capabilities, written by the AAWSAP scientists and engineers, was more than 140 pages long.  None of these papers or reports have ever been seen by Congress or the public \cite{Knapp:2023}.

In conjunction with the NYT article, Luis Elizondo worked to have several infrared videos of UAP taken by the US Navy publicly released \cite{Warrick:2017}.  Later, it was revealed that the US Navy has had ongoing \cite{Rogoway:2019}, and at times daily \cite{Stieb:2019}, encounters with UAP operating with impunity in restricted airspaces and harassing naval pilots during military exercises and wartime operations \cite{Keyhoe+Cecotti:2021,Mizokami:2021,Silva:2020,Stieb:2019}.

In response, the United States Navy has changed its procedures for reporting such encounters \cite{Bender:2019}, and enlisted Congress to take action \cite{Bender:2019b, Golgowski:2019, Lutz:2019, Knapp+Adams:2019}.  The Pentagon officially released three UAP videos in April 2020, confirming that UAPs regularly operate in restricted airspace \cite{Conte:2020, DefGov:2020, Yuhas:2020}.  In June 2020, the Senate Intelligence Committee voted to require the US Department of Defense and the US Intelligence Community to detect, track, compile, catalog, and analyze information on UAPs through a program called the Unidentified Aerial Phenomena Task Force (UAPTF) \cite{DefGov:2020}.

In June 2021, the UAPTF presented its preliminary report to Congress, reconfirming that these UAP are not of American origin and are unlikely to come from any other country.  The possibility was left open that they are craft of potentially non-human origin.  In response, US Senator Kirsten Gillibrand (D-NY) introduced the UAP amendment into the FY22 National Defense Authorization Act (NDAA 2022) to establish a formal office to report and respond to UAPs and provide the scientific capabilities needed to track and share data, investigate sightings, and develop a response to this growing security threat.  This led to the legislation 50 U.S.C. \S 3373 (NDAA 2022 and NDAA 2023), which authorized the establishment of the All-Domain Anomaly Resolution Office (AARO), which is directed to continue the duties of the UAPTF.  This includes ``developing procedures to synchronize and standardize the collection, reporting, and analysis of incidents, including adverse physiological effects, regarding unidentified anomalous phenomena across the Department of Defense and the intelligence community'', ``developing the processes and procedures to ensure that such incidents from each component of the Department and each element of the intelligence community are reported and stored in an appropriate manner that allows for the integration of analysis of such information'', ``evaluating links between unidentified anomalous phenomena and adversarial foreign governments, other foreign governments, or nonstate actors'', assessing possible threats, coordinating with other federal departments and agencies, ``consulting with allies and partners of the United States to better assess the nature and extent of unidentified anomalous phenomena'', and ``preparing reports for Congress, in both classified and unclassified form'' \cite{AARO:2022}.

Although people are generally aware that there have been several programs run by the US government to study UFOs (Unidentified Flying Objects) and UAP, it is not generally known precisely what was done and, more importantly, precisely how little was done.  Moreover, it is generally believed that serious scientists, academics, and especially astronomers do not believe in the reality of UFOs, have not seen them, and certainly do not study them.  By going into some detail on efforts to study UAP in the late 1940s and early 1950s, we learn that some prominent and serious scientists \emph{have} been involved, that they \emph{have} witnessed UFOs of many types, and that they \emph{took the matter quite seriously}.  Such details add context Sturrock's (Sec. \ref{sec:Sturrock}) 1977 survey of members of the American Astronomical Society \cite{Sturrock:1994:part_1} in which he found that out of the 1356 respondents, 23\% stated that UFOs should certainly be studied and 30\% said that UFOs should probably be studied as opposed to 17\% saying that UFOs should probably not be studied and 3\% that said that UFOs should certainly not be studied.  Moreover, Sturrock found that 62 of the respondents had witnessed or obtained an instrumental record of something that they could not explain, and that of the respondent witnesses, 63\% of them were night-sky observers.

This paper provides some background on UAP, followed by a detailed exposition of previous and concurrent efforts to scientifically study UAP in the expectation that a careful examination of these efforts will provide a great deal of information relevant to current and future efforts.  By focusing on what scientific studies of UFOs have taken place, future studies can draw on their experience, which includes both their successes and failures, and can better equip themselves to select the best instruments, design the best observation strategies, and identify the most promising locations to study.

Since many countries, research groups, and individuals have participated in the scientific study of UAP for almost a century, there are several ways in which the information in this article could be presented.  Unfortunately, there is no single presentation style that appears to be optimal.  In some situations, the chronology of events and efforts is central, suggesting that a presentation based on chronology is obligatory.  However, at other times, once efforts became more focused, it appears to be best to separate out efforts based on the nations, research groups, or the individuals involved.  As a result, the presentation in this paper is mixed to a great degree.  At some times, especially early on and during wartime, the interactions among different nations dominated the efforts to understand these phenomena.  For this reason, the paper begins with a chronological, or historical, treatment which, as efforts shift to become more isolated, gives way to a presentation based on the nations, groups, or individuals involved.  An appreciation of both of these differing perspectives is necessary to understand the importance and impact of efforts to study UAP in the past century.

\section{What are UAP?}
The acronym UAP can stand for Unidentified Anomalous Phenomenon or Phenomena, although the A can also stand for Aerial or Aerospace.  The US Congress recently redefined the acronym UAP as the more descriptive Unidentified
Aerospace-Undersea Phenomena~\cite[p.12]{Congress:2022}, but this has since migrated back to the Unidentified Anomalous Phenomena. The term aerospace broadens the study of UAP to include the Earth's atmosphere and outer space; whereas the term undersea extends it to underwater and oceanic domains. The importance of the multi-medium nature of UAPs is perhaps best illustrated by the fact the successor to the UAP Task Force is the Pentagon's \textit{All-domain} Anomaly Resolution Office (AARO)~\cite{Defense:2022}, which was established by the US Defense secretary to provide ongoing reports to Congress~\cite{ODNI}.

The term UAP refers to an object / phenomenon that cannot be immediately recognized as prosaic, \textit{e.g.}~for aerial~phenomena: human-made craft, flying animals, or other well-known natural phenomena. ``Unidentified'' only means that one does not know~what~something is---at least initially---until additional analysis is possible. Since this is essentially a study of unknowns, it could be more helpful to define a UAP in terms of the relevant characteristics identified by the Pentagon's AATIP Program.  These characteristics are known as The Five Observables: \cite{Elizondo:2019, Elizondo:2024}
\begin{enumerate}
  \item Positive Lift without Flight Surfaces
  \item Sudden/Instantaneous Acceleration
  \item Hypersonic Velocity Without Signatures
  \item Trans-Medium Travel
  \item Low Observability or Cloaking.
\end{enumerate}
In some close encounter scenarios, a sixth, albeit more disturbing, observable is sometimes noted: \cite{Elizondo:2019, Elizondo:2024}
\begin {enumerate}
  \setcounter{enumi}{5}
  \item{Biological Effects on Humans and Animals} \label{observable:5}
\end{enumerate}
While many UAP are ultimately identifiable and of mundane origin, some remain unidentified. It is only this unidentified residuum that is of interest.
The degree of identification depends on the amount of detail that can be provided by the various sensing methods used as well as the investigative and scientific resources brought in.

There is little doubt that the majority of UAP are misidentifications, but anywhere between 4-40\% remain unidentified after careful investigations~\cite{Davidson:1966,Clemence:1969,Cometa:1999,AirForce:2003}, depending upon the sources and quality of the reports. Despite this, there exist hard data that demonstrate unreasonably high speeds (above Mach~40-60) and accelerations (hundreds to thousands of times \textit{g}) \cite{Oberth:1954, Hill:1995, Poher:2005, Hill:2014, Knuth+etal:2019, Coumbe:2022}, without the corresponding sonic booms or fireballs; as well as the emission of luminous power sometimes exceeding 100 MW (MegaWatts) \cite{Maccabee:1999,Vallee:2000}. Such observations represent cases of interest as they may require more exotic explanations, such as hypotheses involving novel physics or at least new engineering. However, more data are needed to characterize these anomalous objects and definitively rule out observational errors~\cite{Markowitz:1967,Loeb+Kirkpatrick:2023}.

Many different phenomena are likely being included under the UAP umbrella.
While making an exhaustive list isn't feasible, the possible prosaic explanations of UAP include, but are not limited to: 
\begin{enumerate}
    \item human-made crafts, such as airplanes, helicopters, drones, balloons, satellites including space stations, like the International Space Station (ISS), para-gliders / para-sailors, and marine vessels,
    \item Fauna, such as birds, bats, and insects,
    \item clouds (especially lenticular clouds), atmospheric optical effects, such as sun dogs, solar / lunar halos, Fata Morgana optical illusions and other types of mirages,
    \item celestial bodies, like the Moon, Venus or other planets, meteors, and comets,
    \item Heretofore non-discovered or sparsely-studied atmospheric phenomena may be involved \textit{e.g.}~ball lightning \cite{Cen+etal:2014, Singer:2012} or earthquake lights \cite{Freund:2003,St-Laurent+etal:2006,Freund:2019}. In the recent past, both have been considered illusory and the subject of ``pseudoscience'', despite well-documented observations to the contrary~\cite{Stenhoff:1999}.
\end{enumerate}
 A similar list of prosaic phenomena can be found in the UAlbany-UAPx paper by Szydagis et al. in this issue \cite{Szydagis+etal:2024}, and a detailed table of categorized phenomena can be found in the Galileo Project paper by Watters et al. \cite[Table 1]{Watters+etal:2023}.

\section{Government Efforts to Study UAP}
This paper focuses on the historical efforts to scientifically study UAP.  Recent articles by Ailleris \cite{Ailleris:2011,Ailleris:2024}, Stahlman \cite{Stahlman:2024}, Villaroel \& Krisciunas \cite{Villarroel+Krisciunas:2024}, and Watters et al. \cite{Watters+etal:2023} provide excellent summaries of such efforts.
For those interested in the history of the role of government in this rich and involved topic, we recommend the detailed text by Swords et al. \citep{Swords+etal:2012}, as the governmental aspects of our exposition are guided by their work.  As much of the relevant information exists in the gray (non-academic, non-peer-reviewed) literature, it is important to summarize relevant details, if even to dispel some of the common misconceptions about UAP.

Our exposition begins with its focus on chronology, as the early efforts were international in scope. UAP studies have a rich history dating back to antiquity \cite{Wittmann:1968,Stothers:2007,Vallee+Aubeck:2010,Roberts:1967,Dolan:2025}, and include the airship wave of the late 1800s and early 1900s \cite{Grove:1970,Clarke:1999,Bartholomew:2000}.

We begin our chronological exposition with the so-called ``ghost flyers'' sighted over Scandinavia in the 1930s.  The reason for this is to emphasize several points: that UAP are not a singular phenomenon, but are instead a class of phenomena; that these phenomena did not all begin in 1947; and that they are not a uniquely American phenomenon. For an historical overview, see Eghigian \cite{Eghigian:2024}.

\subsection{Scandinavia}
One of the earliest governmental efforts to study UAP began in late 1933 with a number of sightings of unknown aircraft in the skies above northern Sweden, Norway, and Finland \citep{Swords+etal:2012}.  These were thought to be airplanes involved in smuggling and, as such, had attracted the attention of Swedish customs and the Swedish Air Force, which conducted surveillance.  Together, the unknown aircraft earned the monikers of ``the Ghost Flyer'', ``the Ghost Machine'', and ``The Flying X''.  This activity peaked during the winter of 1933--1934, and by March 1934, 96 reports had been filed with the Swedish military, 157 reports had been filed with the Finnish military, and 234 reports had been filed with the Norwegian military; amounting to a total of 487 official reports \citep{Swords+etal:2012}.

The investigation concluded with the statement that ``There have never been any Ghost Flyers.''  However, General Pontus Reutersw{\"a}rd published his own statement that ``It could not be denied that a violation of our nation's air space has been going on'' \cite{DagensNyheter:1934}, for which he was heavily criticized.  The Swedish military later issued a final report in July 1935 concluding that 42 of the 487 reports were of actual aircraft violating the countries' borders. As a final assessment, it should be noted that very few of the sightings were similar to modern UFO sightings \cite{Swords+etal:2012}.

Reports of unidentified aircraft over Scandinavia returned in the winter of 1936--1937, prompting General Reutersw"ard to write to the Minister of Defense.  However, Social Democrat MP Elof Lindberg insisted that civil experts should investigate the intrusions, rather than the military, which was suspected of embellishing the reports for more resources \cite{PiteaTidningen:1937}.  However, Lindberg's demand for a commission was rejected by the Minister of Defence as there had been few new facts and observations \cite{Swords+etal:2012}.

Scandanavian observations of unidentified aircraft, referred to as ``ghost rockets'', resumed in the late-1940s.  As a result, they had an impact on American UFO investigations at that time \cite{USAFE:1948}.

\subsection{World War II}
Beginning in 1940, pilots from the British Royal Air Force (RAF) began reporting observations of unidentified objects in darkness, as well as unknown aircraft that would follow bombers without attacking them \cite{Rendall:2021}.  Intelligence personnel initially believed that stress caused pilots to imagine these objects.  However, by 1942, there had been a sufficient number of reports that there was some concern that these objects could be experimental devices being tested by the Germans.  There were several incidents in which RAF aircraft opened fire on these unidentified objects with no effect \cite{Rendall:2021}.

During the last years of World War II, Allied pilots, in both the European and Pacific theaters, began reporting unidentified aerial phenomena that in daylight appeared to be small metallic, sometimes translucent, spheres; and at night appeared to be spherical lights ranging from red to yellow in color \cite{SupremeHeadquarters:1944}.  These objects, which earned the nickname ``foo fighters'', would often pace the airplanes, sometimes flying right off their wingtips.  
Reports were made by pilots from the RAF, the Polish Division attached to the RAF \cite{Rendall:2021}, the Royal Australian Air Force (RAAF), the South African Air Force (SAAF) \cite{Orlandi:1994}, and the United States Army Air Force (USAAF) \cite{Orlandi:1994,Rendall:2021,Swords+etal:2012,Weinstein:2009}. 
Again, these were thought to be new Axis weapons \cite{SupremeHeadquarters:1944}.  

Three prominent scientists were enlisted to look into the foo fighter reports: H.P. Robertson (Caltech), Luis Alvarez (UC Berkeley), and David Griggs (UCLA) \cite{Swords+etal:2012}.  Griggs was tasked by General Arnold with investigating particular encounters.  By the end of World War II, Griggs had determined that there was something real to the observations in Europe and in the Pacific and that they were not related to Japanese technology. In addition, he learned that there were cases of aircraft electrical and engine interference associated with foo fighters \cite{Swords+etal:2012}. 
Although foo fighter sightings and encounters were reported to military units and were investigated by these three scientists, it does not appear that the phenomenon was ever systematically studied \cite{Berliner+etal:1995}.

\subsection{United States}
\subsubsection{UFOs and US Weather Bureau Station}
In early 1947, there were a number of sightings made by US Weather Bureau Station observers in Richmond, Virginia, that were of concern to intelligence officials \cite{Swords+etal:2012}.  There were three occasions when after scientists had launched a weather balloon and had begun to track and perform pibal (pilot balloon) observations using theodolites (spotting telescopes) that they observed metallic disc-shaped objects approaching the balloons.  On one occasion, the balloon was at an altitude of 15000 feet, and the disc, elliptical in shape with a flat bottom and a domed top, which appeared to be larger than the balloon, approached the balloon and followed it for about 15 seconds before speeding away.  Another occurrence involved a balloon, at an altitude of 27000 feet, that was similarly approached and followed by a metallic disc.  In all of these occasions, visibility was good.
This series of events affected Air Force officials, who were forced to acknowledge that the observations of these discs were made by the best trained observers to distinguish objects in the sky, using high-quality observation equipment in ideal conditions \cite{Swords+etal:2012}.

\subsubsection{The Wave of 1947}
\begin{figure}
\centering
\makebox{\includegraphics[width=0.7\columnwidth]{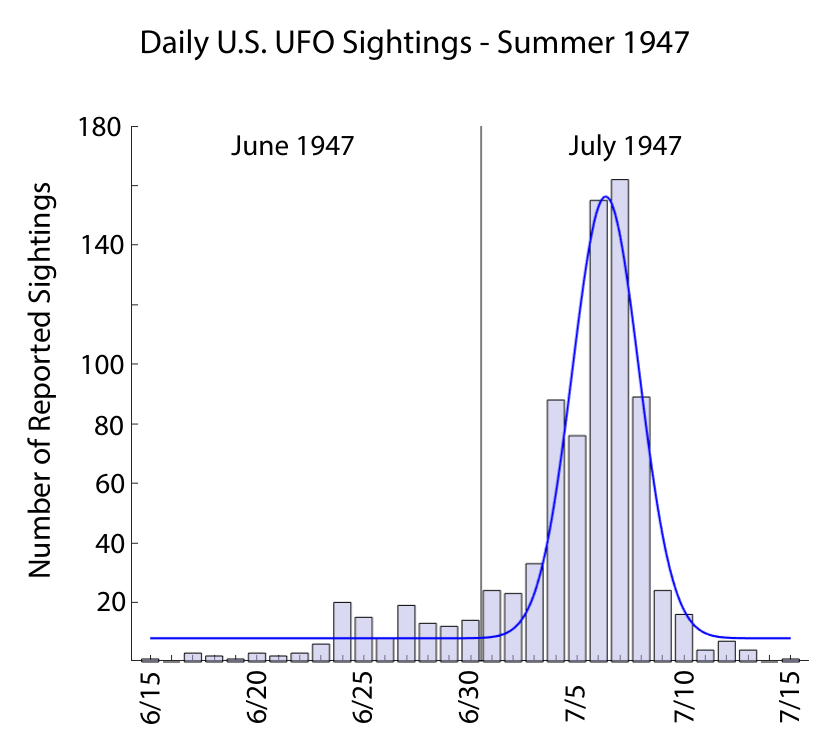}}
\caption{Reported daily UFO sightings count in the US from mid-June through mid-July of 1947 reveals a notable increase centered on July 6, 1947. Data obtained from Bloecher \cite{Bloecher:1967}.  The blue curve represents the best fit Gaussian riding on a constant rate.} \label{fig:1947}
\end{figure}

UFOs entered the social consciousness of Americans during the summer of 1947 when, during the period of a little more than a week in late June and early July, there were a large number of prominent encounters with UFOs in the United States (Fig. \ref{fig:1947}).  The 1947 Wave has been summarized in several texts \cite{Bloecher:1967}\cite{Hall+Connors:1998}, and has been the subject of Jan Aldrich's Project 1947 \cite{Aldrich:Project1947}.  In addition, the NICAP (National Investigations Committee on Aerial Phenomena) Site, coordinated by Francis Ridge, has a comprehensive chronological summary of the 1947 Wave \cite{NICAP:2018}.

\begin{figure}
\centering
\makebox{\includegraphics[width=0.9\columnwidth]{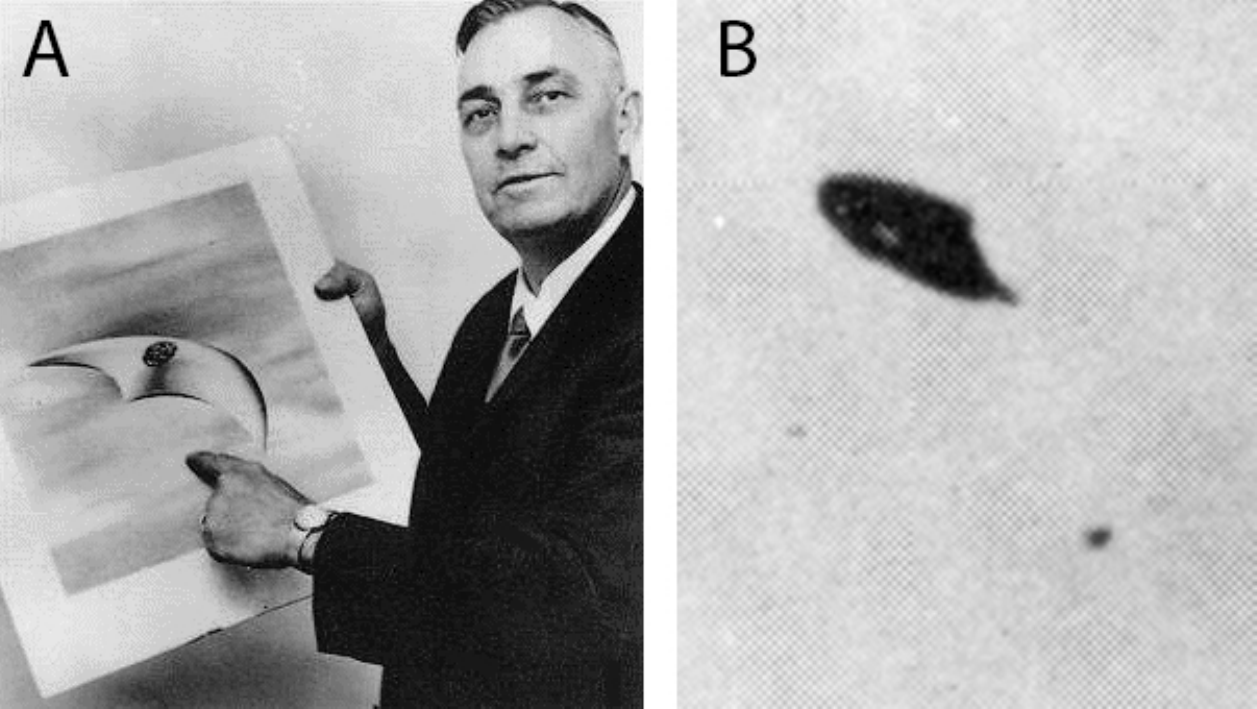}}
\caption{A. Kenneth Arnold showing an artist's rendering of one of the nine objects he observed near Mount Rainier in Washington on June 24, 1947.  B. One of the two photographs of an object observed by William Albert Rhodes on July 7, 1947 in Phoenix, Arizona.  These two objects, observed in different parts of the country within two weeks of one another, exhibit some interesting similarities.} \label{fig:arnold+rhodes}
\end{figure}

The first prominent encounter was Pilot Kenneth Arnold's sighting of a long string of nine ``circular type'' objects near Mount Rainier in Washington on June 24, 1947 \cite{Arnold:1947} (Figure \ref{fig:arnold+rhodes}A), which was independently corroborated by several other witnesses \cite{Hall+Connors:1998}.
This event brought the term ``flying saucer'' into our lexicon \cite{Lacitis:Kenneth-Arnold:2017}.  Arnold estimated the size of the objects to be about 100 feet in diameter, and he used the big sweep 24-hour clock on his instrument panel to time their 1 minute 42 second passage from Mount Rainier to Mount Adams, a distance of about 50 miles. Based on this, he estimated a speed of about 1,760 miles an hour with an uncertainty of a couple hundred miles an hour.  This is significant because this estimated speed is almost three times the world airspeed record of 623.62 mph at the time \cite[p. 274]{Angelucci:1987}.

Data compiled from newspaper articles by Ted Bloecher \cite{Bloecher:1967} reveal that the last week in June and the first week in July 1947 saw an increase in the daily average number of sightings from about $7.20 \pm 6.86$ (consistent with a Poisson distribution) to a nearly Gaussian peak of 155 sightings on July \nth{5} and 162 sightings on July \nth{6}, which represents a significant and rapid increase in the number of daily sightings of about 22 standard deviations above the June mean (Figure \ref{fig:1947}).
Although it is not clear how much of this effect was sociological, there were numerous reliable sightings made by pilots and military personnel, as well as law enforcement officers and academics, presented and discussed by Bloecher \cite{Bloecher:1967}.
We conclude this section by summarizing a number of these sightings.

On July \nth{3}, two Navy Chief Petty Officers, Robert L, Jackson and William Baker, were at the Naval Station in San Diego, California,  where they observed three disc-shaped objects over the Pacific Ocean.  The objects flew in from the West, circled, and then headed back to sea.

On July \nth{4}, Yeoman Frank Ryman, of the US Coast Guard, photographed a disc as it flew over Lake City, Washington. Also on July \nth{4}, Irving C. Allen, Chief of Airports Operations and Management in the \nth{7} Region for the Civil Aeronautics Administration, and his passenger, watched for five minutes, a disc-like object fly north across his flight path near Moscow, Idaho, as they were flying from Coeur d'Alene to Lewistown, Idaho.

United Airlines Flight Trip 105 Captain Emil J. Smith and First Officer Ralph Stevens, along with Flight Attendant Miss Marty Morrow, watched a loose formation of nine flying discs at the same level as their plane as they approached their cruising altitude of 7000 feet over Emmett, Idaho, on July \nth{4} \cite{Bloecher:1967}.  Three weeks later, on July \nth{28}, Captain Charles F. Gibian and First Officer Jack Harvey, flying the same United Airlines Flight Trip, observed an unidentified object at about 9000 feet altitude ``weaving,'' as if ``it were going through choppy air.''

On July \nth{5}, TWA pilot Captain John L. Dobberteen and copilot Frank Corwin made observations of a spinning propeller-like object moving at an estimated speed of 200 mph over Neapolis, Ohio.  

On July \nth{7} around 4:00 pm in Phoenix, Arizona, William Albert Rhodes observed and photographed an elliptical-shaped object, which appears to be similar to the objects observed by Kenneth Arnold (Figure \ref{fig:arnold+rhodes}B).  The object, which made a sound similar to that of an approaching jet, spiraled down from 5000 feet to 2000 feet and then ascended at a 45$^\circ$ angle with an estimated speed of 400 to 600 miles per hour.

On July \nth{8}, there were three independent events that involved the observation of objects of disc or sphere shape over Muroc Air Base, California. One of these events was witnessed by test pilot Major J. C. Wise as he prepared to perform a test of the XP-84 jet.  The object was spherical in shape and was flying against the wind at an estimated speed of at least 200 mph.
Also, on July \nth{8}, more than 100 Navy men at Pearl Harbor, Hawaii, watched a long silvery oblong-shaped object fly westward toward Honolulu.

At 1:00pm on July \nth{8} on Catalina Island, off the coast of Laguna Beach California, three veterans of the US Army Air Corps, Bob Jung, Kenneth Johnson and Alvio Russo, along with hundreds of other witnesses, observed six discs, in a formation of two groups of three, approaching the island from the direction of Los Angeles to the northeast, flying directly over the city of Avalon, and disappearing over the East Peak to the south of Avalon Bay.  The discs flew at high speed, estimated by two of the veterans to be about 850 miles per hour.  Jung, an aerial photographer, was able to photograph one of the discs. \cite{Catalina:1947,Bloecher:1967}

\subsubsection{The US Government's Response}
Documents, not intended for the public, have revealed that US officials were seriously concerned about this sudden increase in UFO sightings \cite{Berliner+etal:1995}:
\begin{displayquote}
    This ``flying saucer'' situation is not all imaginary or seeing too much in natural phenomena. Something really is flying around.\\
    - July 30, 1947 \cite{AFBIR-CO:1947}
\end{displayquote}
\begin{displayquote}
    The phenomena reported is something real and not visionary or fictitious.\\
    - Sept. 23, 1947 \cite{Twining:1947}
\end{displayquote}
\begin{displayquote}
    It is the considered opinion of some elements that the object [sic] may in fact represent an interplanetary craft of some kind.\\
    - Oct. 28 1947 \cite{Shulgen:1947}
\end{displayquote}
By the end of December 1947 it was ordered that USAF Air Materiel Command at Wright-Patterson Air Force Base initiate Project SIGN to establish the nature of these UFOs \cite{Cabell:1948}.

\subsubsection{Project SIGN} \label{sec:SIGN}
Project SIGN, which was established in January 1948, was the first UFO investigation of the USAF, which involved both the Air Force Office of Intelligence (AFOIN) and the Air Materiel Command (AMC) of the USAF at Wright-Patterson Air Force Base.   AFOIN consisted of two distinct executive branches, the Air Force Office of Air Intelligence (AFOAI) and the Air Force Office of Intelligence Requirements (AFOIR), which had two different perspectives on the flying disc phenomenon \cite{Swords+etal:2012}.  AFOIR was closer to the problem and having met witnesses and familiarized themselves with first-hand reports, they were convinced of the physical reality of the phenomenon \cite{Swords+etal:2012}.  AFOAI took a more hands-off approach and focused on the intelligence aspects of the problem, which led to concern about a potential Soviet threat \cite{Swords+etal:2012}.  Meanwhile, the AMC at Wright-Patterson AFB was better equipped to incorporate Project SIGN as a special project in which engineers and technical support personnel worked toward a physical understanding of the phenomenon \cite{Swords+etal:2012}.  Naturally, this led to two very different interpretations of the situation.

Project SIGN did not collect their own data on UFOs. Instead, they collected several hundred reports from both governmental and non-governmental sources \cite{Berliner+etal:1995}.  During the course of the project in 1948, several cases were reported that seemed to defy straightforward explanation.  
For example, on 7 May 1948, witnesses in Memphis, Tennessee, watched an estimated 50 or 60 metallic objects racing across the sky mainly in a straight line, but exhibited some unusual zigzagging \cite{BlueBook:Roll2,Swords+etal:2012}. Several months later, on 24 July, Eastern Airlines Flight 576, having left Houston Texas, was over Montgomery, Alabama, when Captain C.S. Chiles and Captain John B. Whitted spotted an incoming object shaped like an airplane fuselage with no wings or tail \cite{BlueBook:Rolls2+3,Swords+etal:2012}. They estimated the object to be about 100 feet in length with a barrel diameter about three times as large as a Boeing B-29 Superfortress. The object passed near the plane, slightly higher and to the right angling gently upward and away \cite{Swords+etal:2012}.

Unexplained sightings, such as these, were complicated by the fact that the Brigadier General Erik H. Nelson (who, at the time, was a technical advisor for Scandinavian Airlines) revealed that there had been unidentified aircraft observed in Scandinavia again, which were not only rocket-shaped, but also in the shape of discs and spheres \cite{McCoy:1948}.  When the Americans inquired about this, the Swedish Air Intelligence Service stated that \cite{USAFE:1948}
\begin{displayquote}
    some reliable and fully technically qualified people have reached the conclusion that ``these phenomena are obviously the result of a high technical skill which cannot be credited to any presently known culture on earth.''
\end{displayquote}

In November 1948, Project SIGN was required to produce an estimate of the situation, which stated that there was no indication that these UFOs were of domestic origin, and that while such objects could be made to fly, the power required was not attainable with current technology.  The memo went on to advance the extra-terrestrial hypothesis while indicating that there was no tangible evidence for this possibility \cite{Swords+etal:2012}.  The Pentagon pushed back against these suggestions.  The original estimate was ordered destroyed, although records were kept \cite{BlueBook:Rolls3+4,Swords+etal:2012}.  It should be noted that Eghigian casts doubt on this narrative \cite{Eghigian:2024}, pointing out that there is little to no evidence of such a memo.

The final SIGN report claimed to explain most reports \cite{Berliner+etal:1995}.  Moreover, the final report did not include the extraterrestrial hypothesis, the main contributors to Project SIGN were reassigned, and the project name was changed to Project Grudge \cite{Swords+etal:2012}.

\subsubsection{Green Fireballs and Project Twinkle}
In late November of 1948, residents of Albuquerque, New Mexico started reporting green fireballs, streaks, or flares in the sky.  At first, Air Force Intelligence at Kirtland AFB assumed that these were flares, but the number of reports and the brilliance of the lights began to increase.  On 5 December at 9:27pm, Captain Goede and his crew, flying a US Air Force C47 transport plane, saw a bright green ball of fire rise from the eastern slopes of the Sandia mountains, arch upward, and then level out \cite{Ruppelt:1956}.  As they had seen a similar fireball twenty-two minutes earlier near Las Vegas, New Mexico, they decided to report the sightings.

Not long thereafter, at 9:35pm, the pilot of a DC-3, Pioneer Airlines Flight 63, contacted Kirtland AFB reporting a large fireball just east of Las Vegas, New Mexico.  The pilot and co-pilot first saw the fireball approaching their plane head-on as they were heading westbound toward Las Vegas, New Mexico.  They saw that it was too low and had too flat of a trajectory to be a meteor, and watched the object change colors from orange red to green as it approached their plane.  As the fireball approached, its apparent size was larger than that of the full moon.  Fearing a head-on collision, the pilot took evasive maneuvers and the fireball began descending and became fainter until it disappeared \cite{Ruppelt:1956}.

The military became concerned about the number of sightings and their proximity to sensitive nuclear installations and began an investigation.  Given the similarity of the reports to what one expects from meteors, Kirtland Intelligence Officers contacted astronomer and meteor pioneer, Lincoln La Paz from the University of New Mexico.  Beginning with the events of December 5th, La Paz began locating and interviewing witnesses with the aim of reconstructing the trajectories of the fireballs.  If they were meteors, the trajectories would lead him to impact sites where the meteoritic material could be collected.  La Paz estimated that eight separate green fireballs were witnessed that evening.  Despite plotting their trajectories and estimating the positions of the impact sites, no meteorites or evidence of impacts were ever found, leading La Paz to suspect that these green fireballs were not meteorites.

Throughout December 1948 and January 1949, the number of green fireballs observed in New Mexico continued to increase and were observed by Air Intelligence Officers at Kirtland AFB, Air Defense Command officials, a number of distinguished scientists at Los Alamos, as well as La Paz, himself.
In addition, astronomer Clyde Tombaugh, the discoverer of Pluto (and more than 15 asteroids), had observed a total of three green fireballs during the wave, which he claimed ``were unusual in behavior compared to normal green fireballs'' \cite{Ledger:2004}.

In mid-February 1949, a conference was organized with Joseph Kaplan (an expert on upper-atmospheric physics) \cite{Faison:1991}, Edward Teller (of hydrogen bomb fame) and La Paz, along with military officials and scientists from Los Alamos \cite{Ruppelt:1956}.  The conference was convened to discuss the nature of the green fireballs---there was no need to discuss the reality of the phenomenon, since most everyone involved had seen a green fireball.  La Paz insisted that the green fireballs could not be meteors because their trajectories were too flat, their color too green, and that no meteoritic material had been recovered.  Teller maintained that since no sound, or sonic boom, was associated with them, they could not be meteors but must be some kind of electro-optical effect \cite{Swords+etal:2012} \footnote{It should be noted that this is the expected reasoning from a physicist who is not expecting extremely anomalous phenomena. Physicists used similar reasoning in these two papers: \cite{Loeb:2022,Loeb+Kirkpatrick:2023}}.
The conference ended with the consensus being that the green fireballs were some kind of natural non-meteoritic phenomenon.  

By spring 1949, because of ongoing concerns, Kaplan was sent to New Mexico with the news that a network of observation posts would be established along with an effort to sample the air following a green fireball overflight \cite{Rees:1949}.  While it is not clear that this plan was fully enacted, on June 24th, a green fireball flyover provided the opportunity for air sampling, which was successful.  La Paz found the sample to be abnormal as it contained copper, which could explain the green glow of the fireball.  However, meteors have almost no copper, and this suggested to La Paz that these were manufactured missiles of some kind \cite{Swords+etal:2012}.  The investigation was then passed to the Air Force's Cambridge Research Laboratory \cite{Ruppelt:1956,Swords+etal:2012}.

Another meeting involving La Paz, Kaplan, Teller, and Stanislaw Ulam, along with military officials and FBI agents, was arranged in Los Alamos on October 14.  Again, there was no agreed upon explanation for the green fireballs, but their concentration around Los Alamos and Sandia Labs was not only not natural, but ominous.  By February 1950, they established Project Twinkle to study the green fireballs \cite{Ruppelt:1956,Swords+etal:2012}.

Project Twinkle aimed to employ three 35-mm cinetheodolite stations near White Sands, New Mexico, but an insufficient budget and unavailability of cinetheodolites resulted in only one station being constructed, which was moved around, always missing the phenomenon.  Project Twinkle concluded in 1951 having learned essentially nothing \cite{Elterman:1951}.  However, the final project report appears to gloss over an important set of observations \cite{Sparks:1999,Maccabee:2002}.  Specifically, the details of the report of four objects photographed during the test of an MX 776A missile on April 27, 1950 \cite{Mitchell:1950} are missing, which states:
\begin{quote}
...\\
2. Film from station P10 was read, resulting in azimuth[\textit{sic}] and elevation angles being recorded on four objects.  In addition, size of image on film was recorded.

3. From this information, together with a single azimuth angle from station M7, the following conclusions were drawn:\\
a. The objects were at an altitude of approximately 150,000 feet.\\
b. The objects were over the Holloman Range between the base and Tularosa Peak.\\
c. The objects were approximately 30 feet in diameter.\\
d. The objects were traveling at an undeterminable, yet high, speed.
\end{quote}
Instead, the Project Twinkle Final Report merely states \cite{Elterman:1951}:
\begin{quote}
Some photographic activity occurred on 27 April and 24 May, but simultaneous sightings by both cameras were not made, so that no information was gained.
\end{quote}

La Paz continued to believe that these green fireballs were not natural.  Ruppelt, having interviewed La Paz, writes \cite{Ruppelt:1956}:
\begin{quote}
Then there were other points of dissimilarity between a meteor and the green fireballs. The trajectory of the fireballs was too flat.  La Paz explained that a meteor doesn't necessarily have to arch down across the sky, its trajectory can appear to be flat, but not as flat as that of the green fireballs. Then there was the size. Almost
always such descriptive words as ``terrifying,'' ``as big as the moon,'' and ``blinding'' had been used to describe the fireballs. Meteors just aren't this big and bright.\\
...\\
A meteorite[\textit{sic}] is accompanied by sound and shock waves that break windows and causes cattle to stampede. Yet in every case of a green fireball sighting, the observers reported that they did not hear any sound.\\
...\\
But the biggest mystery of all was the fact that no particles of a green fireball had ever been found. If they were meteorites, La Paz was positive that he would have found one. He'd missed very few times in the cases of known meteorites.
\end{quote}

It is doubtful that the green fireballs were meteors for another important reason. Meteor storms are global events that occur when the Earth, in its orbit, passes through an orbiting cometary or asteroid debris cloud.  As such, if the green fireballs were meteors, then they should have been observed worldwide {}---{} certainly not limited to a handful of sensitive nuclear sites in New Mexico, and the meteors would have only lasted a few days {}---{} not months.

Although Project Twinkle was conceived as a data collection effort, it was both underfunded and haphazard in its design.  Moving the single camera used in the study meant that the camera kept missing the phenomenon.
As a result, it provided virtually no useful information about the phenomena it was intended to study.  

The UFO activity at both the White Sands Missile Base and Los Alamos in New Mexico resulted in the formation of informal groups of scientists, in both places, to study the phenomena \cite{Swords+etal:2012}.  One member wrote: \cite{Newburger:1949}
\begin{displayquote}
    ...a group of our physicists (who) set up watches to observe and record the mysterious green fireballs.  As watching progressed, we armed ourselves with a camera and grating to try to photograph the spectrum of the light.  We set up a Doppler meteor detector ... and a low frequency electromagnetic listening and recording device.
\end{displayquote}
The group consisted of ten physicists, many of whom had witnessed these objects themselves \cite{Swords+etal:2012}.  Fred Kalbach described their efforts to James McDonald, an atmospheric physicist at the University of Arizona who studied UFOs, saying: \cite{Kalbach:1970}
\begin{displayquote}
The sightings were so numerous that a group of us staff members produced a reporting form and encouraged those who saw these events to fill out the checklist.  Over a period of say three months, we collected possibly 50 reports.\\
...\\
I have personally attempted to follow up on some of the reports of others only to conclude that there are many things which competent observers have seen which cannot be explained in terms of our present knowledge.
\end{displayquote}

Such efforts eventually paid off. In October 1949, Harley Marshall had just finished his work at the dome at Mt. Palomar Observatory in California \cite{Swords+etal:2012}.  He was driving to the powerhouse to check the weather instrumentation and the Naval Electronic Laboratory (NEL) Geiger counter, when he noticed nine highly reflective circular objects flying in a group of three triangular formations.  They moved at great speed, but without sound or signs of exhaust.  When Marshall arrived at the powerhouse, he noticed that the Geiger counter had risen to a sharp peak and had since been declining.  He wondered whether the recorded radiation spike was due to the flyover that he observed \cite{Swords+etal:2012}.

To this day, it is not clear what the green fireballs were or if they had any relation to the cylinders, discs, or spheres that had been widely observed.  As such, they are an example of the fact that UAPs are a class of unknown phenomena, and not a single thing \cite{Knuth:2023}.  For this reason, the instruments used to study them need to be sufficiently diverse to be able to provide useful information about a wide class of phenomena.

Moreover, it is a widely held belief that scientists, astronomers in particular, do not witness UFOs or UAP.  The events and efforts discussed in this section make it clear that this is not only untrue, but many of the respected scientists who witnessed these phenomena made serious attempts to study them.  In addition to illustrating both the seriousness and pervasiveness of these phenomena, knowledge of these events is critical, as they inform us of what types of equipment have been useful in studying UAP.

\begin{figure}
    \centering
    \includegraphics[width=1\linewidth]{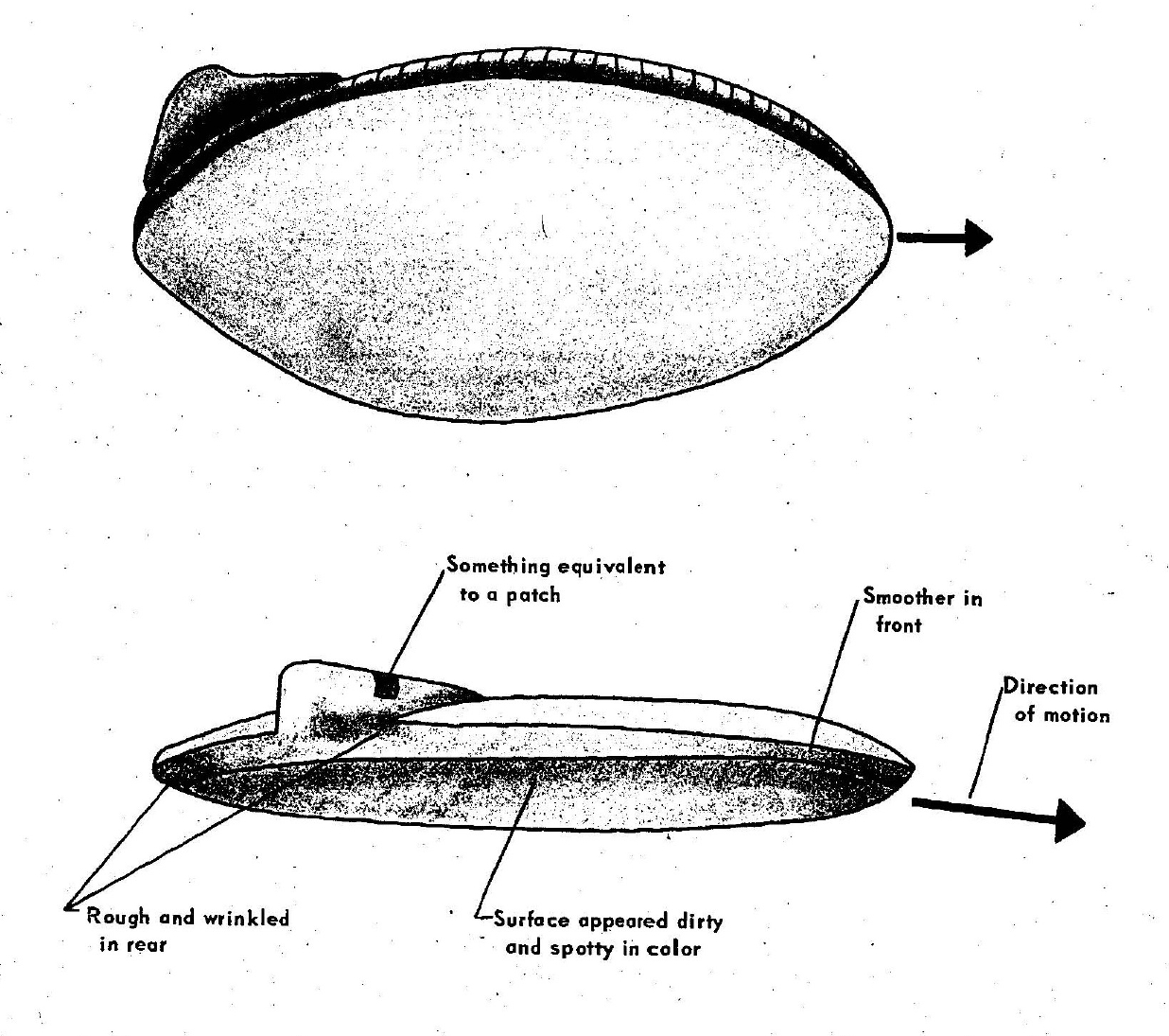}
    \caption{An illustration of the disk observed by five adults while boating on the Rogue River in Oregon in May 1949.  The observers included two researchers who worked for the National Advisory Committee for Aeronautics (NACA); one of whom worked at the Supersonic Wind Tunnel at Moffett Field, California, and the other who was a draftsman; both of whom were familiar with aerodynamic design.  Illustration from Project Blue Book Special Report 14, page 86 \cite{HeadquartersUSAF:1949, AFForm112:1949}\cite[pp. 863-879, 890-3]{BlueBook:Roll5}.}
    \label{fig:rogue-river-ufo}
\end{figure}

\subsubsection{Project Grudge} \label{sec:GRUDGE}
Project Grudge replaced Project SIGN in April 1949.  In most ways, it was a continuation of Project SIGN, with the focus on collecting and analyzing UFO reports.  However, Project Grudge was assigned new personnel along with an understanding to not consider the extra-terrestrial hypothesis.

In December 1949, Donald Keyhoe, a well-known aviation writer and retired major, published an article on UFOs based on his private studies of UFOs and his military contacts.  In this article, Keyhoe claimed that UFOs were alien spacecraft and that the US government was keeping that information secret.  Since the Air Force was more concerned about the public's perceptions, to demonstrate that UFOs were not a serious concern, they reduced Project Grudge to a routine intelligence effort \cite{Berliner+etal:1995}.

To highlight the Air Force's hostility to UFO cases, it is worth discussing a sighting that occurred in May 1949 on the Rogue River in Oregon \cite{HeadquartersUSAF:1949, AFForm112:1949}\cite[pp. 863-879, 890-3]{BlueBook:Roll5}\cite{Swords+etal:2012}.  A group of six adults were on vacation in a boat on the river when at about 5pm (it was still daylight) they observed a round object that looked like a mirror on edge.  The witnesses had binoculars that were passed around so that they all observed the object well.  The object had a diameter of about 25--30 feet, was flat on the bottom with an edge like a coin, and had a slightly convex top with what appeared to be a fin-like protrusion on one end (Fig. \ref{fig:rogue-river-ufo}).   Two of the witnesses worked for the National Advisory Committee for Aeronautics (NACA, which was the predecessor of NASA).  Mr. Don Heaphy was a mechanic at the $1 \times 3$ - foot Supersonic Wind Tunnel at the Ames Laboratory at Moffett Field, California, and the other was Mr. Gilbert Rivera, a draftsman in the Drafting Section of NACA at Ames Laboratory \cite{HeadquartersUSAF:1949}.  As such, both witnesses were uniquely familiar with aerodynamic designs. The object was observed to accelerate from a ``leisurely pace'' to jet speeds, but there was no jet stream or contrails, and the object made no noise.  At one point the object just turned, but the observers noted that it was strange how the object did not need to bank to accomplish the turn.  The project report card summarized the event, stating: ``No data presented to indicate object could have NOT been an aircraft. Conclusion: AIRCRAFT'' \cite{Swords+etal:2012}.

\subsubsection{Project Blue Book} \label{sec:BlueBook}
Project Blue Book took over in March 1952, where Project Grudge left off.  A brief period of serious investigation was instituted under its new head, Captain Edward Ruppelt \cite{Ruppelt:1956}, but eventually, as with Project Grudge, the success metrics appear to have focused on explaining away sightings as prosaic and reassuring the public that there was nothing to worry about.  Sightings and project documents were collected and archived.  They are available under the Freedom of Information Act at the National Archives \cite{NationalArchives:2024}.

\addcontentsline{toc}{subsubsection}{\hspace{0.5cm}The Robertson Panel}
\paragraph*{The Robertson Panel}
\label{sec:RobertsonPanel}
    \mbox{}\\   
    Swords et al. \cite{Swords+etal:2012} describe the CIA's involvement with UFOs as being both controversial and inadequately documented.  By spring of 1952, before the 1952 summer ``flap'' of increased UFO activity, Frederick Durant from the CIA's Office of Scientific Intelligence (CIA-OSI) collected information from the Pentagon UFO Chief Dewey Fournet and Project Blue Book Chief Edward Ruppelt on how the US Air Force was handling the UFO problem.  As a result, when after the Washington D.C. sightings in late July \cite{Maccabee:2018}, President Truman requested that the CIA be involved, the CIA was prepared with recommendations to the President the next day \cite{Clark:1952}.  Swords et al. \cite{Swords+etal:2012} detail the reconstructed efforts of the CIA in mid-1952 and summarize the situation, in part, as such:
    \begin{itemize}
  \item The CIA criticized the Air Force for poorly handling the UFO problem by trying to find prosaic explanations on a case-by-case basis, rather than trying to understand the phenomenon as a whole.
  \item The CIA determined that the phenomenon was ``frequent and worldwide'' with about 20\% of the cases considered unexplained. 
  \item Consultation with three scientists led them to believe that the solution to the UFO problem was ``just beyond the frontiers of present knowledge.''  There was some thought that the concentration of radioactive waste could play a role, suggesting that they may have been aware of the higher density of sightings at places, such as nuclear sites, such as Oak Ridge National Laboratories.  The perception was that the phenomenon was interesting and worthy of scientific study.
  \item There was concern of hysterical panic in America and it was thought that there should be a national policy on what the public could be told to minimize panic.
\end{itemize}

By November 1952, the CIA-OSI group led by Assistant Director Chadwell had come to the conclusion that there should be a UFO research project focused on determining what UFOs were \cite{Chadwell:1952a, Reber:1952, Chadwell:1952b}.  It was obvious to them that something important was going on, which required immediate attention, especially because many UFO sightings involved objects at very high altitudes and over sensitive military installations. CIA Director Smith and Assistant Director Chadwell decided to produce a National Security Council Intelligence Directive (NSCID) that would have the CIA-OSI create a scientific study of UFOs.

However, behind the scenes, two things were happening \cite{Swords+etal:2012}.  A group of MIT scientists at Lincoln Lab had become interested in UFOs because they demonstrated how poorly we were at detecting potentially dangerous airspace violations.  Their lead scientist, physicist Julius Stratton, met with Chadwell to offer their services in taking on all UFO case analyses to aim at better understanding both the phenomenon and national defense.  This incursion by a non-Air Force-controlled scientific group appears to have alarmed the Air Force's Pentagon divisions, and the USAF began countermeasures.

Exactly how the USAF managed this is not clear, as it is the poorest documented area, but it did happen \cite{Swords+etal:2012}.  Due to Air Force influence, the thrust of the Chadwell panel meeting was altered dramatically from a workshop-like affair to judge the needs and scientific value of researching UFO incursions to an assessment of the phenomenon's national security implications only.  This is why science-oriented personnel, such as J. Allen Hynek, Major Dewey Fournet of the Pentagon, Captain Ruppelt of Project Blue Book, and the photo-analysists of the Navy's photo-interpreting group came to the meeting loaded with evidence-oriented materials and left stunned that nothing they said was ``heard''.  Hynek noted: \cite{Emenegger:1974}
\begin{displayquote}
    ... one could argue that they deliberately chose high scientific-establishment men, men who were terribly terribly busy, could obviously not spend a lot of time examining things and had no intention of doing their homework, but of simply passing judgment on the basis of their previous scientific experience in, and only in, their present scientific framework.
\end{displayquote}

Nothing of scientific value came from the committee, which lived forward in history as ``The Robertson Panel'', under the name of the chairman, Howard P. Robertson, who Harvard astronomer Donald Menzel (Sec. \ref{sec:Menzel}), friend of Robertson and Panel member, described as biased against the topic from the beginning \cite{Menzel:1961}.

The Panel did have recommendations.  They were to reduce the seriousness with which people took the subject, to take the excitement and belief out of it, and to do it because people reacted badly to such events and that foreign adversaries could use that emotional weakness in the public (and even some members of the military) to their aggressive advantage.  As a result, the Air Force immediately changed policies in accordance with these recommendations.
The final report had two areas of focus: dismissing the phenomenon and making recommendations on how to control the public perception of the phenomenon \cite{Swords+etal:2012}.

In December 1953, Joint Army-Navy-Air Force Regulation number 146 (JANAP 146) made it a crime for military personnel to discuss classified UFO reports with unauthorized personnel \cite{JANAP-146}.  In February 1953, the Air Force issued Regulation 200-2 \cite{AFR:200-2}, which forced base commanders to only discuss UFO reports that had been explained, while classifying those that had not been explained.  In this way, the public perception was that the UFO phenomenon could be easily explained.


\addcontentsline{toc}{subsubsection}{\hspace{0.5cm}The Colorado Project and the Condon Report}
\paragraph*{The Colorado Project and the Condon Report}
\label{sec:ColoradoProject}
    \mbox{}\\   
When the United States was at the early stages of the large wave of UFO sightings that occurred in 1966, two nearly coincidental events occurred in Michigan, which achieved national attention.  Due to a variety of blunders, the Director of Project Blue Book, Major Hector Quintanilla, pushed Hynek to issue a statement debunking the sightings.  Hynek described the sightings as ``swamp gas'', which appeared to denigrate the Michigan citizenry involved.  This angered them and their prominent governmental representative, Speaker of the House, Gerald Ford.  Ford decided to make a public affair of his displeasure, and the Air Force was humiliated right up to the Secretary of the Air Force, Harold Brown.  Brown, having been criticized in this way, had difficulty regaining credibility.

The problem was softened by the Air Force's suggestion that they would be willing to cooperate with some independent research institution to explore the subject and report back to the public and Congress.  Thus began the Colorado Project.  The project was almost the complete reverse of a scientific research project and a grant.  The ``suggestion'' came from the military---not a scientific organization.  Moreover, the chief researcher, Edward Condon, knew little to nothing about the subject aside from being a world expert seeking a grant.  None of the chief researchers knew anything in depth about the subject.  In fact, those who could have been classified as experts were instead specifically excluded as being biased.  

As a result, the project had no research plan or really any idea of how to do the job.  After failing to ``interest'' around a dozen academic institutions (which gave the Air Force blunt refusals), the University of Colorado was somehow talked into taking on the project.  The sociology of all of this is rather complicated and is described directly from the Colorado reports in the review paper on the subject by Swords \cite{Swords:1995/96}.

Condon would only agree to lend his name as Chief Investigator if someone else could be found to do the actual administrative work and keep him informed.  The administrator was Walter Robert Orr's assistant director, Robert Low, at the High Altitude Observatory at Boulder, Colorado.  Low took on the real task of directing and did a surprisingly good job of creating a plan.

Due to a near circus of blunders and personality problems on all sides, none of Low's plans came to fruition.  Despite this, many aspects of the phenomenon were analyzed (most notably photographic cases by William Hartmann and radar cases by Gordon Thayer), which indicated that there was much unexplained and researchable about UFOs \cite{Swords:1995/96}.

Over time, the project fell apart. One must read the details to believe them \cite{Swords:1995/96}.  James Wadsworth, who was a graduate student member of the primary staff, commented: \cite{Wadsworth:1967}
\begin{displayquote}
Most of the project heads have duties in their departments and are only part-time on this.  Thank God that is the case as most of them contribute generously with the axe and have little positive to offer, much less enthusiasm. I feel like each general meeting sets the whole project back.  You would not believe the chicken-shit security-notched academic egotism that goes on.

It is as though the first concern is to protect themselves from getting tainted by the quasiscientific animal known as UFO.  By the time they have succeeded at this, their value as open-minded scientists has suffered greatly.  They are too busy maintaining a role to let loose what little creativity they have.
\end{displayquote}

Every prominently involved researcher (other than two) was of the opinion that UFOs were worthy of serious study.  Those ten persons went on the record at various times stating so. Nevertheless, Condon wrote oppositely in his conclusions and was quite emotionally opposite in his interactions with scientists.  Although not being able to ``unscientifically'' vent in print, Condon in his correspondence and other interactions makes his overemotionalism against the UFO subject undeniably clear. 
Almost everything about the structure of the organization, the choice of personnel, and the behaviors of all involved are classic examples of how not to do science.  For example, there was never even a published list of the project staff \cite{Swords+etal:2012}.

After thousands of reports had been collected, the Condon report \cite{Condon:1969}, which has been highly criticized, concluded that UFOs did not constitute a direct threat to national security and that the scientific study of UFOs was unlikely to be productive \cite{Alexander:2011}.  The Air Force used this conclusion to close Project Blue Book in 1969.

In general, the situation is best described by the excerpt from Berliner et al. \cite{Berliner+etal:1995}:
\begin{displayquote}
    While the U.S. Air Force unquestionably had the capability to investigate UFOs scientifically, there is no evidence that it has ever done so. Published reports and related documents suggest studies were hastily done, each time forced by short-term political considerations and public pressures rather than scientific inquiry.  The resulting studies were superficial at best, inept at worst.
\end{displayquote}

\subsubsection{American Association for the Advancement of Science Conference}
The same year that Project Blue Book was shut down, the American Association for the Advancement of Science (AAAS) held a symposium on Unidentified Flying Objects during their annual meeting in Boston, Massachusetts, USA in December 26--31, 1969 \cite{McMahon:1969}.  The Unidentified Flying Objects Symposium was organized by Thornton Page (NASA Manned Spacecraft Center, Houston TX USA), Philip Morrison (Massachusetts Institute of Technology (MIT), Cambridge MA USA), Walter Orr Roberts (University Corporation for Atmospheric Research, Boulder CO USA), and Carl Sagan (Cornell University, Ithaca NY USA).

The meeting was summarized as \cite{AAAS:1969} :
\begin{displayquote}
The symposium is intended to demonstrate the application of scientific methodology to a contemporary controversy and to acquaint scientists with the wide variety of facts and interpretations. It is not expected that any firm conclusion will be reached about `the correct interpretation' Presentations and discussion should be of interest to astronomers, physicists, meteorologists, sociologists, psychologists, and educators.
\end{displayquote}
The 15 papers from the symposium were published in the book ``UFO's---A Scientific Debate'' by Carl Sagan and Thornton Page \cite{Sagan+Page:1972}.  It was at this meeting that James McDonald presented his well-known paper ``Science in Default'' \cite{McDonald:1972}.  An audio recording of McDonald's talk can be heard here \cite{McDonald:1969:audio}.

\subsubsection{The Pocantico Meeting}
In the almost thirty years that had passed since James McDonald's presentation on ``Science in Default'', virtually no progress had been made to understand that recalcitrant UFO phenomenon that simply and frustratingly, refused to go away.  Laurance Rockefeller initiated a study of the UFO phenomenon led by Stanford astrophysicist Peter Sturrock (Sec. \ref{sec:Sturrock}) and a review panel of nine impartial scientists, which took the form of a scientific workshop at the Pocantico Conference Center in Tarrytown, New York (30 miles north of New York City) from September 29 to October 3, 1997 \cite{Sturrock:1999}.  

As we will describe, the Pocantico workshop stands as a refreshing and unique example of how the scientific community can be brought in, and introduced to this complicated UFO problem.  The workshop was structured so that there was a Review Panel of nine minimally-biased scientists who were introduced to the UFO research performed by scientific investigators.  Ultimately, the investigators did their best in presenting a scientific case to the Review Panel, and the Review Panel gave a fair and rational response to the presentations made by the investigators.  The result was that the Review Panel agreed that the UFO problem was a topic worthy of scientific study, and was in need of institutional support.

Sturrock described the Review Panel: \cite[p. 113]{Sturrock:1999}
\begin{displayquote}
    The nine scientists who agreed to serve on the review panel were clearly not close-minded, or they would have never have arrived at Pocantico. They were all curious to learn something about this strange subject that has been in scientific limbo for fifty years.  On the other hand, scientists are human beings, and no human being is free from prejudice. To the best of his or her ability, a scientist must strive to separate his evaluation of evidence from whatever prejudice he may have: this is not always easy, and some scientists are in this respect more successful than others.\\

    ... since it was impossible to recruit scientists with no bias whatever concerning the UFO question, I chose to select scientists whose bias would tend to be the opposite of the investigators.  the investigators clearly believed that the UFO problem deserved study... To the best of my knowledge, no member of the review panel had come to the same conclusion or, indeed, to any relevant conclusion.
\end{displayquote}

The Scientific Steering Committee consisted of homas E. Holzer, Robert Jahn, David E. Pritchard, Harold E. Puthoff, Yervant Terzian, and Charles R. Tolbert \cite[p.xi]{Sturrock:1999}.  Sturrock thanked the following scientists (the investigators):\footnote{The astute reader will note that Drs. Haines, Rodeghier, Strand, Swords, and Vall\'{e}e are all coauthors of this present work.} Richard F. Haines (Sec. \ref{sec:Haines}), Illobrand Von Ludwiger, Mark Rodeghier, John F. Schuessler, Erling Strand, Michael D. Swords, Jacques F. Vall\'{e}e (Sec. \ref{sec:Vallee}), and Jean-Jacques Velasco for providing a crash-course on the UFO topic, and the following scientists: Von R. Eshleman, Thomas E. Holzer, J.~R. (Randy) Jokipii, Fran\c{c}ois Louange, H.~J. (Jay) Melosh, James J. Papike, Guenther Reitz, Charles R. Tolbert, and Bernard Vayret for serving on the Review Panel and making a ``brave attempt to assess the topic in that very short time'' \cite[p.xi]{Sturrock:1999}.

Sturrock described the meeting: \cite[p.114]{Sturrock:1999}
\begin{displayquote}
    The proceedings began on Tuesday morning.  No one knew quite what to expect.  The investigators no doubt hoped for sincere appreciation of the results of their long efforts under difficult circumstances to address a recalcitrant problem.  The panel members probably hoped for clear, crisp presentations ... The moderators hoped that they could keep things moving... The director hoped that during the exchange each side would quickly develop an appreciation for and understanding of the other side, and that the panel members would rapidly assimilate the information that the investigators planned to present.

    No one got everything they wanted.  On the first day some got much less than they wanted.  The panel members seemed to feel that the presentations were poorly prepared and poorly presented, and the investigators may have felt that the panel members were overly critical and overly academic. ...

    At the end of the first day, the investigators were invited to leave the meeting room, so that the panel could meet in executive session... the first crisis occurred. The panel members could and did speak their minds, and it was immediately very clear that some members were concerned about what they had gotten themselves into.

    ...

    If every participant at the meeting had then gone off to his own hotel for the evening and night, the second day might have begun on a rather sour note.  Fortunately, all the participants spent a very pleasant evening together, and we convened on Wednesday morning somewhat sadder but definitely wiser than we had been when we all arrived on Monday afternoon.  No one bailed out.

    ...

    The review picked up speed on Wednesday and got into stride on Thursday.  the panel members had probably become somewhat more realistic in their expectations and were beginning to appreciate the complexity of the UFO problem.  The investigators had come to realize that they had a tough audience, and they attempted to rise to the challenge.
\end{displayquote}

The Review Panel summarized their conclusions and recommendations: \cite[pp. 120--122]{Sturrock:1999}
\begin{displayquote}
    The thrust of these presentations was that at least some of the phenomena are not easily explainable.  The panel focused on incidents involving some form of physical evidence, with clear recognition of the dangers of relying wholly on the testimony of witnesses and of the importance of physical measurements for distinguishing among hypotheses.

    It was clear that at least a few reported incidents might have involved rare but significant phenomena such as electrical activity high above thunderstorms (e.g. sprites) or rare cases of radar ducting.  On the other hand, the review panel was not convinced that any of the evidence involved currently unknown physical processes or pointed to the involvement of an extraterrestrial intelligence.  A few cases may have had their origin in secret military activities. 

    The history of earth science includes several examples of the final acceptance of phenomena originally dismissed as folk tales: two centuries ago, meteorites (then regarded as stones falling from the sky) were in this category.  the reality of ephemeral phenomena such as ball lightning and sprites was questioned until recently.

    It may therefore be valuable to carefully evaluate UFO reports to extract information about unusual phenomena currently unknown to science.  However, to be credible to the scientific community, such evaluation must take place with a spirit of objectivity and a willingness to evaluate rival hypotheses.

    It appears that most current UFO investigations are carried out at a level of rigor that is not consistent with prevailing standards of scientific research.  However, the panel acknowledged the initiative and dedication of those investigators who made presentations at this workshop, both for their efforts to apply the tools of science to a complex problem long neglected by the academic community, and for the diligence in archiving and analyzing relevant observational data.

    The panel further concluded that further analysis of the evidence presented at the workshop is unlikely to elucidate the cause or causes of the reports.  However, the panel considers that new data, scientifically acquired and analyzed (especially of well-documented, recurrent events), could yield useful information.  In this case, physical scientists would have an opportunity to contribute to the resolution of the UFO problem.
\end{displayquote}
The panel concluded with a list of recommendations \cite[p. 122]{Sturrock:1999}, which include:
\begin{displayquote}
    \begin{itemize}
      \item The UFO problem is not a simple one, and it is unlikely that there is any simple universal answer.
      \item Whenever there are unexplained observations, there is the possibility that scientists will learn something new by studying thise observations.
      \item Scientists should concentrate on cases which include as much independent physical evidence as possible and strong witness testimony.
      \item Some form of formal regular contact between the UFO community and physical scientists could be productive.
      \item It is desirable that there be institutional support for research in this area.
      \item The GEPAN/SEPRA project of CNES... in France has since 1977 provided a valuable model for a modest but effective organization for collecting and analyzing UFO observations and related data.
      \item Reflecting on evidence presented at the workshop that some witnesses of UFO events have suffered radiation-type injuries, the panel draws the attention of the medical community to a possible health risk associated with UFO events.
    \end{itemize}
\end{displayquote}

The meeting ultimately resulted in the publication of the book ``\textit{The UFO Engima: A New Review of the Physical Evidence}'', which provides a unique summary of the topic.  The volume consists of five main parts, which review the History of the UFO topic, cover twelve Presentations made at Pocantico, present several Reflections by the Review Panel on what was learned, as well as summaries of case Material.  The book is a valuable resource especially since the twelve chapters on presentations cover topics that include photographic evidence, luminosity estimates, radar evidence, the Hessdalen Project (Sec. \ref{sec:Hessdalen}), vehicle interference, aircraft equipment malfunction, apparent gravitational or inertial effects, ground traces, injuries to vegetation, physiological effects on witnesses, and analyis of debris \cite{Sturrock:1999}.  

\subsubsection{United States: The AATIP Program} \label{sec:AATIP}
There is some evidence that in the years after Project Blue Book, the United States Government continued to have programs in place that continued to collect and internally communicate information on UFOs.  For example, on Sept. 19, 1976 a UFO encounter in Tehran, Iran involving two Imperial Iranian Air Force F-4 Phantom II jets, that were dispatched to investigate an unknown object and had lost instrumentation and communications as they approached the object as well as a temporary weapons systems failure prompted a report to be sent to the US Joint Chiefs of Staff \cite{NSA:1976}.  Such reports have an origin, which has a mandate to collect the information and produce them.

On Dec. 16, 2017, the NYT published an article that revealed the existence of a \$22 million USD Pentagon program, called the Advanced Aerospace Threat Identification Program (AATIP), which investigated reports of unidentified flying objects \cite{Cooper+Blumenthal+Kean:2017}.  The program was initiated by the then US Senate Majority Leader Harry Reid (D-Nevada) in 2007 under the name of the Advanced Aerospace Weapon Systems Applications Program (AAWSAP) to study UAP \cite{Knapp+Adams:2018}.  The AATIP Program was supported by military intelligence official, Luis Elizondo \cite{Cooper+Blumenthal+Kean:2017}.  Elizondo strongly supported the program saying:
\begin{displayquote}
    AATIP (Advanced Aviation Threat Identification Program) did find a lot of stuff. This wasn’t just a one-off looking at the Nimitz incident. There were many incidents we looked at, and we looked at them on a continuing basis.
\end{displayquote}

During his time at AATIP, Elizondo worked to release three US Navy Forward-Looking InfraRed (FLIR) videos of UAPs \cite{Cooper+Blumenthal+Kean:2019}, along with Christopher Mellon, the former Deputy Assistant Secretary of Defense for Intelligence, who provided the audio of the pilot's communications:
\begin{description}
   \item[FLIR1 (The Nimitz Encounter)] \hfill \\ which documents the encounter of an F-18 jet from the \textit{Nimitz} Carrier Group in the Pacific Ocean off the coast of San Diego, California in 2004 with a white cylindrical, or Tic-Tac-candy-shaped, object. \cite{Powell+etal:2019, Knuth+etal:2019}\\ \url{https://www.youtube.com/watch?v=6rWOtrke0HY}
   \item[Gimbal] \hfill \\ which documents the encounter of an F/A-18 Super Hornet jet from the \textit{Roosevelt} Carrier Group with one of a number of objects encountered over the Atlantic Ocean off the coast of Florida USA in 2015.\\
   \url{https://www.youtube.com/watch?v=tf1uLwUTDA0}
   \item[Go Fast] \hfill \\ which documents the encounter of an F/A-18 Super Hornet jet from the \textit{Roosevelt} Carrier Group an object encountered over the Atlantic Ocean off the coast of Florida USA in 2015.\\ 
   \url{https://youtu.be/wxVRg7LLaQA?si=kkqIHBcJjlcfN-Y9}
\end{description}

During its existence, the main contractor for the AATIP program was Bigelow Aerospace through Bigelow Aerospace Advanced Space Studies (BAASS), and its predecessor National Institute for Discovery Science (NIDS), which were listed as the place of contact for commercial pilot UFO reports \cite{Greenewald:2018} freeing the Federal Aviation Administration (FAA) from having direct involvement with UFOs \cite{Hanks:2021}. In addition to collecting reports, BAASS collected data on UFOs for AATIP on Robert Bigelow's ranch near Vernal, Utah in the Uintah Basin.
Mr. Bigelow noted in an interview that:
\begin{displayquote}
    Internationally, we are the most backward country in the world on this issue... Our scientists are scared of being ostracized, and our media is scared of the stigma. China and Russia are much more open and work on this with huge organizations within their countries. Smaller countries like Belgium, France, England and South American countries like Chile are more open, too. They are proactive and willing to discuss this topic, rather than being held back by a juvenile taboo.
\end{displayquote}
The scientists working for BAASS were involved in producing a number of Defense Intelligence Reference Documents (DIRDs) \cite{Greenewald:2023} including topics such as materials \cite{DIRD:MetallicGlasses:2009}, advanced propulsion \cite{Puthoff:2012}, and space access \cite{DIRD:SpaceAccess:2010}\cite{DIRD:Wormholes:2010}.  Harold Puthoff, the Chief Scientist for BAASS described the situation as: \cite{Knapp+Adams:2018}
\begin{displayquote}
You’ve got these advanced aerospace vehicles flying around, that we don’t know where they come from, what the intent is, possibly off-world even...
\end{displayquote}

The funding for AATIP ran out in 2012, and Elizondo resigned in October 2017 in protest of ``what he characterized as excessive secrecy and internal opposition.'' Elizondo emphasized that there needed to be more serious attention to the:
\begin{displayquote}
... the many accounts from the Navy and other services of unusual aerial systems interfering with military weapon platforms and displaying beyond-next-generation capabilities.
\end{displayquote}
The AATIP Program was followed by the UAP Task Force (UAPTF) as discussed in the Introduction (Sec. \ref{sec:intro}).

\subsection{Canada}
\subsubsection{Project Magnet}
In 1950, Canada embarked on its first official investigation into UFOs, driven by the vision of Wilbert Smith, a senior radio engineer from the Department of Transport. Smith had a bold idea: that the Earth's magnetic field might hold the key to advanced propulsion technology, potentially used by mysterious flying objects commonly referred to as ``flying saucers.''

Smith's project, known as Project Magnet, was officially approved in December 1950, with the backing of Commander Edwards, the Deputy Minister of Transport for Air Services. At its core, Project Magnet sought to explore geomagnetic phenomena, not just to understand the behavior of the Earth's magnetic field but also to investigate its potential role in the propulsion systems of UAP. Smith's research was spurred by reports of UAP sightings and his belief in the possibility of extracting usable technology from this enigma, technology that might include the harnessing of geomagnetic energy.

Project Magnet was not merely theoretical. Smith set up a dedicated observatory at Shirley Bay \cite{Greer:1953}, about 20 miles outside Ottawa, equipped with an array of scientific instruments: a gravimeter to measure the Earth’s gravity, a magnetometer to detect magnetic field variations, and even an ionospheric reactor to study atmospheric conditions. This facility, dubbed the ``Canadian Flying Saucer Observatory'' by the press, became a hotspot for media attention. The observatory aimed to measure any magnetic, radio noise, or even gravitational anomalies associated with UAP sightings, in the hope of collecting concrete scientific evidence.

Smith’s work was ambitious, but his unconventional research attracted skepticism and mounting pressure from within the Department of Transport. Although he maintained that his work adhered to rigorous scientific principles, his growing emphasis on the extraterrestrial hypothesis raised eyebrows among his colleagues. The credibility of the project was further strained as the media coverage increased, leading to public interest and a flood of inquiries to the department. This negative publicity only heightened tensions between Smith and his superiors, who had begun to regret supporting such a controversial endeavor.

Despite Smith’s dedication and the meticulous experiments conducted at Shirley Bay, Project Magnet was officially terminated in mid-1954. The Department of Transport cited adverse publicity and a perception that Smith's investigations had drifted outside the scope of its mandate. However, Smith's efforts left a lasting impact, highlighting the complexities of governmental engagement with the UFO phenomenon and setting the stage for ongoing debates about the nature of UFOs and the pursuit of scientific truth.

Smith’s story serves as a testament to the challenges faced by pioneers trying to bring legitimacy to the study of UAP and the delicate balance between scientific inquiry and institutional support. Project Magnet stands as a historical case of how curiosity and hope for technological advancement met the bureaucratic realities of a skeptical world, laying the groundwork for future discussions on the trust gap between governments and the public in the search for extraterrestrial life \cite{Hayes:2019}.

\subsubsection{Sky Canada Project}
The Sky Canada Project, led by the Office of the Chief
Science Advisor (OCSA) of Canada, was launched in the Fall of 2022 to study and evaluate how UAP sightings are managed in Canada \cite{SkyCanada:2025}.  
A preliminary report was published in January 2025 \cite{SkyCanadaPreview:2025} with a full report expected to be publsihed in March 2025.

Sky Canada works to collect information from relevant federal departments and agencies, experts, non-government organizations, and a number of countries, including the G7 nations and members of the Five Eyes.  These federal departments include the Canadian Coast Guard, the Canadian Nuclear Safety Commission, the Canadian Space Agency, the Department of National Defence (DND), the Defence Research and Development Canada (DRDC), GEIPAN (France), the Library and Archives Canada (LAC), the Meteorological Service of Canada (MSC) at Environment and Climate Change Canada (ECCC), MUFON Canada, the National Research Council of Canada (NRC), NAV CANADA, the Ontario Provincial Police (OPP), the Royal Canadian Mounted Police (RCMP), S\^{u}ret\'{e} du Qu\'{e}bec (SQ), and Transport Canada \cite{SkyCanadaPreview:2025}.

The preliminary report notes that Canadians report 600 to 1,000 sightings annually, with one in four respondents claiming to have personally witnessed a UAP in their lifetime.  Sky Canada recommends that 1) data on UAPs be made publicly available to support research, 2) periodic surveys be conducted to gauge public perception and to improve services, 3) participatory science programs should be developed to enable volunteer involvement in UAP studies, and 4) publicly accessible digital tools should be made available to aid in data collection. 
 Furthermore, Sky Canada recommends developing ``partnerships with international entities dedicated to UAPs, such as AARO and NASA (U.S.A.), GEIPAN (France) and SEFAA (Chile), to share data, methodologies and best practices in UAP research and investigation'' \cite{SkyCanada:2025}.

The Sky Canada preliminary report concludes by stating: \cite{SkyCanadaPreview:2025}
\begin{displayquote}
    Adopting a science-based, collaborative approach will help
address public concerns, demystify UAPs, and potentially
reveal valuable insights into aerial phenomena that are
currently unexplained.
\end{displayquote}

\subsection{France}
\subsubsection{1940-1968: Airspace Vulnerability and Scientific Interest}
The sightings of ``ghost rockets'' in Scandinavia had also captured the attention of the French.  On May 13, 1946, the then president of the provisional government of the Republic, Georges Bidault \cite{Mesnard+Gonzalez:1996} addressed the highest authorities of French Defense. Not only was this document transmitted to the highest Defense authorities, showing the interest in the subject, but its recovery by the American military services also shows that the USA was interested in the French position.
The document reads: 
\begin{displayquote}
... faced with the result of numerous observations made, and in particular those of 9 and 10 July (more than 250 in Sweden --- a number which appears quite high and which must include engines counted several times) it is impossible to doubt that they are projectiles. The Swedish and Finnish staff are now absolutely convinced; the certain proof which would constitute an almost intact projectile has nevertheless not yet been found.
\end{displayquote}
The favored hypothesis at the time was that these ``ghost rockets'' were examples of Soviet rocket fire.  The characteristics of these devices, reported by numerous witnesses and difficult to explain prosaically by the Scandinavian authorities, were of interest \cite{Roussel+Buchi:1976} to the Bureau Scientifique de l'Arm\'{e}e de l'Air (Air Force Scientific Bureau) \cite{Bourdais:2000-2001,Sigma2TechnicalCommitte:2021}. which was created in 1945.  The Bureau was called upon in 1951 to set up a file on UFO sightings following a wave of reports that same year.

Meanwhile, echoes of the summer of 1947 wave of UFO sightings in the USA had reached the French press \cite{ScienceEttechniques:1947}.
As soon as the press picked up on the 1947 wave in the United States, and then on the waves of 1952 and 1954, a number of players from civil society, the armed forces \cite{Plantier:1953, Plantier:1958} and sciences became closely interested in the subject. They set up UAP research associations, such as the International Commission of Scientific Inquiry in 1951, later renamed Ouranos, created by Marc Thirouin, with the help of another important figure on the subject: the writer Henry-Ren\'{e} Guieu, who became head of its investigative section \cite{Canuti:2019}. These groups structured the French landscape of UAP studies. 

While the aim of this section is to present French initiatives on the subject, we can see that from the outset, UAP-related issues were perceived and shown to be international. The documents will mainly be restricted to France, but these international interactions must be taken into account. 

It should be noted that, in the 1950s, the subject was dealt with in line with the zeitgeist.  Thus, in the early years, the subject was first studied by proponents of euhemerism (myths are accounts of actual events), of which the aforementioned Marc Thirouin was also a figure during the interwar period. It wasn't until the late 1960s that a split took place between those motivated by a more scientific approach and others led by a more literary and historical approach, but generating rejection because of its proximity to esotericism.  

Jean-Ren\'{e} (Jimmy) Guieu played a role in popularizing the subject. A prolific writer, he only published two works on the subject of UFOs \cite{Guieu:1954,Guieu:1956}, but used witness observation reports as a source of inspiration for his science fiction works, a possible explanation for the connections often made in sociology between testimony and fiction work \cite{Meheust:1978}. A journalist, he hosted numerous sections on the subject on RMC and France 3 --- one of the state channels. 

Aim\'{e} Michel is one of the key figures in the birth of the study of UAP in France. Working for the national radio station since 1946, he later joined the Ouranos group. He published two reference works on the subject in 1954 and 1958 \cite{Michel:1954,Michel:1958}. In Myst\'{e}rieux Objets C\'{e}lestes (Mysterious Celestial Objects) \cite{Michel:1958}, he mentions that although he had heard of the phantom rockets of 1946 and the Mount Rainier sighting in 1947, it was the discovery of a file on unusual phenomena reported by stations, shown by the deputy director of the weather service's technical services, engineer Robert Clausse, that convinced him around 1950. He then researched who was dealing with the subject in France and discovered articles by Captain Cl\'{e}rouin, head of the French Air Force intelligence service, and Lieutenant Plantier, already mentioned above. Through the former, he met Jean Latappy, who had begun work on the subject before the Second World War. Latappy proposed an initial classification system based on the object's appearance and behavior.  Each class was associated with a specific symbol, and these were then placed on a map of France locating the observation. Thus, the parade of maps and symbols made it possible to model UFO sightings in France.

Following his discussions with these pioneers of the subject, Aim\'{e} Michel wrote his first book. Readers then contacted him, including the astronomer Pierre Gu\'{e}rin and Professor R\'{e}my Chauvin. In his second book, he describes the influence of Charles Fort concerning possible contacts with the ``Others'' and Edward Ruppelt, who allowed him and his network to understand the concept of ``wave'' during the increase in the number of cases in 1954. Aim\'{e} Michel describes how direct witnesses push those close to them, including filmmaker Jean Cocteau and General Chassin, commander of NATO Air Defense in the central European sector.

From the study of the testimonies, Michel postulated that objects, whatever they may be, emit high-intensity magnetic fields. He also noted that automatic astronomical instruments seem to capture inexplicable luminous objects. He noted that some reports appear to align geographically. 

At the beginning of the 1960s, he joined another informal research group: the Invisible College, along with Jacques Vall\'{e}e, Pierre Gu\'{e}rin, R\'{e}my Chauvin, Olivier de Costa Beauregard, and also Claude Poher (\ref{sec:Poher}). The nickname was coined by J Allen Hynek, ``in reference to the secretive group of `natural philosophers' who fought against the established dogmas of the church in the mid-seventeenth century to acquire knowledge through experimental investigation.'' \cite{Vallee:1975}

Jacques Vall\'{e}e is certainly the most emblematic personality in the study of UAP, both in France and abroad. From 1962, he tried to find models corresponding to the statistics of UFO sightings. His research was published in 1966 in ``Les ph\'{e}nom\`{e}nes insolites de l’Espace'' with Janine Vall\'{e}e \cite{Vallee:1966:phenomenes}. He also developed a UAP classification method that included both effects on the environment and witnesses.

The combination of these two sets bears a striking resemblance to the findings of research carried out by other groups, sometimes more than half a century later \cite{Elizondo:2024}, and in other countries. The authors also propose to generalize the use of statistical analysis and to install observation stations capable of recording the polarizations and the spectrum of light collected by photographing these objects. In the event that the phenomenon is too rare for it to be effective, they propose sending automatic probes to search for traces of civilizations in the solar system. 

In 1962, after the end of the pioneering organizations interested in the subject, including Jacques Baccard's Scientific Research Center, created in 1949, and the Ouranos Commission \cite{Canuti:2019}, the engineer Rene Fou\'{e}r\'{e} and his wife established the Groupement d’Etudes des Ph\'{e}nomènes A\'{e}riens (Aerial Phenomena Study Group), or GEPA, one of the first organizations devoted to the subject, which brought together the military and scientists. 

Thus, from 1962, the landscape for the study of UAP in France was established. Military observations were transmitted to army offices. In society, citizen investigators took up the problem to answer questions from the public and inventory cases \cite{Quincy:1950}, and sometimes came together as ufological associations. Military personnel and scientists interested in the subject would also mingle with these organizations. These organizations were sometimes public, going so far as to obtain the title of press organ, or private, allowing the exchange of information without fear of being mocked. 

\subsubsection{1968-1977: Creation of a Public Structure and Private Organizations}
Between 1962 and 1968, numerous figures on the subject emerged, multiplying the number of structures and research groups. According to several sources \cite{Vallee:1992, Wiroth:2016, Bourdais:2000-2001}, General De Gaulle gave his approval to the creation of an investigation group, concerned by a case of mass observation in 1954 in Tananarive \cite{GEPA:1964}, on the island of Madagascar. The creation of the investigation group was ultimately aborted in 1968, but would remain the first attempt on the part of the executive to have a structure specifically responsible for the subject. The same year, Air Force Colonel George Marey described the type of investigation carried out by the prospective-air office \cite{Marey:1968}, responsible for investigating espionage attempts by rival actors.

Since 1952, the \textit{gendarmes} (French military police) have also been participating in these field investigations, the most notable of which concerns the case of Valensole, in 1965 \cite{CNES:2015}. In 1974, journalist Jean-Claude Bourret interviewed Minister of the Armed Forces Robert Galley on the subject of UFOs. Jean-Claude Bourret is a figure in French journalism and had a column on the national radio France Inter on the same subject, which was quoted by Robert Galley during the interview \cite{Bourret:1974}.

The impact that Jean-Claude Bourret had on media coverage was significant; even today, he is one of the rare public figures on the subject in France. He notes \cite{Cometa:1999} moreover that the aversion of certain French scientists to UFOs is directly linked to the publication by the United States of the Condon report (see Sec. \ref{sec:BlueBook}).

In this interview between Jean-Claude Bourret and Robert Galley, the Minister of the Armed Forces declared \cite{Bourret:1974}: 
\begin{displayquote}
We have had a certain number of radar observations. In particular, in the 1950s, we had an inexplicable and still unexplained radar echo lasting for ten minutes.
\end{displayquote}
Asked about the close encounters between French citizens and the pilots of these machines, Robert Galley declared \cite{Bourret:1974}:
\begin{displayquote}
I must say that if your listeners could see the accumulation of information coming from the air gendarmerie, the mobile gendarmerie, the gendarmerie in charge of territorial investigations which was transmitted to the CNES by us, it is actually quite disturbing.
\end{displayquote}

These actions continued in the 1970s with the Bureau Prospectives et Etudes de l’Etat Major de l’Arm\'{e}e de l’Air (French Air Force Forward Studies Office) or BPE/EMAA, which was already responsible for processing observations made by pilots at the end of the second world war \cite{Cometa:1999}.

The engineer Claude Poher (see Section \ref{sec:Poher}), quoted in the interview by the Minister of the Armed Forces, is another of the great figures who structured the field of scientific research on UFOs. A member of the Collège Invisible and GEPA, in 1969 he was appointed Director of the Fusées-Sonde Division at CNES.

Unlike other national space agencies, CNES is a public establishment of an industrial and commercial nature. France is the fourth largest space power in the world in terms of budget, behind the United States, China and Japan, and ahead of Russia, Germany and Italy. Created by General De Gaulle in 1961, CNES made France the third country to launch a satellite into space in 1965, after the USSR and the United States. This agency is based on three different ministries, the Ministry of the Economy, Finance and Industrial and Digital Sovereignty, the Ministry of the Armed Forces and the Ministry of Higher Education and Research.

Claude Poher published in the journal ``Les Lumi\`{e}res Dans La Nuit'' a statistical study of UFO sightings in 1972 \cite{Poher:1972}. Created in 1958 by Raymond Veillith, this journal was one of the main platforms allowing researchers, witnesses, scientists, and soldiers to publish their work. It still exists today under the direction of Jean-Louis Lagneau.

According to Thomas Wiroth \cite{Wiroth:2016}, Poher stated that according to his study:
\begin{displayquote}
    UFO sightings were a worldwide phenomenon; the phenomenon presented exactly the properties expected for actually observed manifestations; the features described were distinct from anything known.
\end{displayquote}

Another interesting event that occurred in 1974 was the dissemination of recommendations by members of the Institut des Hautes \'{E}tudes de D\'{e}fense Nationale (IHEDN) proposing the creation of a UFO study office. These first recommendations, disseminated under the direction of General Blanchard \cite{Bourdais:2000-2001}, were confirmed by a report in 1977, disseminated by the Association of Auditors, under the leadership of General Richard \cite{AAIHEDN:1977}\cite[p.300]{Velasco+Montigiani:2007}. 

\subsubsection{1977-2024: Evolution of the study of UAP}
General Richard’s willingness to advance the topic, along with a renewed public interest due to a wave of observations at the beginning of the 1970s, as well as the work of the members of the Invisible College, constituted a spark: in 1977, CNES general manager Yves Sillard created \textit{Groupe d'\'{E}tudes et d'Informations sur les Ph\'{e}nom\`{e}nes A\'{e}rospatiaux Non-identifi\'{e}s} (the Unidentified Aerospace Phenomena Study Group, or GEPAN, later referred to as GEIPAN), to be led by Claude Poher. A scientific council, responsible for overseeing GEPAN, was associated with it, chaired by the president of CNES. Between 1977 and 1978, Claude Poher presented 2 studies, totaling nearly a thousand pages.

Following the departure of Claude Poher in 1978, Alain Esterle took the helm of GEPAN. Under his direction, the organization published numerous technical documents \cite{GEPAN:1981-A, GEPAN:1981-B, GEPAN:1981-C, GEPAN:1981-E, GEPAN:1981-F, Audrerie+etal:1981, Esterle+etal:1981, Jimenez:1981, Besse+etal:1982}, information notes \cite{GEPAN:1982-A, Esterle:1980, GEPAN:1981-Note-C} and working documents \cite{GEPAN:1982-A-working, GEPAN:1982-B-working, Zappoli:1982}. Alain Esterle left the group in 1983 and was replaced by Jean-Jacques Velasco, who led it for 20 years. Under his direction, GEPAN was renamed the Atmospheric Reentry Phenomena Expertise Service (SEPRA) in 1988 \cite{Velasco:2010}. The structure operated with a network of investigators making it possible to meet witnesses after they have given testimony to the gendarmerie, and after an initial field investigation by the \textit{gendarmes}. The GEPAN and then SEPRA investigators had specific equipment, including the SIMOVNI developed in part by Mr. Velasco, making it possible to simulate in the field what the witness observed, as well as an inclinometer to measure the elevation of the object in relation to the visual field of the witness. Although GEPAN enjoyed a certain level of independence \cite{Esterle:1981}, CNES decided to withdraw the organization’s ability to conduct further scientific studies, after some failed to achieve their goals \cite[p.98--99]{Rossoni+etal:2007}. It seemed that all GEPAN could do now was collect data. The scientific programs were now ordered by the CNES Scientific Programs Committee. However, even though GEPAN / SEPRA could not finance them, it could help other research actions \cite{Aguavo+etal:1983}.

In February 1995, General Denis Letty organized a conference at the Air School on unidentified aerospace phenomena. At the same time, he contacted General Bernard Norlain, then director of IHEDN, and obtained his support in the creation of a ``Comit\'{e} des Etudes Approfondies'' (Committee of In-depth Studies) (COMETA), on the subject of UAP. In 1999, the result of the work of this committee formed under the associative regime was published as a report after validation by the authorities. Bringing together military personnel, defense engineers and scientists, the COMETA association delivered its work under the title ``UFOs and Defense, what should we prepare for'' \cite{Cometa:1999}. Based on case analyses by French and international pilots, but also observations by witnesses, including cases of close encounters, the report highlights national security, but also technological and societal issues.

The COMETA Report notably lists technologies that could explain machine performance, witness paralysis, stopping of engines of land vehicles, propulsion systems that allow movement on an interstellar scale - noting that new technologies could be developed in the future that would better explain the observed kinetics of these devices \cite{Cometa:1999}.

The report addresses the different hypotheses, considering, for example, attempts to explain observations such as collective hallucinations as unscientific. It also lists the possibilities of technological breakthroughs by rival nations, attempts at disinformation, holographic image emissions, and unknown natural phenomena as not sufficient to explain the performance of UAPs and being unrealistic. The document discusses the hypothesis of an extraterrestrial origin at length, considering it as the least improbable \cite{Cometa:1999}. 

According to Yves Couprie and Egon Kragel \cite{Couprie+Kragel:2010}, the COMETA report was submitted by Jean-Jacques Velasco to Prime Minister Lionel Jospin in 1999. On 16 July, it was broadcast in a special edition of the weekly magazine VSD with a circulation of 70,000 copies, thanks to the journalist Bernard Thouanel \cite{Canuti:2007}. 

In 1998, Jacques Vall\'{e}e (Sec. \ref{sec:Vallee}) wrote an article on the composition of materials supposedly found near UAP \cite{Vallee:1998}.

In February 2000, CNES decided to remove the monitoring of atmospheric re-entries from the mission of SEPRA (renamed for the occasion Service d'Expertise des Ph\'{e}nom\`{e}nes Rares A\'{e}rospatiaux), whose mission was to refocus on collecting information on UAP sightings and building up a database of eyewitness accounts.

CNES then decided to follow the recommendations of an external audit of SEPRA, the LOUANGE \cite{SEPRA:2001} report, commissioned by the CNES Presidency in 2001, and to reorganize SEPRA by creating the Groupe d'Etudes et d'Informations sur les Ph\'{e}nom\`{e}nes A\'{e}rospatiaux Non identifi\'{e}s (Group for Studies and Information on Unidentified Aerospace Phenomena) or GEIPAN in September 2005, while a Comit\'{e} de Pilotage des Etudes et de l'Information sur les Ph\'{e}nom\`{e}nes A\'{e}rospatiaux Non identifi\'{e}s (Steering Committee for Studies and Information on Unidentified Aerospace Phenomena) or COPEIPAN was set up to succeed the GEPAN and SEPRA Scientific Committees. GEIPAN's mission is to collect, analyze, archive, and publish aerospace observations and phenomena not explained by witnesses. The information mission gives GEIPAN a new central dimension.

Under the impetus of its first director, Jacques Patenet, GEIPAN launched a vast project to sort and digitize its archives, as well as a methodological overhaul aimed at presenting the public with complete and coherent files. In 2009, with 30\% of the investigation files ready for publication on the GEIPAN \cite{GEIPAN:2024} website created for the occasion, the group was authorized to open its archives to the public, an unprecedented initiative that marked a major step forward in the transparency and accessibility of data on unidentified aerospace phenomena. This groundbreaking decision quickly attracted worldwide media attention, testifying to the widespread interest in UFOs and UAP.

In the early days, the department enjoyed good support from CNES management and most of the employees had a positive view of it. However, public attitudes varied, from opposition to strong interest, and even increased demand for in-depth investigation of the UFO phenomenon. 

During this period, GEIPAN continued to evolve, organizing the management of testimonies and improving procedures for collecting them by telephone or through statements made in the gendarmerie and later by email. The service continues to be structured, replacing the \textit{Inspecteurs de Premier Niveau} (First Level Inspectors) (IPN) with a corps of volunteer investigators drawn from the general public. Recruitment, training, and operating rules were rapidly implemented, resulting in a national network of about 15 investigators who could be commissioned by GEIPAN to carry out investigations.

Building on the initiatives taken by GEPAN, CNES / GEPAN has developed and formalized strong partnerships with local authorities to facilitate field investigations and obtain accurate and reliable information. The partnerships initiated from the outset with the Gendarmerie Nationale have been extended to include several national institutions such as METEOFRANCE, the \textit{Centre National des Op\'{e}rations A\'{e}riennes} (National Air Operations Center) or NOA of the French Air Force, the \textit{Centre Op\'{e}rationnel de la Surveillance de l'Espace} (Space Surveillance Operations Center) or COSE at the CNES, the DGAC for domestic air traffic, the CNRS for observations related to astronomy, lightning and plasmas, as well as university experts in scientific, cognitive and clinical psychology.

GEIPAN has its own investigative tools, such as image processing software to authenticate photographs and videos supplied by witnesses \cite{IPACO:2024}, and interfacing algorithms dedicated to extracting information from these media.

GEIPAN also coordinates a network of multidisciplinary national experts in the fields of aviation, aerospace, and psychology, who are regularly consulted in difficult investigations, or to give their opinion on the classification to be assigned to a case after investigation.

This classification of UAP after investigation has been slightly revised since the inception of GEPAN, and has remained stable since 2005:
\begin{enumerate}
    \item phenomenon perfectly identified
    \item probably identified but lacking evidence
    \item phenomenon not identified due to lack of data
    \item phenomenon not identified after investigation
\end{enumerate}

GEIPAN is considered an expert service of CNES, which reports to the CNES \textit{Technique et Num\'{e}rique} executive department. The general public's interest in UFOs and UAP makes it a showcase for the media. The early integration of the CNES Communications Department into COPEIPAN has boosted GEIPAN's visibility, while ensuring that the media authorized to communicate with the group are properly channeled. The media have also been encouraged to turn to the website and its informative and educational features, which have gradually reinforced the dissemination of rigorous and transparent information on UAP. GEIPAN also provides information to the general public and students through public conferences.

In April 2007, Vall\'{e}e published an update to his case classification method \cite{Vallee:2007}. Based on his work in 1962 \cite{Vallee:1963}, it is the result of nearly four decades of research on the extraction of usable information from witness testimony. Unlike other methods of classification, which focus on the witness, the one presented in this article seeks to categorize the behavior of the object perceived by the witness. 

In June 2007, the French Aeronautical and Astronautical Association (3AF) created a UAP commission under the chairmanship of Alain Boudier. The commission was structured under the name Sigma in May 2008, and produced its first internal report in 2010. The chairmanship changed in 2013 to Luc Dini \cite{Sigma2:2024}. 

In 2014, GEIPAN organized a workshop \cite{GEIPAN:2014} on methods and tools that are likely to improve the Collection and Analysis of Information on Unidentified Aerospace Phenomena (CAIPAN). During the \nth{8} and \nth{9} of July, ufologists and scientists exchanged their expertise, both astronomical and psychological. The subjects presented were, among others, survey methodologies among air personnel and air traffic operators, statistical studies based on observation reports, witness interview techniques, the use of sound data, video expertise, the establishment of databases, surveillance of the sky and space by the Air Force, the use of sky observation cameras within the FRIPON network, and the capabilities of detections of UAP.

In 2021, the 3AF-Sigma 2 technical commission published a report \cite{Sigma2:2024} of 377 pages of technical analyzes on UAP cases, and has since organized annual conferences on the subject. More information about this group and their work will be presented later in the article. 

In March 2021, the Defense Historical Service announced the opening of the UFO archives \cite{ServiceHistoriqueDefense:2021}, made up of archives of investigations carried out by the Gendarmerie between 1954 and 1983.

In 2022, CNES organized a second CAIPAN conference \cite{GEIPAN:2022}. This conference covered two themes: the methods and surveys on the one hand, and on the other hand, the tools and knowledge \cite{Friscourt:2022}. During this conference, Jacques Vall\'{e}e proposed new avenues of study, including the examination of debris recovered using multiplexed ion ray imaging technology developed by Garry Nolan \cite{Nolan+Vallee:2022}, but also the use of artificial intelligence to search for matches in UAP case databases.

On the \nth{10} of March 2023, The Debrief published the reply of a spokeswoman from ONERA (Office nationale d'\'{e}tudes et de recherches a\'{e}rospatiales) (National Office for Aerospace Studies and Research) to a question on the subject of UAPs: \cite{Friscourt:2023}
\begin{displayquote}
As far as ONERA is concerned, its research work is focused on identified aerial phenomena for defense purposes essentially...\\
We have programs to develop instruments for this purpose, which could potentially pick up unidentified phenomena or detect `things’ moving in unexpected orbits, but that would be the end of it.
\end{displayquote}

On June 9, 2023, Jean-Claude Bourret and physicist Patrick Marquet published a book: \textit{Les Ovnis voyagent dans le temps} (``UFOs travel in time'') in which the authors support a hypothesis explaining the anomalies observed by UAP witnesses through the use of distortions of space-time. In this book interview, the reader finds a bibliography of scientific articles supporting this possibility \cite{Bourret+Marquet:2023}.

On the \nth{11} of October 2023, responding to a citizen question via the Agora relay platform, the Minister Delegate in charge of Industry Roland Lescure declared: \cite{Friscourt:2023:Minister}
\begin{displayquote}
We have approximately 700 declarations per year of these UAP. Of these 700, there are 97(\%) which are very well identified, it could be a satellite, it could be a reflection of the moon, it could be, frankly, someone who had a few drinks and declared something that didn't exist.
There are 3\% who are not identified today. They are - they are not UFOs, they are not Martians but they are phenomena which are classified for reasons of national defense. So let me reassure you: we are completely transparent.
\end{displayquote}
The minister's office then informed the Parisien journalist Ga\"{e}l Lombart that the minister made a mistake during his declarations. The perfectly identified cases represent 24.5\% of cases, and not 97\%, and very abnormal cases are not under Defense secrecy \cite{Lombart:2023}.

On the \nth{2} of July 2024, IHEDN organized a round table on UAP to provide an overview of its international developments and its challenges \cite{Tronche:2024}. 

The Sentinel Lab initiative plans to launch, in January 2025, a mobile UAP detection platform, which has been in the testing phase since the summer of 2024 \cite{Sentinel:lab}.  As we have seen, the desire to record data scientifically goes back to the Second World War, and while the deployment of mobile investigation platforms has a long tradition in the territory, the development of new surveillance capabilities using wide-field cameras (high sensitivity and high image refresh rate at a lower cost) represents an interesting and important development.

This brief chronology only reflects a fraction of the work that has been carried out in France.

\subsection{Russia}
In an interesting symmetry with the US, the scientific study of UAPs in the Soviet Union and then in Russia went through a series of advances and setbacks, depending on the political considerations of the state and the scientific consensus of the time.

According to cosmologist Felix Ziegel \cite{Ziegel}, one of the pillars \cite{Ziegel:1968} of the scientific study of UAPs, it began in 1946 with the proposal by engineer A.~P. Kasantzev that the Tunguska event in 1908 could correspond to the crash of an extraterrestrial craft, triggering fact-finding missions to the site.

In 1956, Yu.~A. Fomin, senior lecturer at the Department of Automation at the Moscow Technological Institute of the Food Industry, began collecting the first reports of sightings over Soviet territory. However, in 1960, the Moscow Planetarium told civilians asking about their UAP sightings that the phenomenon they observed was ``apparently related to one of the experiments carried out to measure the density of the atmosphere at high altitudes, with the
launch of a sodium cloud''.

In 1961, in an article in the national daily \textit{Pravda} written by the physicist Lev Artsimovich, he declared that UAPs ``exist to the same extent as reflections on water or a rainbow in the sky, i.e. only as a play of light in the atmosphere. Everything else is either self- deception or a deliberate falsification of facts.''
As a result, Fomin was expelled from the All-Union Society for the Dissemination of Political and Scientific Knowledge, which put the brakes on the study of UAPs. 

In 1967, B.~P. Konstantinov requested and received reports of UAP sightings from the Minister of Civil Aviation, E.F. Loginov, to
accompany the publication of a book on the habitability of space. The media coverage rekindled interest in UAPs, and a UAP study group was established, electing Major General P.~A. Stolyarov as its head. Its aim was to collect reports of sightings of UAPs and to create a public committee to study UAPs. This was achieved the same year with the support of the authorities. Following an appeal to the public, an influx of reports arrived, but shortly afterwards the president of their supervisory authority decided to dissolve the UAP study structure. A few days later, speakers from the Moscow planetarium gave presentations explaining that UAPs were nonexistent and that the study groups on the subject were dangerous. It should be noted that at the same time, the United States was also trying to find out what the USSR knew about UAPs \cite{CIA:1967}.

Following a request by Lev Arsimovich, the Department of General and Applied Physics of the USSR Academy of Sciences published a resolution condemning the study of UAPs.  This cycle of efforts by individuals to establish a scientific study structure with the backing of the authorities, the setting up of a group, and its subsequent dissolution, often by state action, was to be repeated many times. 

According to the 2021 Sigma 2 Report \cite{Sigma2TechnicalCommitte:2021}, Russian studies over the period 1950--1990 show a peak in research in the period 1970--1986 with a civilian program (SETKA-AN) for the study of phenomena called `\emph{anomalies}' and a military program (SETKA-MO) for defense work concerning the sensitivity of defense systems to these phenomena or the use of research for the purpose of military technological fallout.

The Russian-American agreement concluded in 1971 \cite{Bureau-of-International-Security-and-Nonproliferation:1971} relating to the risks of accidental launching of nuclear weapons specifies both the detection of unidentified (unknown) objects and the risks induced by possible interference affecting missile launch facilities. 

Without mentioning the causes of such phenomena, this demonstrates a mutual concern with regard to risks that have nothing to do with the risks of characterized intrusion into the airspace or of a preventive strike of the installations, but rather with the need to put in place alert and mutual information procedures (``red telephone'') from the beginning of the 1970s. This concerns the risks affecting the launch bases, which could lead to firing missiles caused by a misinterpretation linked to the presence of an unknown object, or by the effect of interference.

The initial meeting of the civil SETKA-AN and SETKA-MO research programs (one on physics, the other on military effects and applications) took place on October 18, 1978 at the Academy of
Sciences with its main protagonists, among which were:
\begin{itemize}
\item The Izmiran Institute, at the Academy of Sciences, in charge of terrestrial magnetism and the diffusion of radio waves: Prof. Vladimir Vasilyevich Migulin and head of research Yury Victorovich Platov, Izmiran being the leading institute for the academy of sciences.
\item The Institute of Space Studies of the USSR Academy of Sciences, represented by Prof. Georgiy Stepanovich Narimanov and Inna Gennadyevna Petrovskaya.
\item The Moscow Institute of Technology, represented by Prof. Rem Gennadiyevich Varlamov.
\item The Department of General Physics and Astronomy of the USSR Academy of Sciences, represented by Prof. A.~N. Makarov.
\item The Schternberg State Institute of Astronomy represented by Prof. Lev Mironovich Gindilis, who can be found cited in the archives of GEIPAN \cite{Esterle:1980}.
\item The NI-22 or 67947 military research unit: M.~M. Victor Petrovich Balashov and Vladimir Ivanovich Volga.
\item The air defense forces represented by Colonel Zaytsev.
\end{itemize}

Among the military works, the role of the Soviet navy seems to have been important in view of the very numerous observations of flying or aquatic objects (nicknamed Quakers) brought together by their reinforced observation network between 1977 and 1980.

The Soviet Navy was also involved from 1977 (07/10/1977) in research on abnormal phenomena. It issued guidelines for investigating and gathering information on UFOs to fleets and flotillas, by order of Admiral Smironov, Deputy Chief of Staff of the Navy. This was then extended on January 20, 1978, by Vice Admiral Ivanov, Chief of the Intelligence Directorate of the Navy, who gave instructions to the Oceanographic Commission, with the assistance of V.~G. Ajaja. According to the KGB, mention is made of cases of material collection with particular properties on certain sites such as Dalnegorsk. This work seems to have stopped or to have been greatly slowed down after 1990.

In 1979, a study \cite{Gindilis+etal:1980} carried out by the PR 473 Space Research Institute, part of the USSR Academy of Sciences, and written by L.M. Gindilis, L.G. Petrovskaya and D.A. Menkov, presented a statistical study of UFO sightings. The study concludes: \cite{Gindilis+etal:1980} 
\begin{displayquote}
    A certain portion of the observations may be due to various technical experiments in the atmosphere and space near the Earth, to observations of space technology objects in particular. However the kinematic characteristics exclude the possibility of such an explanation for at least one third of the cases.
\end{displayquote}

Meetings on UAP events between Russian and Chinese experts in the early 1990s were mentioned in CIA archives on UAP published in 2016 and the holding of academic conferences around 1994.

Observation on kinematics has since been reinforced by other strange characteristics such as electromagnetic emissions at certain frequencies of artificial origin. They have been noted in the past by the Americans and by the Russians in their respective reports.

In 2023, the Keldysh Institute of Applied Mathematics announced the launch of a UAP study program \cite{Bronson:2023}. The Institute was founded in 1953 and was involved in ``national projects of space exploration, atomic and thermonuclear application'' \cite{Keldysh:2024}. Witnesses of unusual objects or phenomena were asked to send videos and photos. Some papers related to UAP research can also be found in scientific journals \cite{Petrovich:2015}.

\subsection{China}
The studies of UAP have been quite similar in China. Depending on the period, the associations collecting testimonies, directly under the authority of the party or the military forces, grew in importance until the authorities decided to dissolve them. Culturally, folklore can often be related to UAP observations \cite{Vallee:1969:Magonia}, and, much like Russia, the media is particularly quick to cover national and international news on the subject as long as it does not involve national security.

One notable exception occurred in 2015. A Chinese initiative, supported by Russia and led by Jin Fin, was created, which brought together an international panel of ufologists. The initiative was financed by China and relayed in the West by Roberto Pinotti and Donald Schmitt. Its first event was the ``Five Continent Forum'', in Moscow, in 2018. Its goal was the creation of an international working group on UAP with national representatives and the signing of an oath of allegiance to the organization stating that the ``UAP Phenomenon was real and that we were dealing with an extraterrestrial / non-human intelligence.'' Shortly after, Chinese commitments failed to be achieved and the group continued its work independently under the name International Coalition for Extraterrestrial Research (ICER) \cite{ICER:2023}.

Another exception occurred in 2021. Shortly before the release of the US Department of Defense 2021 UAP report, in the English-language Chinese media, the South China Morning Post (SCMP), Stephen Chen wrote \cite{Chen:2021}:
\begin{displayquote}
According to Wuhan-based researcher Chen Li from the Air Force Early Warning Academy, human analysts have been overwhelmed in recent years by the rapidly mounting sighting reports from a wide range of military and civilian sources across the country.\\
...\\
The PLA’s [(People's Liberation Army)] task force dedicated to the unknown objects increasingly relies on AI technology to analyze its data, according to Chen’s report, which is in line with several other military studies published in domestic journals, most recently in August last year.\\
...\\
According to Chen and his colleagues, the PLA has a three-tier reporting system to handle unknown aerial objects. The base level, which includes military radar stations, air force pilots, police stations, weather stations, and Chinese Academy of Sciences observatories, is responsible for gathering as much raw data as possible.

The information is processed in mid-tier by the PLA’s regional military command which conducts preliminary analysis and transfers the data to a national database.

With the help of AI, PLA headquarters assigns a ``threat index'' to each object based on its behaviour, frequency of occurrence, aerodynamic design, radioactivity, possible make and materials,
along with any other information.

The AI can pull together other information which may help determine an object’s purpose.
\end{displayquote}

Interestingly, the author adds: \cite{Chen:2021}
\begin{displayquote}
The only officially confirmed UFO sighting in China occurred on a military airbase in Cangzhou, Hebei province, on October 19, 1998. According to a report in the Hebei Daily, the official newspaper of the
province that neighbors Beijing, two military jets were ordered to intercept a low-flying object that suddenly appeared above the airbase. 

The object looked like a ``\emph{short-legged mushroom}'', with two beams of light shooting down from its belly. When the jets approached, the object climbed with ``ghostlike'' speed to an altitude of more than 20,000 meters, before disappearing from radar and visual contact.
\end{displayquote}

Since then, SCMP \cite{SCMP:2023, Nicholson:2023, Ryall:2024} has steadily covered the topic, showing not only China's interest in UAPs but also a desire to communicate to the international audience that it is interested in this topic. The February 2023 ``balloons'' event in the United States only rekindled that interest \cite{Lo:2023, Bloomberg:2024, Chen:2024}.

\subsection{Governmental Studies Summary}
It really is not possible to describe all the efforts to study UAP in the last century in this paper.  We covered some of the most prominent, influential, and representative examples, such as those efforts made by the United States, France, at least one effort in Canada, as well as having summarized the activities of Russia and China.  As a result, we neglected to discuss the efforts made by the United Kingdom, Australia and New Zealand, as well as Asian countries, such as Japan and Korea, and Central and South American countries, such as Argentina, Brazil, Chile, Costa Rica, Columbia, Mexico, Paraguay, Peru, and Uruguay, several of whom have governments that have dedicated themselves to an unrivaled level of transparency on this topic.  For example, open official efforts such as Centro de Identificaci\'{o}n Aeroespacial (CIAE) in Argentina, which was created in 2019 after a restructuring of the earlier Comisi\'{o}n de Estudio de Fen\'{o}menos Aeroespaciales (CEFAE), have been in place for more than a decade \cite{MinisterioDeDefensa:Argentina:2024}.  Similarly, the Chilean government has been extremely transparent, having both declassified and released documents \cite{Greenewald:Chile:2023} and military videos \cite{Kean:2017}.

There are many details that the above presentation has omitted, in part because there is much that is not known.  For example, whereas many of the early US government programs to study UAP were focused on compiling and investigating sightings, there is some evidence that there were government-supported activities to develop new methods to scientifically collect independent data on UFOs.  For example, an article from a 1952 issue of \textit{Look} magazine notes that under the direction of the Air Force Air Technical Intelligence Command, a UCLA physicist was developing a camera system using a diffraction grating to capture both imagery and spectra of UFOs, and that this camera system was to be used in conjunction with cine-theodolites, radar detection systems, and Navy sonar systems to collect hard scientific data on UFOs that would be presented to a board of scientists for study \cite{Moskin:1952}.  

The mention of Navy sonar systems in the 1952 \textit{Look} magazine article, above, suggests that as early as the 1950s, there was an awareness of the submarine nature of UFOs long before transmedium capabilities were noted by the AATIP program. This neglected aspect of UFOs had been indicated and discussed by some UFOlogists \cite{Sanderson:1970, Feindt:2016, Dennett:2018, Dolan:2025} and emphasized in the recent paper by Admiral Gallaudet \cite{Gallaudet:2024}.  This important characteristic of UAPs is summarized in Section \ref{sec:trans-medium}.

\section{UAP and Nuclear Weapons} \label{sec:nukes}
One of the most disturbing and important aspects of UAP is the fact that they have been observed for decades to be active around nuclear power plants and nuclear weapon sites.  Robert Hastings has been studying UFO incursions at nuclear sites for more than fifty years (since 1973), having interviewed over 150 former and retired military personnel involved in such cases in addition to having identified and acquired supporting declassified and FOIAed (Freedom of Information Act) government documents from the U.S. Air Force, FBI, and CIA, which confirm the relationship between UFOs and nuclear weapons.  Hasting's research \cite{Hastings:2017} provides important details that highlight the extent and significance of these events.

\begin{figure}
\centering
\makebox{\includegraphics[width=0.95\columnwidth]{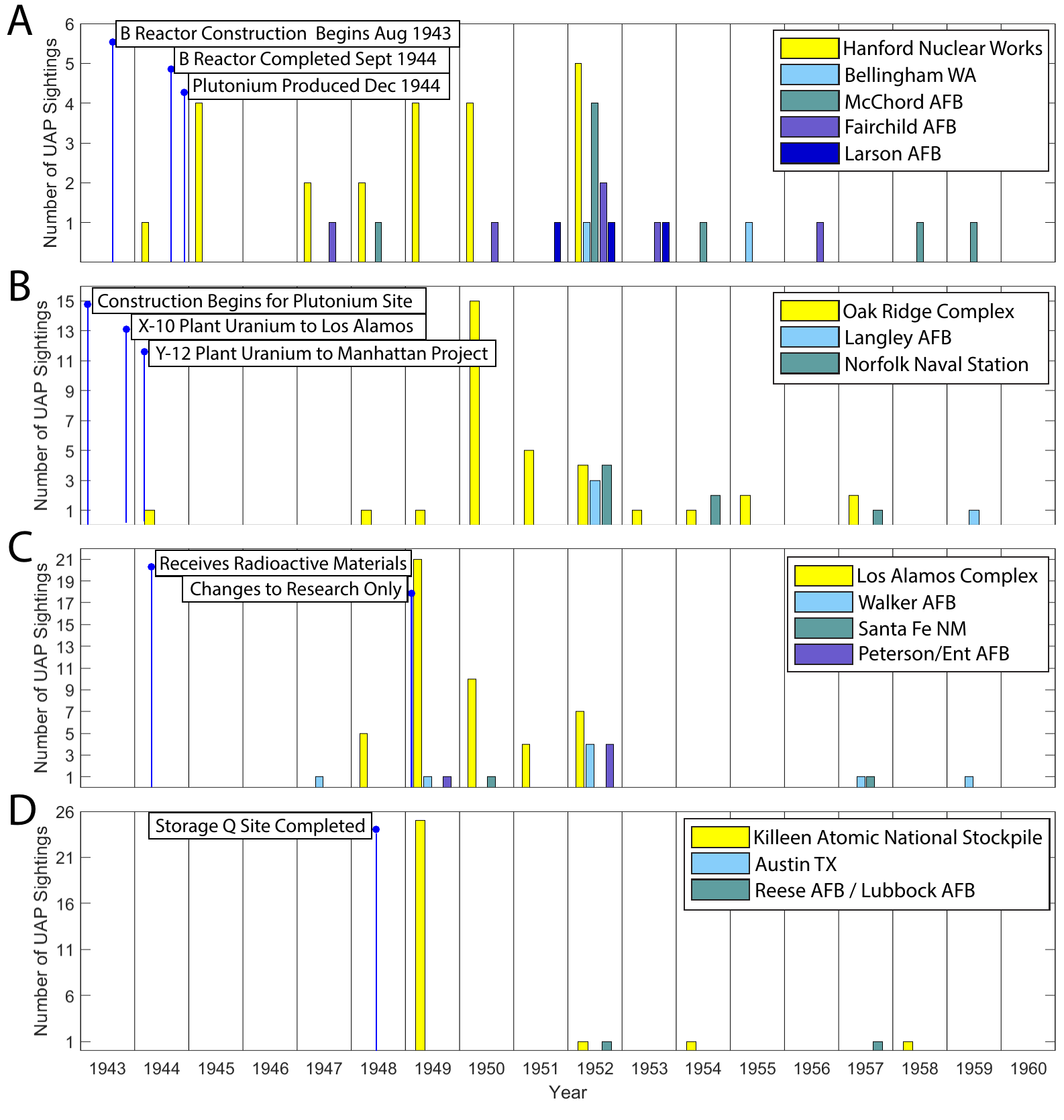}}
\caption{This figure shows histograms of UAP sightings at four different nuclear weapon complexes (top, yellow) and their nearby control sites from 1943 to 1960 (Adapted from: \cite{Hancock+etal:2023:pattern}).  A. The Hanford Nuclear Complex in Pasco Washington was a production facility for radioactive material.  The control sites were the nearby Bellingham, Washington population center, McChord Air Force Base, Fairchild Air Force Base, and Larson Air Force Base.  B. The Oak Ridge Complex in Oak Ridge Tennessee was also a radioactive material production site.  Its control sites were the nearby Langley Air Force Base and the Norfolk Naval Station.  C.  The Los Alamos Complex in Los Alamos, New Mexico was a nuclear weapon assembly facility.  Its control sites were the Santa Fe, New Mexico population center, Walker Air Force Base and Peterson/Ent Air Force Base.  D. The Killeen Atomic National Stockpile was a nuclear weapon storage area near Austin Texas.  Its control sites were the Austin population center and the Reese and Lubbock Air Force Bases.} \label{fig:atomic-sites}
\end{figure}

For example, a January 1949 memorandum to Edgar J. Hoover, the Director of the U.S. Federal Bureau of Investigation (FBI), titled ``Protection of Vital Installations'' \cite{FBIMemorandum:1949} is focused on ``unidentified aircraft'' observed at vital military facilities:
\begin{quote}
At Weekly Intelligence Conferences of G-2, OHI, OSI, and F.B.I., in the Fourth Army Area, Officers of the G-2, Fourth Army have discussed the matter of `` `Unidentified Aircraft' or `Unidentified Aerial Phenomena' otherwise known as `Flying Discs', `Flying Saucers', and `Balls of Fire'.  \underline{This matter is considered top secret} \underline{by Intelligence Officers of both the Army and the Air Forces.}\\
$\cdots$\\
During the past two months various sightings of unexplained phenomena have been reported in the vicinity of the A.E.C. [Atomic Energy Commission] Installation at Los Alamos, New Mexico, where these phenomena now appear to be concentrated.  During December 1948 on the 5th, 6th, 7th, 8th, 11th, 13[th], 14th, 20th, and 25th sightings of unexplained phenomena were made near Los Alamos by Special Agents of the Office of Special Investigation; Airline Pilots; Military Pilots; Los Alamos Security Inspectors, and private citizens.  On January 6, 1949, another similar object was sighted in the same area.
\end{quote}

Details also exist on sightings at Hanford Nuclear Works (Fig. \ref{fig:atomic-sites} A).  In early 1945, Clarence R. ``Bud'' Clem, a Lieutenant Junior Grade in the US Naval Reserves, was an F6F Hellcat pilot and was deployed to Naval Air Station (NAS) Pasco, Washington, for ground support training in March 1945.  Lt. Clem writes \cite[pp.43-44]{Hastings:2017}:
\begin{quote}
One night, shortly after the evening meal $\ldots$ We learned that an unidentified `bogey' was over the Hanford Ordinance Works, according to the radar operator located on an auxiliary field just across the Columbia River from Hanford reservation.

We had been instructed upon arrival [at NAS Pasco] that the Hanford Ordinance Works was Top Secret and no flights over any part were permitted. $\ldots$ The duty officer stated that something was in the sky over the area and wanted someone to investigate.  A plane was [already] armed and warmed-up on the tarmac.  Brown stated that he would go and Neal was to stand-by in another plane, in case of trouble.  I was to join the [controller] in the tower and communicate info from radar to the pilots.

Brown quickly found the object, a bright ball of fire, and took chase.  But he could not close, even with water injection that gave a quick boost in speed.  The object headed out NW towards Seattle and was quickly lost by radar.  Brown returned to base and we three returned to the club, still shaking and wondering what we had encountered.
\end{quote}

UFO historian Jan Aldrich provided Hastings with documents obtained by his research group, Project 1947, from the Headquarters Fourth Air Force.  One of these documents, dated January 23, 1945 and directed to the Commanding General of the Army Air Forces and the Assistant Chief of Air Staff Training, places the events described above by Lt. Clem in a greater context \cite[p.45]{Hastings:2017}\cite{HeadquartersFourthAirForce:1945}:
\begin{quote}
Resulting from an unidentified aircraft flying over the Hanford Engineering Company Plant at Pasco, Wash. on at least three nights in the past month (this company is involved in undisclosed projects for the War and Navy Departments) this HQ was requested by [Western Defense Command], about ten days ago, to move one [battery] of searchlights from Seattle to the Pasco plant.  The Thirteenth Naval District has made arrangements for Naval Air Station, Pasco, to employ both radar and fighter aircraft in attempting interception of these unidentified aircraft.  The airspace over the Hanford Company is both a Danger area and a Restricted area.  Our battery of searchlights has been in place since 15 January; one incident has occurred since that date in which a brief radar contact was made---attempted night interception again failed.
\end{quote}

Killeen Atomic National Stockpile, also known as Site Baker, was one of the most sensitive atomic weapon sites in the US.  Hastings \cite{Hastings:2017} quotes researcher Loren Gross \cite{Gross:1990}:
\begin{quote}
After becoming a true bomb factory, Sandia shipped assembled bombs to Camp Hood, Texas where there was a secure storage site guarded by the 12th Armored Infantry Battalion under the command of the Fourth Army
$\ldots$
It is suggested that the `green fireballs' which appeared over Sandia in late 1948 bear a direct relationship to a sudden ramp-up of American nuclear weapon production.  [Similarly,] in March 1949, when strange `flares' appeared around the `Q' area at Camp Hood, it is suggested that this interest by UFOs was triggered by the recent arrival of the first shipment of atomic bombs which was stored as America's first stockpile.
\end{quote}

These strange flares were first observed on March 6th 1949 by Army security patrols near the Q Area, which housed the nuclear weapons.  On the evening of March 8th, there were seven sightings with multiple observers across the base allowing for precise triangulation.  Capt. McCulloch and his party observed the lights and recognizing that it was not a flare, he placed the entire base on alert.  On reviewing the sightings and their triangulations, Fourth Army Intelligence Personnel in San Antonio became alarmed noting that the lights had done a ``good job'' bracketing off the Q Area \cite[p.257--259]{Clark:1998}.

Some of the strangest encounters occurred more than a month later when on April 27, 1949 at 9:20pm two members of a patrol station southeast of Killeen base witnessed a small inch and a half diameter (4 cm) blinking violet light\footnote{This account of a small violet light is strikingly similar to the luminous violet ball that Mary Kingsley encountered in West Africa in 1893 (Sec. \ref{sec:HistoricalUSOs}).} hovering about six feet off the ground and only about ten to twelve feet away \cite[p.259]{Clark:1998}.  They watched the light for about a minute until the light left by ascending through tree branches.  Five minutes later, four Army men located about two miles away observed a small bright light with a two to four inch metallic cone attached.  The light and cone approached them at about 60 to 70 mph making a closest approach of about 150 feet.  Numerous other similar encounters occurred on the base in days that followed.  On one occasion, eight to ten lights appeared together with the light-and-cone among them \cite[p.259]{Clark:1998}.

The most remarkable events involve UAPs hovering over nuclear ICBM missile silos, in some cases while the missiles are brought offline, while not commanded to do so.  The most famous of these events are the 1967 incursions at the nuclear weapon sites near Malmstrom Air Force Base (AFB), in Great Falls, Montana, which coincided with the anomalous shutdown of almost twenty nuclear missiles \cite{Klotz+Salas:2005, Hastings:2017, Salas:2023}.  Even more disturbing, journalist George Knapp testified before the US Congress \cite{Knapp:2023} that Russian Col. Boris Sokolov shared information about a similar event that took place at a Russian ICBM base in Ukraine, where
\begin{displayquote}
    UFOs appeared over the base, performed astonishing maneuvers in front of stunned eyewitnesses and then somehow took control of the launch system. The missiles were aimed at the US and were suddenly fired up. Launch control codes were somehow entered, and the base was unable to stop what could have initiated World War 3. Then, just as suddenly, the UFOs disappeared, and the launch-control system shut down.
\end{displayquote}

\begin{figure}
\centering
\makebox{\includegraphics[width=0.37\columnwidth]{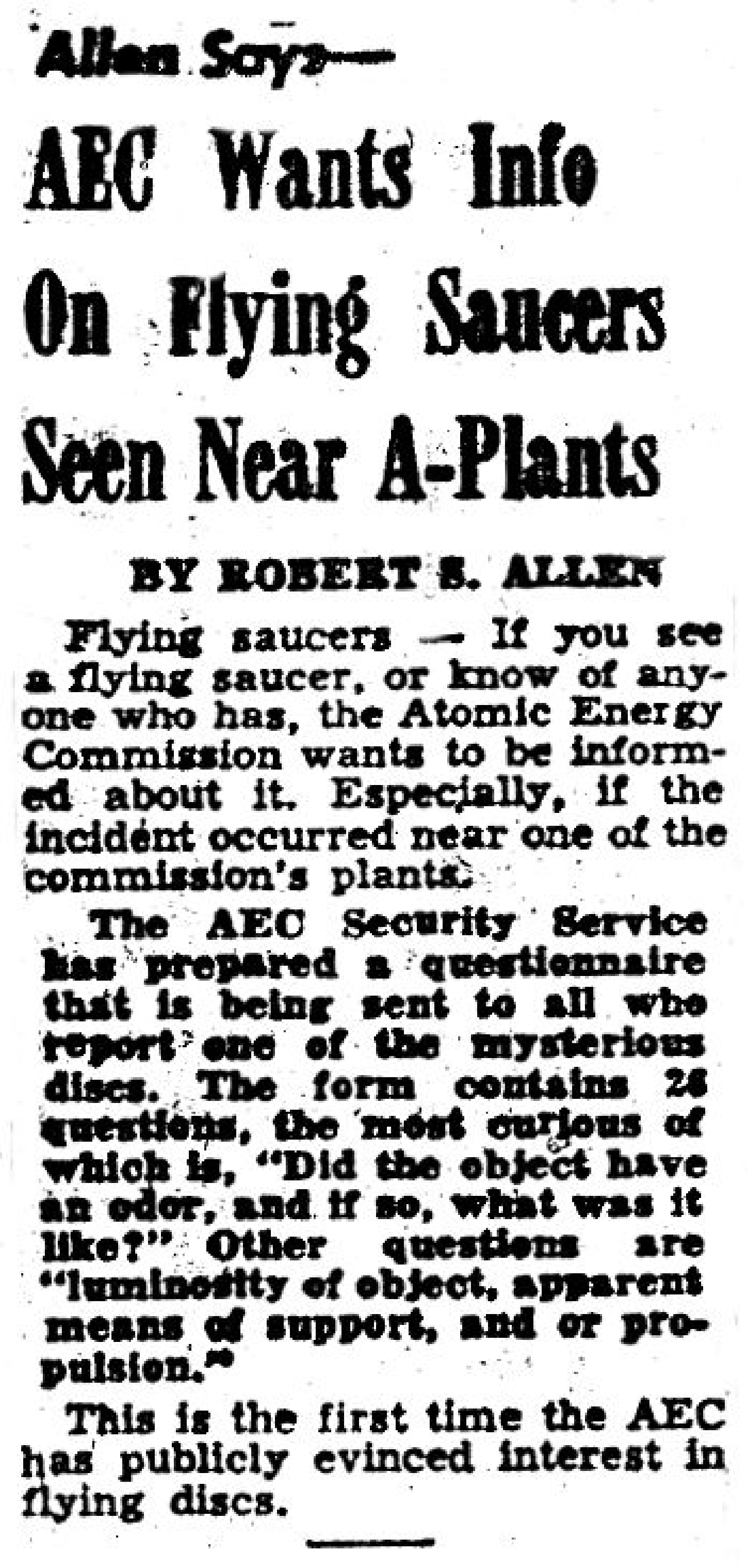}}
\caption{A newspaper article from the OakRidger (12 October 1950) announcing that the Atomic Energy Commission is seeking information about flying saucers observed near our nuclear energy sites.  This article was published in 1950, which was the year that exhibited the maximum number of UAP sightings at the Oak Ridge Nuclear Complex (see Fig \ref{fig:atomic-sites}B).} \label{fig:oak-ridger}
\end{figure}

There is an excellent series of studies performed by Larry J. Hancock, Ian M. Porritt, Sean Grosvenor, Larry Cates, and Ike Okafor of the Scientific Coalition for UAP Studies (SCU) (Sec. \ref{sec:SCU}) dealing with the relationship between UAP and the 1945-1975 United States Atomic Warfare Complex \cite{Hancock+etal:2023:pattern, Hancock+etal:2023:indications}, and the US military in general \cite{Hancock+etal:2024:activities}.
This study compiles the number of yearly UFO sightings from 1943 to 1975 associated with military nuclear weapons complexes in the United States.  The primary data source was the Brad Sparks Catalog \cite{Sparks:2020}, which consists of incidents that were officially reported to, and investigated by, the United States Air Force UFO programs SIGN (Sec. \ref{sec:SIGN}), GRUDGE (Sec. \ref{sec:GRUDGE}), and BLUEBOOK (Sec. \ref{sec:BlueBook}).  Only incidents that were classified as unidentified were selected for analysis.  The first part of the study focuses on three study site classes, each associated with different aspects of nuclear weapons production.  These are (1) production plants for radioactive materials, (2) nuclear weapon assembly facilities, and (3) nuclear weapon stockpile sites.  It is important to note that the layout and construction of each type of facility and the resources available to them were distinct.  Control sites were selected to be nearby civilian population centers and high-security non-nuclear military bases.

The data from four such sites and their control sites, illustrated in Figure \ref{fig:atomic-sites}, demonstrate that \cite{Hancock+etal:2023:pattern} 
\begin{displayquote}
elevated UAP activity was found at all three atomic site classes and was most noticeable in the earliest facility in each class. UAP activity began during the construction phase for some sites and escalated when the site became operational. Elevated activity at study sites occurred in a ``window'' between 1948-1951, continued through the national spike in UAP reporting in 1952, then dramatically decreased, never to repeat the ``window'' levels during the remainder of the study period.
\end{displayquote}
For example, the year 1950 saw the greatest number of UAP observations at the Oak Ridge Complex in Oak Ridge Tennessee (Fig. \ref{fig:atomic-sites}B).  This is highlighted by the October 12, 1950 newspaper article in the OakRidger (Fig. \ref{fig:oak-ridger}) announcing that the Atomic Energy Commission (AEC) is seeking information about flying saucers observed near our nuclear sites.

The SCU team's findings concluded that these UAPs appeared to be involved in intelligent and focused activity, which is possibly best described as surveillance.  This is an important fact to recognize and appreciate, as the sciences are usually focused on studying passive physical phenomena---not phenomena exhibiting intelligence and intention.

\section{Physical Evidence}

The fact that there have been relatively few efforts thus far to collect data on UAP in the field means that most UFO/UAP cases rely on witness testimony.  However, there have been some cases in which physical evidence could be studied afterward \cite{Vallee:1998}.

\subsection{UFO Crashes}
Most people are aware of purported UFO crashes and alleged government retrievals \cite{Randle:2010, Svozil:2023}, such as the crash at Roswell, New Mexico, USA, \cite{Randle+Schmitt:1994, Berliner+Friedman:2004} and the often conflated San Agustin Plains, New Mexico crash \cite{Eberhart+etal:1992, Campbell:2013}, both of which occurred in the first week of July 1947 during the statistically significant increase in US sightings (Figure \ref{fig:1947}).  However, such events typically leave very little physical evidence, if any, that can be studied \cite{Perrone:2025, Szydagis:2025:parts, Yates:2025}.  Despite this, knowledge of real crashes, if any, would prove to be extremely valuable information \cite{Szydagis:2025}.

\subsubsection{The Ubatuba Incident, 1957, Ubatuba Brazil}
The Ubatuba incident is one of the most fascinating events that involved the catastrophic explosion of a UFO resulting in debris that was collected \cite{Sued:1957, Lorenzen:1962, Vallee:1998, Sturrock:2001, Powell+etal:2022}.  However, the most difficult aspect of this case was the fact that the identities of the witnesses are unknown, which compromises the veracity of the account.

The incident was first mentioned in an article titled ``A Fragment from a Flying Disk!'' in the 14 Sept. 1957 issue of \textit{El Globo}, a Rio de Janeiro newspaper \cite{Sued:1957, Sturrock:2001}.  Pieces of a flying disk were supplied to Ibrahim Sued of \textit{El Globo} along with a letter stating: \cite{Sued:1957, Sturrock:2001}

\begin{displayquote}
    ... I was fishing together with some friends, at a place close to the town of Ubatuba, Sao Paulo, when I sighted a flying disc. It approached the beach at unbelievable speed and an accident, i.e. a crash into the sea seemed imminent. At the last moment, however, when it was almost striking the waters, it made a   sharp turn upward and climbed rapidly on a fantastic impulse. We followed the spectacle with our eyes, startled, when we saw the disc explode in flames. It disintegrated into thousands of fiery fragments, which fell sparkling with magnificent brightness. They looked like fireworks, despite the time of the accident, at noon, i.e. at midday. Most of these fragments, almost all, fell into the sea. But a number of small pieces fell close to the beach and we picked up a large amount of this material - which was as light as paper. I am enclosing a sample of it. I don't know anyone that could be trusted to whom I might send it for analysis. 
\end{displayquote}

The Ubatuba samples have been extensively studied by more than a dozen laboratories \cite{Lorenzen:1962, Powell+etal:2022}.  The first studies, conducted by Dr. Olavo Fontes and performed at the National Department of Mineral Production in Brazil, determined that the samples were pure magnesium \cite{Fontes:1962, Sturrock:2001}.  

The samples were the only physical evidence examined by Condon's Colorado Project (see Sec. \ref{sec:ColoradoProject}).  The Colorado Project used direct gamma spectrometry and half-life measurement to identify manganese (Mn), aluminum (Al), zinc (Zn), mercury (Hg), and chromium (Cr). After radiochemical separation of the elements, gamma spectroscopy revealed the presence of copper (Cu), barium (Ba), and strontium (Sr) \cite[pp. 94--97]{Condon:1969}.  Dr. Roy Craig, working with the Colorado Project, used the facilities of the Alcohol and Tobacco Division National Office Laboratory to perform neutron activation to measure the isotopic abundance of \ce{^{26}Mg}. They found that the abundance of \ce{^{26}Mg} was $14.3\% \pm 0.7\%$, which they claimed was in agreement with the terrestrial value, which is between $10.99\%$ and $11.03\%$ \cite{USGS:2006}.

The paper by Powell et al. \cite{Powell+etal:2022} summarizes chemical tests performed over several decades from the 1960s through the 1980s, as well as their work with two independent laboratories, Cerium Laboratories in Austin, Texas and ICP and ICP-MS Services in Cleveland, Ohio, using High-Resolution Inductively Coupled Plasma Mass Spectrometry (HR-ICPMS) to examine the isotope ratios of several elements, such as strontium, copper, zinc, and barium, in addition to magnesium.  The Cleveland lab showed that the magnesium isotope ratios were consistent with the terrestrial ratios.  However, the Austin lab is believed to have been in error as the Mg ratios they obtained, with the same sample, differed from terrestrial values.  In this study, no conclusions were reached on whether the isotope ratios of the trace elements differed from the terrestrial values \cite{Powell+etal:2022}.  The paper concludes with a useful discussion of the difficulty of obtaining and interpreting isotope ratios of these types of samples.

\begin{figure}
\centering
\makebox{\includegraphics[width=0.8\columnwidth]{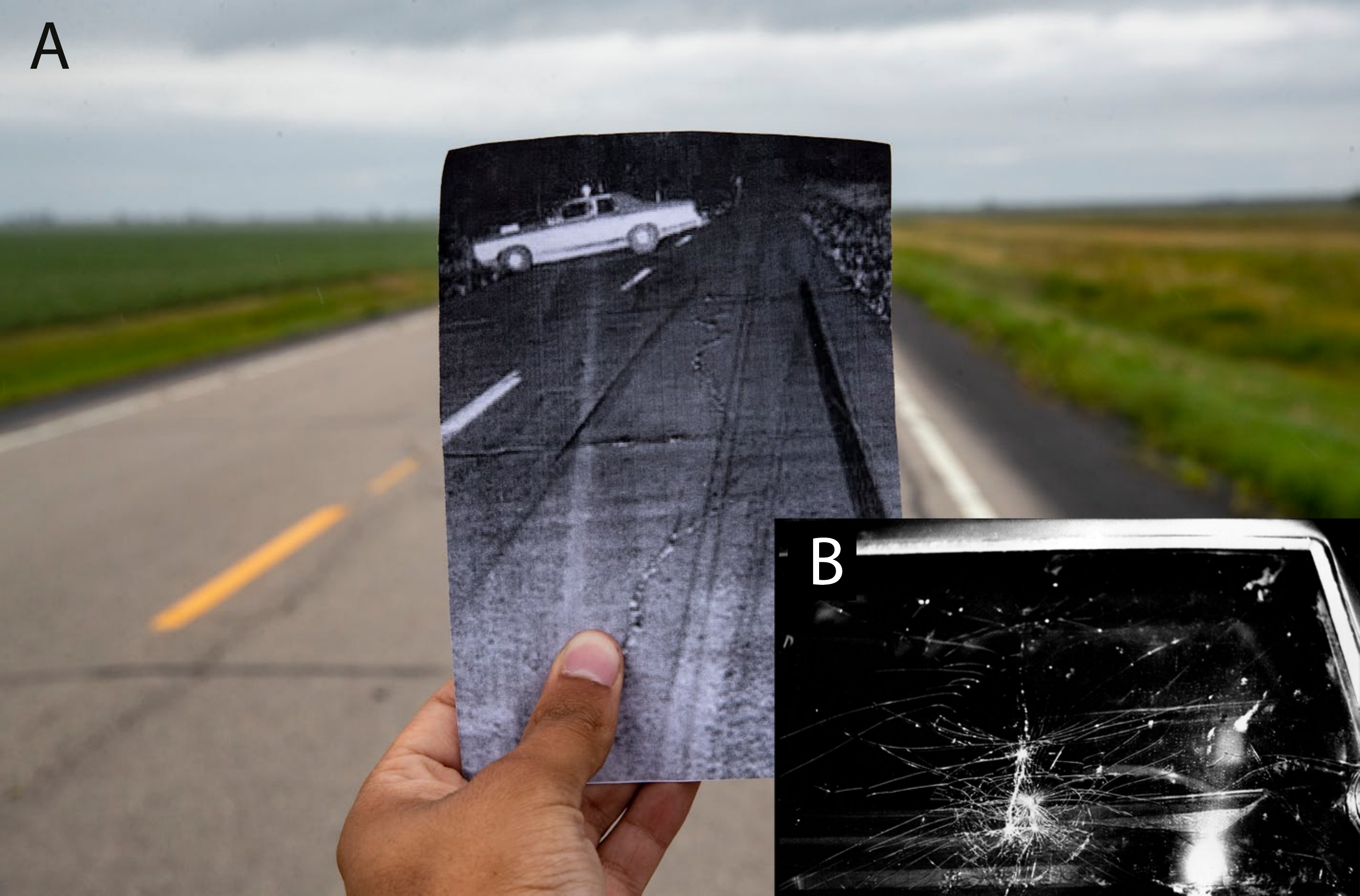}}
\caption{A. This figure shows a black-and-white photo of Deputy Johnson's car, after it was hit by a UAP, being held up against the present-day location of the event (Credit: Alex Kormann/The Minnesota Star Tribune). B. (Inset) A photograph of the shattered windshield of the car. (Credit: Regene Radniecki/The Minnesota Star Tribune).} \label{fig:val-johnson}
\end{figure}

\subsection{Collisions with UAP}
There have been less dramatic incidents involving collisions that resulted in physical evidence.  
\subsubsection{Deputy Val Johnson UFO collision, 1979, Minnesota, USA}
One of the most remarkable incidents involves a collision between a deputy sheriff's car and a UAP.  Deputy Sheriff Val Johnson of Marshall County, Minnesota, US, was driving on the country roads west of the small town of Stephen, Minnesota very early in the morning on Aug. 27, 1979.  He saw a light in the sky that appeared to be the landing light of an aircraft.  He turned from County Road 5 onto Highway 220 to investigate. Deputy Johnson described the event: \cite{Pearson:2025}
\begin{displayquote}
It sat there and appeared to be stationary. But when I got closer, boom, it was right there, just right now. I heard glass breaking, saw the inside of the car light up real bright with white light. It was very, very extremely bright. That’s all I can remember.
\end{displayquote}

Deputy Johnson lost consciousness, and about 40 minutes later when he came to, he radioed for help \cite{Pearson:2025}
\begin{displayquote}
Something just hit my car. I don’t know how to explain it. Strange. Something attacked my car. ... I can’t see very well. I don’t think I’m losing any blood anywhere.
\end{displayquote}

When Deputy Greg Winskowski arrived on the scene, he found Deputy Johnson's car perpendicular to the road and damaged with a dented hood, a broken windshield, 
 a broken headlight, and a broken red light on the stop of the car (Fig. \ref{fig:val-johnson}).  The car's antenna was also strangely bent at a 90 degree angle.  Deputy Winskowski asked him what kind of vehicle hit him.  Deputy Johnson replied \cite{Pearson:2025}
\begin{displayquote}
It wasn’t a vehicle, Greg. I don’t know what the hell it was.
\end{displayquote}

The car's clock and the deputy's watch had both stopped for 14 minutes before starting up again.  Deputy Johnson was taken to the hospital and treated for ``welding burns'' to his eyes.  The sheriff went to the County Commissioners and convinced them that because the event was so strange, it would be better to keep the car as it is rather than repair it.  The car is currently housed at the Marshall County Historical Society Museum in Warren, Minnesota.

\begin{figure}
\centering
\makebox{\includegraphics[width=0.8\columnwidth]{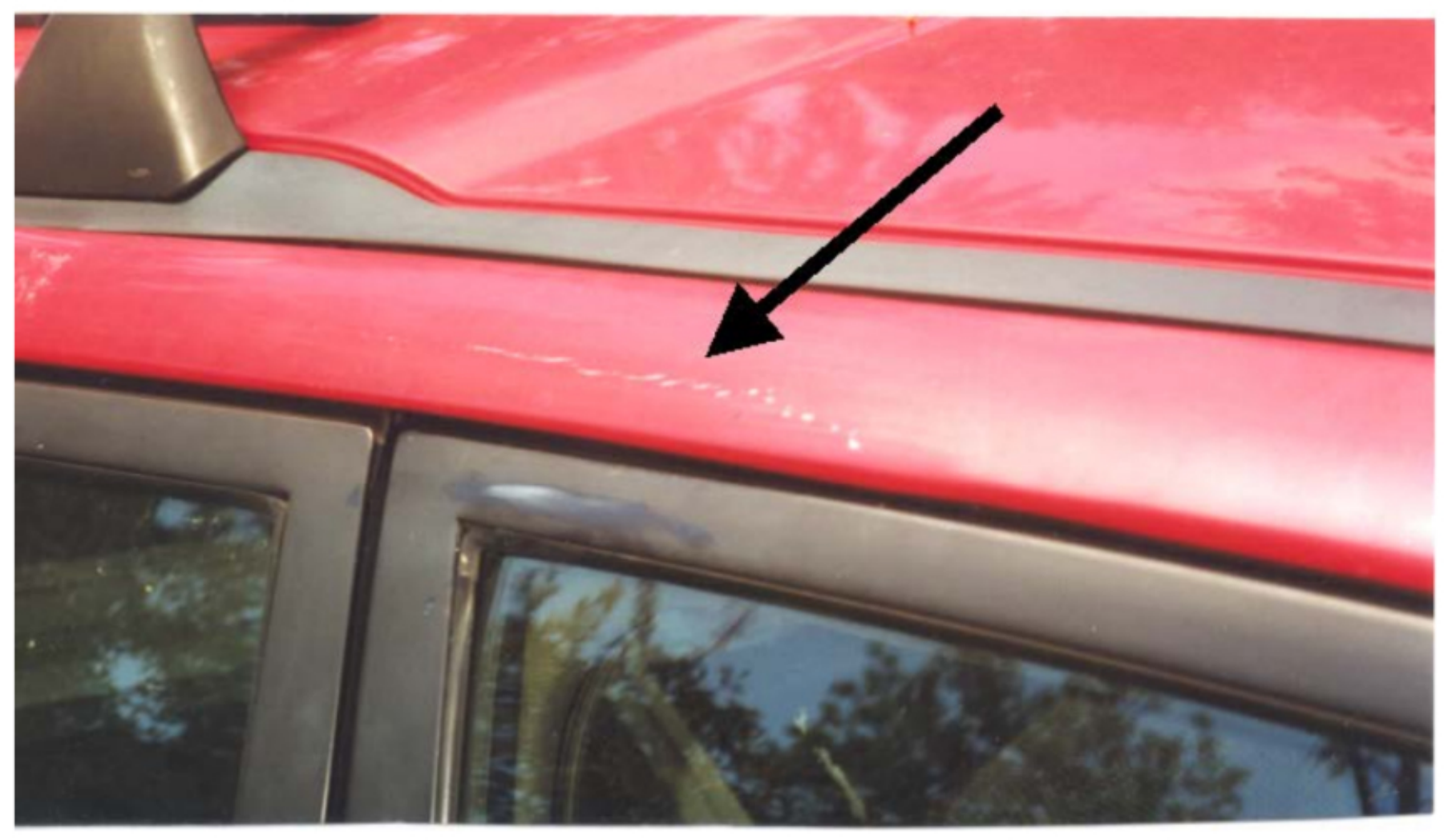}}
\caption{This figure shows the light-colored residue left at the impact site of the UAP along the roof of the car. (Source: Phyllis Budinger, Frontier Analysis, Ltd.)} \label{fig:michigan-collision}
\end{figure}

\subsubsection{Mt. Clemens UFO collision, 1998, Michigan, USA}
A similar collision occurred in the early evening of 27 November 1998 as it was getting dark when a woman was driving a 1992 Ford Escort station wagon east on 25 Mile Road north of Mt. Clemens, Michigan, US.  She noticed an intense, white basketball-sized light rapidly and silently approaching the car.  The object hit the side of the roof of the car with a thump.  It is not known what happened to the object after hitting the car, but it was found that the collision left a cream-colored residue at the impact site (Figure \ref{fig:michigan-collision}).

The residue was collected by scraping the site and wrapping the 2\unit{mg} sample in a polyethylene plastic wrap.  The sample was submitted to Frontier Analysis, Ltd. and analyzed by chemist Phyllis Budinger (see Sec. \ref{sec:Budinger}) \cite{Budinger-UT011:2001}.  Budinger obtained infrared microscopic spectra and determined that the sample consisted of car finish polymers, kaolin mineral (an aluminum silicate), celluloidal material, and metal oxide.  Although the metal oxide was not specifically identified, its spectral bands were closest to manganese oxide. Aluminum oxide was ruled out.  There was not enough of the sample to perform a scanning electron microscopy with energy-dispersive X-ray spectroscopy (SEM/EDS) test to conclusively identify the metal oxide.  However, the conclusion was that the metal oxide must have come from the UAP \cite{Budinger-UT011:2001}.

\subsection{UAP Deposits}
UAP landings \cite{Vallee:1969:landings} are often accompanied by damage to foliage (by breakage or burning) and sometimes leave behind chemical residue. In addition, there have been situations in which UAP drop, dump or deposit material, which has been analyzed \cite{Vallee:1998}.

\begin{figure}
\centering
\makebox{\includegraphics[width=0.9\columnwidth]{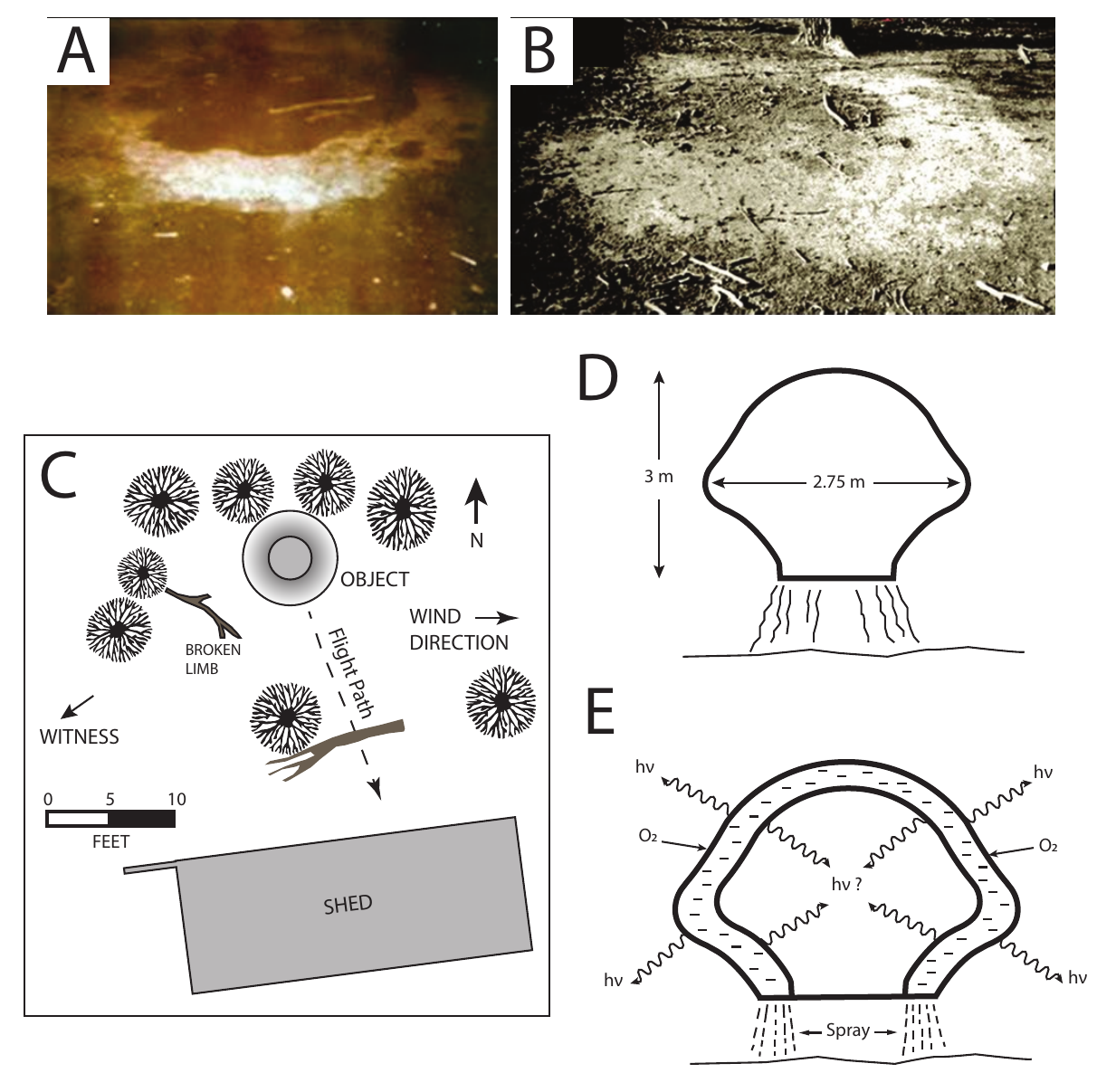}}
\caption{A. The Polaroid photograph of the ring of material taken by Mrs. Johnson just after the UFO left.  No flash was used as the material emitted sufficient light for a photograph. B. The photograph taken by the Sheriff the next day (19 hours after the event). Some spreading, probably due to wind, of the material toward the southeast is visible.  C. A map of the area adapted from \cite{HumanitiesKansas:2022}. D. A sketch of the UFO, adapted from \cite{Faruk:2014}. The entire surface of the UFO glowed with blue, red, and orange light.  The space below the UFO was illuminated with a shimmering material that hit the ground.  E. A reproduction of the hypothesized chemistry associated with the UFO, adapted from \cite{Faruk:2014}.  The surface of the UFO could have been covered with an aqueous solution of chemiluminescent material interacting with the oxygen in the air.  This would explain the blue glow emitted from the surface of the UFO.}\label{fig:delphos}
\end{figure}

\subsubsection{Delphos, Kansas USA UFO Landing, 1971}
One of the most famous landings occurred on 2 November 1971 at about 7:00 pm on a farm in Delphos, Kansas, US.  Sixteen-year-old Ronald Johnson was tending sheep in a shed and heard a rumbling noise like a washing machine out of balance.  He stepped out of the shed to look and saw a brilliantly illuminated mushroom-shaped object hovering below a tree about 75 feet away from him (Figure \ref{fig:delphos}C).  The object was approximately 10\unit{feet} (3\unit{m}) high and had a diameter of 9\unit{feet} (2.75\unit{m}) \cite{Faruk:2014, Faruk:2021} (Figure \ref{fig:delphos}D).  The object's entire surface emitted a multicolor glow with red, blue, and orange light.  The object hovered about 2 to 5 feet above the ground, and the space below the object was illuminated with shimmering material that fell from the object onto the ground below.  It was extremely bright and it hurt his eyes to look at it.  His eyes were sore and he suffered from headaches for several days. 

After several minutes, the object moved upward and flew low over the shed at which point it began making a high-pitched sound like a jet engine.  Ronald temporarily lost his vision at this point, and when his vision was restored a short time later, he could see the bright light speeding off to the south-east.  Ronald went and got his parents who made it outside in time to see the bright light in the distance about half the size of the full moon, which was also in the sky.  They described the object as being the ``color of an arc welder''.
The sighting was independently confirmed by reserve police officer Lester Ernsbarger in Minneapolis, Kansas (about 10 miles south of Delphos), who reported a bright light in the northern sky at about 7:30 pm \cite{Faruk:2014, Faruk:2021}.

The Johnson family walked over to the place where the UFO had hovered and saw that there was a ring of soil glowing in the dark.  The trees also had glowing material on them.  Mrs. Johnson ran inside to get her Polaroid 104 camera and took a photograph of the landing site.  No flash was needed for the photograph because the light of the ring was bright enough to read a newspaper (Figure \ref{fig:delphos}A).  The glow lasted until the next night \cite{Faruk:2014, Faruk:2021}.

The Johnsons touched the glowing soil and described it as having a cool, smooth crust-like structure as if it had been crystallized.  In close inspection, it appeared to be moist and pitted with small craters.  Their fingers smelt of an unusual odor, and the material had a numbing effect on their fingers that took days to weeks to wear off \cite{Faruk:2014, Faruk:2021}.

They reported the event to the sheriff and a newspaper reporter.  The sheriff took samples of the ring material and a photograph of the ring (19 hours after the landing), which by this time had seemed to spread out to the south-east possibly due to the wind (Figure \ref{fig:delphos}B).  The affected soil exhibited a strange water repelling property, which lasted for several months \cite{Faruk:2014, Faruk:2021}.  Ted Phillips arrived about a month later to interview the family and investigate the event.  He noted the strange water-repelling property that resulted in the ring being covered by unmelted snow \cite{Phillips:1972a, Phillips:1972b}.

The material was sent to several laboratories for analysis, but given the lack of funding and resources, there were no results.  Dr. Erol A. Faruk (see Sec. \ref{sec:Faruk}), who was in a postdoctoral fellowship with supervisor Dr. B.~W. Bycroft of Nottingham University, England, requested a sample of the material from CUFOS.  Together with the resident chemiluminescence expert, Dr. Frank Palmer, they began to study the samples by recording fluorescence spectra.  Faruk discovered that the ring soil was impregnated by an air-sensitive organic compound that on chromatographic purification revealed that it had all the attributes necessary to generate light, a process known as oxidative chemiluminescence. He was able to use the soil data to propose a viable theory of how the ring had actually formed, essentially confirming the witness' description of the nearby presence of a hovering aerial device of unknown technology and origin \cite{Faruk:2014, Faruk:2021}.

Twenty-seven years later, Phyllis Budinger (Frontier Analysis, Ltd.) obtained soil samples from the CUFOS Chicago office \cite{Budinger-UT001:1999} and utilized more modern techniques to analyze the material and add to Faruk's results.  Budinger agreed that the UFO definitely must have released this material, and she concurred with Faruk that the material was most probably deposited as an aqueous solution.  Budinger noted that even after being in storage for 27 years, the material still coated the soil and exhibited the same hydrophobic behavior that the Johnsons observed.  Budinger found that $5 \pm 2 \%$ of the weight was calcium oxalate, which she suspected was deposited as free oxalic acid, which then was combined with calcium in the ground.  Calcium oxalate and oxalic acid are known skin and eye irritants, which could explain the physical effects felt by the witnesses who touched the material.  The material also contained sulfur and/or mercaptan $(< 0.1\%)$ of the weight, which could explain the foul odor \cite{Budinger-UT001:1999}.  Budinger found that $(85\pm10\%)$ of the weight was fulvic acid, which is a water-soluble humate.  Because highly polar substances with carbonyl and double bonds would chemiluminesce if exposed to an ionizing electric field, one would expect that the fulvic acid and oxalate would chemiluminesce if exposed to an ionizing electric field.  Budinger also noted \cite{Budinger:2025} that insoluble humates were not found in the soil samples, and that when she ``added water to the Delphos soils in a test tube and shook it up only for a minute, the deposit material coating the soil solubilized and the soil was no longer hydrophobic. It was normal.''  

The Delphos landing has similarities to another case also studied by Budinger \cite{Budinger-UT021:2002}.  In May 1995, white circles appeared in freshly plowed fields in New Pine Creek, Oregon, USA.  The area was purported to have a high incidence of UFO sightings, but none was known to be associated with the circles, the origin of which was unknown.  The white material in one field was sampled on 20 September 1995.  The material was observed to be present up to a depth of 18\unit{inches} (almost 50\unit{cm}).   Infrared spectra, obtained using the Harrick SplitPea\textsuperscript{\textregistered} cell on the Nicolet Avatar 360 spectrometer, allowed the material to be identified as calcium oxalate \cite{Budinger-UT021:2002}.  No other components were detected.

\begin{figure}
\centering
\makebox{\includegraphics[width=0.9\columnwidth]{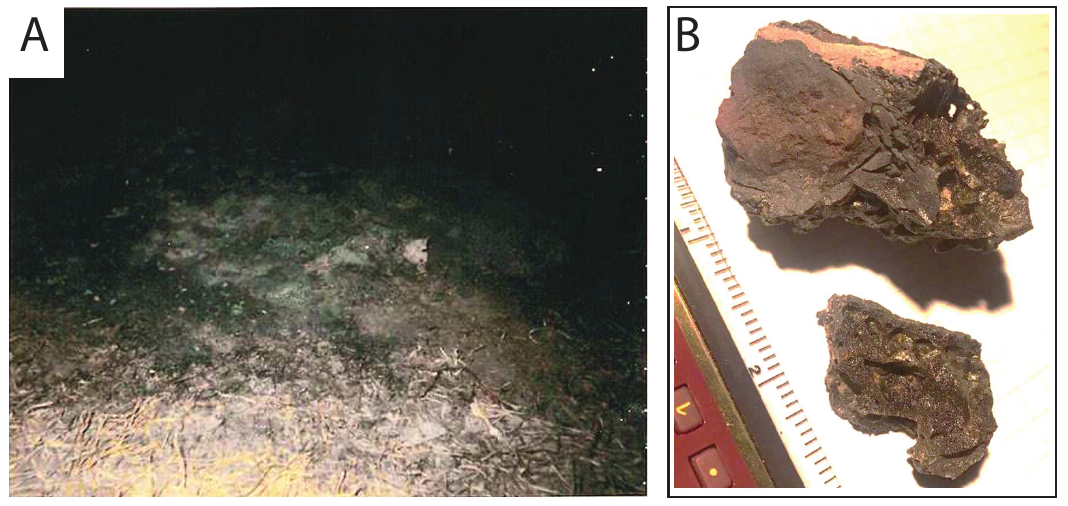}}
\caption{A. An enhanced Polaroid photograph of the deposit site at Council Bluffs, Iowa.  (Credit: Officer Dennis Murphy, source: \cite{Nolan+Vallee:2022})  B. Samples (scale in inches) recovered from Council Bluffs. (Source: \cite{Nolan+Vallee:2022})} \label{fig:council-bluffs}
\end{figure}

\subsubsection{Council Bluffs, Iowa USA}
On 17 December 1971, in Council Bluffs, Iowa, USA, about 11 people, in separate groups, witnessed an event where a hovering object dropped a large mass of molten metal in Big Lake Park where it started a small grass fire \cite{Vallee:1998, Sturrock:1999, Nolan+Vallee:2022}.  A couple was driving east toward downtown Council Bluffs when they spotted an object hovering.  The object was described as ``a big round thing hovering in the sky, below the tree tops'' with ``red lights blinking in sequence around the periphery.''

The object was observed to dump some kind of luminous material.  Another couple, who refused to be identified, saw a bright red object ``rocket to the ground near Big Lake.'' \cite{Vallee:1998, Nolan+Vallee:2022}  At the same time, three people, Kenny Drake, his wife Carol, and nephew Randy James, were driving on North 16\textsuperscript{th} Street when they witnessed a reddish ball at a height of about 500 to 600 feet that fell straight down into Big Lake Park where 8 to 10 foot high flames erupted. \cite{Nolan+Vallee:2022}  

The Drakes called the authorities, and Assistant Fire Chief Jack Moore was dispatched to the scene.  He described the material as ``running, boiling down the edges of the levee.  The center of it was way too hot to touch.''  Despite the freezing cold temperatures, the material stayed warm for approximately two hours \cite{Nolan+Vallee:2022}.  Officer Dennis Murphy arrived and photographed the material (Fig. \ref{fig:council-bluffs}A), which was observed to cover an area of about $4\unit{feet}$ by $6\unit{feet}$ ($1.25\unit{m} \times 1.8\unit{m}$) and $4\unit{inches}$ ($10\unit{cm}$) deep. 

Mr. Robert Allen, who wrote an astronomy column for the local paper and who had served in the Air Force, investigated the site the next day and determined that the material had been traveling from the southwest to the northeast when it hit the ground.
The samples of the deposit were sent to the Ames National Laboratory at Iowa State University, where they were studied by Dr. Robert S. Hansen, director of the Ames Energy and Mineral Resources Research Institute, and Edward DeKalb, of the analytical spectroscopy section, and to Griffin Pipe Products Company \cite{Vallee:1998}, where they were studied by lab technician Jack Coan. 

The Ames analysis found that the material was solid metal and slag: \cite{Vallee:1998}
\begin{displayquote}
    The metal is chiefly iron with very small amounts (less than 1\%) of alloying metals such as nickel and chromium. The slag is a foam material containing metallic iron and aluminum with smaller amounts of magnesium, silicon, and titanium probably present as their oxides.
\end{displayquote}
The inclusions of white ash in the slag were ``principally calcium with some magnesium, again probably as oxides.''

The Ames report ruled out a meteoritic origin, since ``in that case one would expect a much higher nickel content'' \cite{Kayser:1978, Nolan+Vallee:2022}.  They also ruled out the possibility of man-made space hardware: \cite{Kayser:1978, Nolan+Vallee:2022}
\begin{displayquote}
Such hardware usually involved alloys of much higher strength-to-weight ratio, containing high amounts of nickel, chromium, and titanium.  Our first judgment was that the material resembled cast iron.
\end{displayquote}

Lab technician Jack Coan of Griffin Pipe Products ran two exposures using a ``Spectro-Comp'' instrument and spectroscopically identified 18 elements, including carbon (0.70\%), manganese (0.56\%), and silicon (0.52\%) \cite{Nolan+Vallee:2022}.  Similarly, Prof. Frank Kayser, of the metallurgy division at University of Iowa, reported that the material was most likely a carbon steel, and that ``despite the name, carbon steels contain less carbon (about 1.0 to 1.2\%), than cast iron (about 4\%).''  The microstructure suggested that the material had been cast, then subsequently heated to 900 to 1000\unit{C}, and then cooled so that it resembled wrought iron \cite{Kayser:1978, Nolan+Vallee:2022}.  

Allen also corresponded with Air Force Space Systems, who gave four reasons that the material was not man-made space debris: \cite{Vallee:1998}
\begin{enumerate}
\item Reentering spacecraft debris does not impact in a molten state.
\item The large mass of the material did not leave a crater or indentation.
\item The visual sighting was only at 500 to 600 feet, where it would not be glowing.
\item The material lacked any structural indications.
\end{enumerate}

Jacques Vall\'{e}e retained some of the material, and later Garry Nolan (Stanford Univ.) with Sizun Jiang (Stanford Univ.) and Larry Lemke (NASA Ames) reanalyzed the material (Fig. \ref{fig:council-bluffs} B) \cite{Nolan+Vallee:2022}.  Using a NanoSIMS machine (manufactured by Thomson-Syseca), they performed secondary ion mass spectrometry to determine isotopic ratios.  They found that the distribution of titanium isotopes (\ce{^{46}Ti}, \ce{^{47}Ti}, \ce{^{48}Ti} and \ce{^{49}Ti}), iron isotopes (\ce{^{56}Fe} and \ce{^{57}Fe}), and chromium isotopes (\ce{^{52}Cr} and \ce{^{53}Cr}) were consistent with terrestrial material.

Nolan et al. noted that \cite{Nolan+Vallee:2022}
\begin{displayquote}
While this study verified the prior findings in terms of elemental composition and ``natural'' isotope content, we additionally found that there was local homogeneity of the samples to the degree measured, but considerable diversity in the elemental ratios across the subsamples.  This implies that whatever the origin of the sample, it was incompletely mixed at the time of deposition. 
\end{displayquote}
Despite what has been learned from these analyses, the nature and origin of the material remains unknown.

\begin{figure}
\centering
\makebox{\includegraphics[width=1\columnwidth]{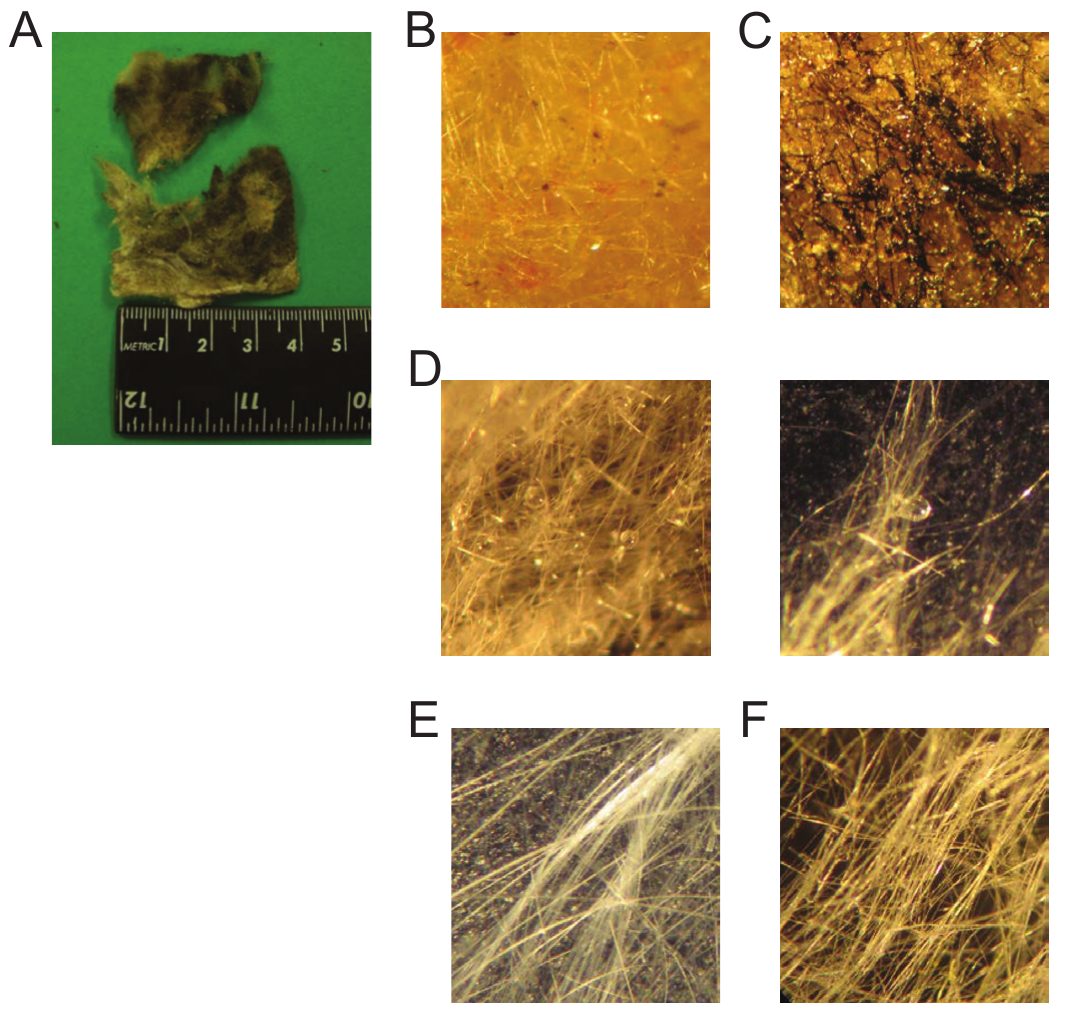}}
\caption{A. A photograph of the fragment that fell flaming from the orange-red orb in Claymont, Delaware, USA, in 2014.  Microscopic examination (using a Canon A520 digital camera interfaced to a Leica GZ6 stereomicroscope) of the fragment shows droplets/particulates of material on a fibrous matrix.  B. A photograph of an unburned portion of the material revealing droplets or spheroids. C. A photograph of a burned portion of the material, which illustrates the charring of the organic material. D. Photographs of the material after being washed with hexane and acetone.  The spheroids are more apparent in these images. E. Glass wool for reference. F. House insulation for reference.  The reference fibers appear longer and thicker, and lack the spheroids on the UAP sample.  (Source: \cite{Budinger-UT087:2014})} \label{fig:delaware-UT087}
\end{figure}

\subsubsection{Red Orb Drops Flaming Material, 2014, Claymont, Delaware, USA}
\begin{figure}
\centering
\makebox{\includegraphics[width=0.9\columnwidth]{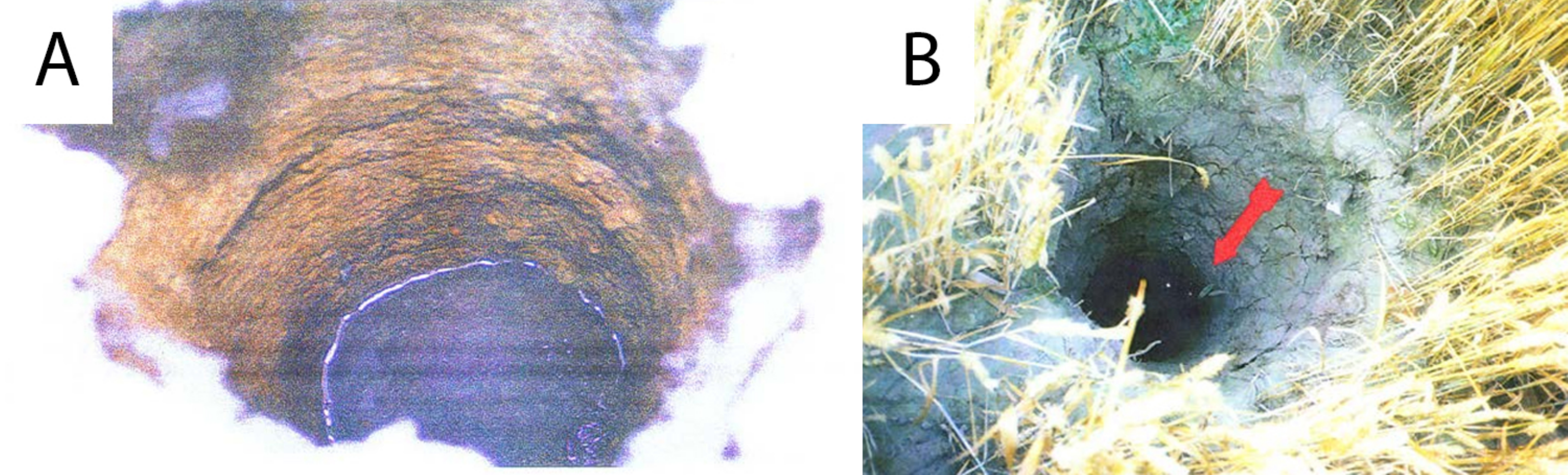}}
\caption{A. A photograph (taken the next day) of the hole drilled by the UFO.  (Credit: Adam Piekut)  B. A photograph of the same hole taken in late April of the same year when the soil sample was taken (location indicated by the red arrow). (Credit: Adam Piekut)} \label{fig:dabrowka-hole}
\end{figure}

Around midnight on 29 July 2014, two tenants witnessed about 24 orange-red UAPs descend on their neighborhood.  One of the objects acted as if it were malfunctioning, dropping a flaming piece of material (Fig. \ref{fig:dabrowka-hole}A), which was recovered and studied by Phyllis Budinger of Frontier Analysis, Ltd.  Budinger's technical report describes the event: \cite{Budinger-UT087:2014}
\begin{displayquote}
    Four witnesses of the event include a man, his wife, and two tenants who rent an upstairs room in their house. The tenants were outside about midnight on June 29th smoking cigarettes when they observed multiple orange-red unidentified aerial objects descend over their neighborhood. At the height of the event there were up to 24 objects. They were silent and floated irregularly. One of the tenants filmed the event using his SamSung Galaxy S5 phone. Toward the end, the objects appeared to slowly and silently move off into the distance in sort of a formation. One orb was still floating behind the others and acting irregularly as if it were malfunctioning. It seemed to drop/jettison something that burnt wildly the whole way down to earth. This was now about 1 a.m. Then the object instantly ``darted'' out of sight going in another direction than the other objects. (It went back ``in the direction it came from.'') The `fireball' dropped straight down, being a fairly windless night, taking about 10 seconds. It landed four houses up the street. The yellow flame reached about 2' [$2\unit{inches} \approx  5\unit{cm}$] in height, which was eventually stomped out. The residual fragment was retrieved and kept in a small Styrofoam container until MUFON investigators arrived.
\end{displayquote}

Fourier transform infrared spectra were acquired using a Thermo Electron Avatar 360 spectrometer with the Smart Herrick diamond sampling accessory \cite{Budinger-UT087:2014}.  The infrared spectra obtained from the `as received' sample revealed that the material consisted mainly of palmitic acid (hexadecanoic acid) and a smaller amount of glassy material in the unburned area (Fig. \ref{fig:dabrowka-hole}B); while the burned area (Fig. \ref{fig:dabrowka-hole}C) only has a small amount of palmitic acid with the glassy material, suggesting that much of the palmitic acid had burned off \cite{Budinger-UT087:2014}.

The sample was washed with hexane followed by acetone to better isolate the constituents \cite{Budinger-UT087:2014} (Fig. \ref{fig:dabrowka-hole}D).  The component isolated using hexane was palmitic acid, which consisted of 32\% of the weight of the sample.  Budinger reasoned that this fractional weight was initially higher because some of the palmitic acid had burned off.  The components isolated using acetone comprised 7\% of the fractional weight and were found to be a mixture of long carbon chain carboxylic acids, which may be impurities associated with palmitic acid.  A trace of an ester-type impurity was also detected.  The remaining insoluble material was glass fiber with trace amounts of quartz \cite{Budinger-UT087:2014}.

Budinger concluded that the fragment did not originate from a Chinese lantern.  Furthermore, she noted that the material has the general appearance of an insulating material (See Figs. \ref{fig:delaware-UT087}E and F for comparison with glass wool and home insulation), but that the presence of palmitic acid is unusual. Perhaps the fact that palmitic acid can be used as a waterproofing agent \cite{Sharif+etal:2022} is important.  Budinger also noted that palmitic acid is a phase change material (PCM) that can be used for thermal energy storage.  In fact, a recent publication focused on the use of palmitic acid/\ce{SiO_2} composites (with \ce{SiO_2} in fibrous glass form) as thermal energy storage \cite{Fang+etal:2011}.

\subsubsection{UFO Hole Drilling, 2003, Dabrowka, Poland}
In the early evening (5:20pm) in Dabrowka, Poland, Zofia Marciniak was in her yard with her dog when she saw a brilliant one-half-meter-tall triangular-shaped object hovering at treetop level about 200\unit{m} away, over the field, next to her home \cite{Talbott-Budinger:2004, Budinger-UT035:2004}.  The triangular object was covered with many red, yellow, green, and blue multicolored lights, much like a Christmas tree.  The object hovered in place for about 15 minutes, and she watched it with curiosity.  After 15 minutes, she had to go into the house to remove something from the stove.  When she returned several minutes later, the UFO was no longer there.  She said that it then reappeared, but the lights were all white and had it moved further away to the right.  The lights then became multi-color again.  The lights then dimmed for about a minute and then the UFO disappeared \cite{Talbott-Budinger:2004, Budinger-UT035:2004}.

Later that night, around 9:00 pm, Mrs. Marciniak called her grandson, Adam, and told him what she saw.  Adam arrived the next day to see if he could find any physical evidence of the UFO's presence.  The field was covered with snow, so they did not expect much.  They found a 30\unit{cm} hole that was more than 5\unit{m} deep and was filled with about 1\unit{m} of standing water (Fig. \ref{fig:dabrowka-hole} A), which was surprising because the temperature was below freezing.  The hole appeared to have been drilled, but there was no evidence of any soil that had been removed from the hole.  In late April that year, Adam took a soil sample of about 0.5\unit{m} down inside the hole (Fig. \ref{fig:dabrowka-hole} B), and a control sample taken from the field about 10\unit{m} away from the hole \cite{Talbott-Budinger:2004, Budinger-UT035:2004}.

The soil sample was studied by Phyllis Budinger (Frontier Analysis, Ltd.). \cite{Budinger-UT035:2004}.  Budinger specifically tested the presence of calcium oxalate, since it had been found in other physical traces left by UFOs, such as in the Delphos case, above \cite{Budinger-UT001:1999}.  Budinger did not detect the presence of calcium oxalate.  Instead, she found that the soil from the sides of the hole was mainly composed of quartz, kaolinite clay, and a smaller amount of calcium carbonate.  She found that the control sample was also mostly quartz but that while there might be traces of other minerals, there was no kaolinite clay or calcium carbonate.  Budinger noted that the control sample should have been taken at the same depth as the surface soil was probably different from the soil 0.5\unit{m} down \cite{Budinger-UT035:2004}.

\subsubsection{Angel Hair}
Occasionally, UAP have been observed to be associated with a sticky white fibrous substance that floats down from the UAP as if it were released or created by the UAP \cite[p. 170]{Aime:1958}\cite{Maney+Hall:1961}\cite[p. 89]{Condon:1969}\cite{Basterfield:2001, Bott:2002}.  This substance is generally called \textit{angel hair}, or \textit{siliceous cotton}. Angel hair resembles the gossamer threads used by spiders in using air currents to travel from one location to another; a process called ballooning or kiting \cite{Glick:1939, Bell+etal:2005, Morley+Robert:2018}.  In fact, several posited examples of angel hair have been found to be spider silk \cite[p. 97]{Hall:1964}\cite[p. 89]{Condon:1969}.  In addition to angel hair, similar evanescent residues, such as a misty, smoke-like deposit, or luminous haze, have been reported \cite[pp. 19--32]{Lorenzen+Lorenzen:1967}\cite[p. 89]{Condon:1969}.  Phyllis Budinger (Frontier Analysis, Ltd.) has studied a number of angel hair samples from different events.  We first consider Budinger's reports on angel hair falls that were not related to any observed UAP \cite{Budinger-UT002:2002, Budinger-UT003:2002, Budinger-UT018:2002, Budinger-UT046:2006}. 

\addcontentsline{toc}{subsubsection}{\hspace{0.5cm}Angel Hair Cases Not Related to UAP}
\paragraph*{Angel Hair Cases Not Related to UAP}
\label{sec:AngelHair-NoUAP}
    \mbox{}\\   
In the case occurring in the northeastern farm country of Colorado, USA, in October 2005 \cite{Budinger-UT046:2006}, long web-like white strands were observed over a wide area (of more than ten miles linear distance along roads) drifting through the air, hanging on power lines, fences, and foliage, and on vehicles.  The material appeared as fine strands that were combined in a rope-like fashion.  Budinger noted that the strands had very fine droplets adhering to the surface, indicative of a natural ester, which would explain their adhesive properties and indicate their source as being biological.  Infrared spectra showed prominent absorption bands due to N-H (3700--3000\unit{cm}$^{-1}$), secondary amide C=O (1650--1600\unit{cm}$^{-1}$) and CNH (1580--1470\unit{cm}$^{-1}$) \cite{Budinger-UT046:2006}.  These strong bands, along with weaker bands, identify the strands as a protein material.  Budinger also noted that there was a very weak ester C=O absorption band (1740\unit{cm}$^{-1}$), as expected from the droplets.  The spectra were more similar to the spectra of silk from spiders, caterpillars, and silkworms, suggesting that the origin was biological.

Three other cases, none of which involved UAP, were similar \cite{Budinger-UT002:2002, Budinger-UT003:2002, Budinger-UT018:2002}.  One set of samples was found in November 1998 in the carport attic of John Timmerman, Lakeview, Ohio, USA. The samples had been preserved there in glass jars and an envelope for several years after the closure of the Chicago office of CUFOS (Center for UFO Studies).  One of the samples came with a handwritten note from late ufologist Paul Cerney, which indicated that the samples were collected in October 1977 in Los Gatos, California, USA \cite{Budinger-UT002:2002}.  The Los Gatos samples were determined to be polymers containing protein amide-type linkages.  The presence of these amide-type linkages rules out spider silk, as these structures are more characteristic of caterpillar silk, such as from silkworms and tent caterpillars.  Trace amounts of a long carbon chain ester and carboxylic acid were detected, but it is unclear whether this was associated with the sample or whether it was a contaminant.  There was no evidence of volatile decomposition products in the samples, which is not unexpected as any volatiles would have dissipated after two decades \cite{Budinger-UT002:2002}.  The material, collected from Sacramento, California, USA in 1999 \cite{Budinger-UT003:2002}, was similar to the Los Gatos material.  The fibers were made of a polymer containing protein-amide type linkages, identical to the Los Gatos material.  However, the fibers in the Sacramento material were coated with droplets that were determined to be a natural long-chain fatty ester-type material.  In addition, the Sacramento material contained volatiles that were determined to be: 2-methyl propane; 2-methyl-1-propene; 2-methyl-1-butene; 2-methyl pentane; 3-methyl pentane; hexane; dimethyl-pentane; two \ce{C_6H_{12}} components (molecular weight = $84\unit{g/mol}$) hydrocarbon structures (specific isomers unidentified); one \ce{C_8H_{16}} component (molecular weight=$112\unit{g/mol}$) (specific isomer unidentified).  There were possibly trace amounts of heavier hydrocarbons, such as two \ce{C_{20}H_{42}} components (eicosane or icosane) and a \ce{C_{23}H_{43}} component.  In addition, trace amounts of carbonyl sulfide (COS) and carbon disulfide (CS2) were indicated \cite{Budinger-UT003:2002}.  The third set of samples from the California Sierras (USA) collected in November 2001 was similar to the other two \cite{Budinger-UT018:2002}.

\addcontentsline{toc}{subsubsection}{\hspace{0.5cm}Angel Hair Cases Related to UAP}
\paragraph*{Angel Hair Cases Related to UAP}
\label{sec:AngelHair-YesUAP}
    \mbox{}\\   
There were two sets of angel hair samples that were associated with UAP.  In the first case, Shenandoah, Iowa, USA, on 4 October 1981, a UAP with the appearance of a silver dollar in the sky was observed in a bright clear sky at an altitude of about $60^\circ$.  The UAP appeared to stand still for about a minute before it suddenly sped up and disappeared.  After this, large `~``globs'' of a white fibrous material were observed to be floating in the sky.  It moved in giant spirals and drifted down covering telephone lines, trees, and bushes everywhere in town, and the surrounding country.'  The fibrous material was found to be a polymer containing protein amide-type linkages, similar to caterpillar silk \cite{Budinger-UT017:2001}.

Another event in Illinois, USA, in 2016, which involved two witnesses, was described by investigator Bill Schroeder \cite{Budinger-UT096:2017}.  Witness 1 was an adult male with a Ph.D.-level education who had been working in the aviation industry for more than 30 years, designing military and civilian aircraft systems.  Witness 2 was an adult female, with a college education, who was a professional artist.  Schroeder described the event as:
\begin{displayquote}
On November 4th, 2016, the Investigator was contacted by witness 1. He advised that he and witness 2 had observed a number of Barbell shaped objects in flight over his dwelling. The object first observed was silver metallic in color and 20 to 30 inches in
length. When first observed the object was at rooftop level or about 35 feet over his head. The object was moving silently and laterally towards a small lake just south of the dwelling. At the time of the sighting witness 1 was in the company of witness 2.
Witness 1 noted that there was no visible means of propulsion observed and no sound detected. As the object arrived to the center of the lake it stopped moving laterally and began to rise. As it started to rise it began exuding a white filament like material. The material was very fine and stretched all the way to the ground. After it had risen to several hundred feet is was joined by more than 12 other like objects that seemed to be circling it as it continued to rise. At this point in time the white material was spewing from the devices in such quantity it looked like rain. The objects rose together into a clear sky until they were no longer visible.

The white material that had fallen remained visible in trees for about 24 hours before  dissipating. When witness 1 contacted the Investigator, he was requested to get samples if  possible. The witnesses were able to obtain some samples of the material. Witness 1 also sent  an email report of the sighting. Witness 2, who is a professional artist, painted a picture that  depicted the sighting. The material was collected using a clean shish-kabob skewer. At the  Investigators request the samples were placed in paper and plastic bags. The samples were sent to the Investigator... 
\end{displayquote}

Analysis performed by Budinger found that the material is biologically-derived silk typical of that produced by spiders.  Celluloid material (possibly from plants) was also found, which could have been a contaminant \cite{Budinger-UT096:2017}.

\begin{figure}
\centering
\makebox{\includegraphics[width=0.9\columnwidth]{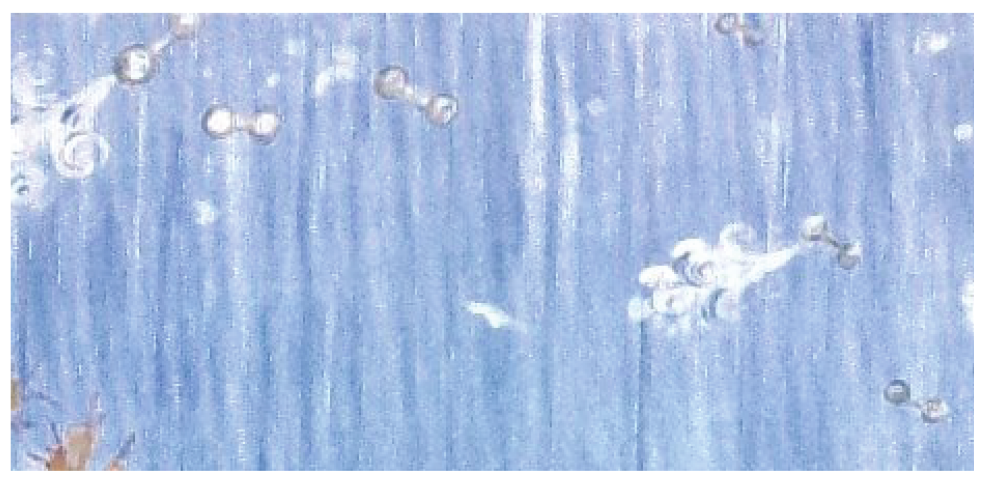}}
\caption{A painting made by Witness 2 of the 2016 Illinois, USA, observation of several small metallic barbell-shaped objects spewing angel hair. (Source: \cite{Budinger-UT096:2017})} \label{fig:angel-hair-w-barbells}
\end{figure}

Another case involves Mrs. Retha Rutherford of Burlington, West Virginia, USA, who on 21 September 2000 found a strange white fibrous material in her yard \cite{Budinger-UT008:2000}. She and her husband took photographs and tried to collect the material, which turned into a clear gelatinous material.  Although no UAP was observed to be associated with the material, Mrs. Rutherford noted that at 7:00 pm the previous night, she heard a loud droning sound, similar to a large airplane, which lasted for about an hour.  The source of the sound was not visually obvious.  Afterward, the dog became sick and vomited, and Mrs. Rutherford suffered a severe sinus attack.  The fact that the sixth of the ``Five Observables'' involves biological effects on humans and animals suggests that the seemingly correlated illness might indicate that a UAP was present, although not visually observed, 

The analysis \cite{Budinger-UT008:2000} showed that the fibrous material was similar to the Los Gatos material \cite{Budinger-UT002:2002} in that it was a protein related to caterpillar silk and was coated in droplets like the Sacramento material \cite{Budinger-UT003:2002}.  However, these samples were unique because of the presence of fatty acid amines.  Budinger notes that the following are `specifically ``suggested''': 4-methyl-pentamide; hexadecanamide; dodecanamide; and N-tetradecanoic acid amide.  Other heavier hydrocarbons were found, such as eicosane (\ce{C_{20}H_{42}}) and 2-methyl hexadecane.  Eicosane was also found in the Sacramento material \cite{Budinger-UT003:2002}.  Budinger hypothesized that these could be related to the alleged gelatinous material and perhaps related to the final degradation products \cite{Budinger-UT008:2000}.

It is strange that these cases of angel hair all seem to involve caterpillar and sometimes spider silk, especially when, in some of the cases, UAP were observed, by multiple witnesses, to be extruding the material.  Although the material is known to have a biological origin, this does not completely rule out the possibility that some UAP can produce this material.  This would make some sense if the UAP were under some form of intelligent control, since scientists and engineers recognize the unique combinations of properties, such as the strength, toughness, extensibility, and energy absorption, of spider and caterpillar silk, and that it is not known how to achieve such a combination of properties with nonprotein-based materials \cite{Dou+etal:2019}.  In the following section, we examine a well-known case in which the angel hair was found to be borosilicate glass.

\begin{figure}
\centering
\makebox{\includegraphics[width=0.9\columnwidth]{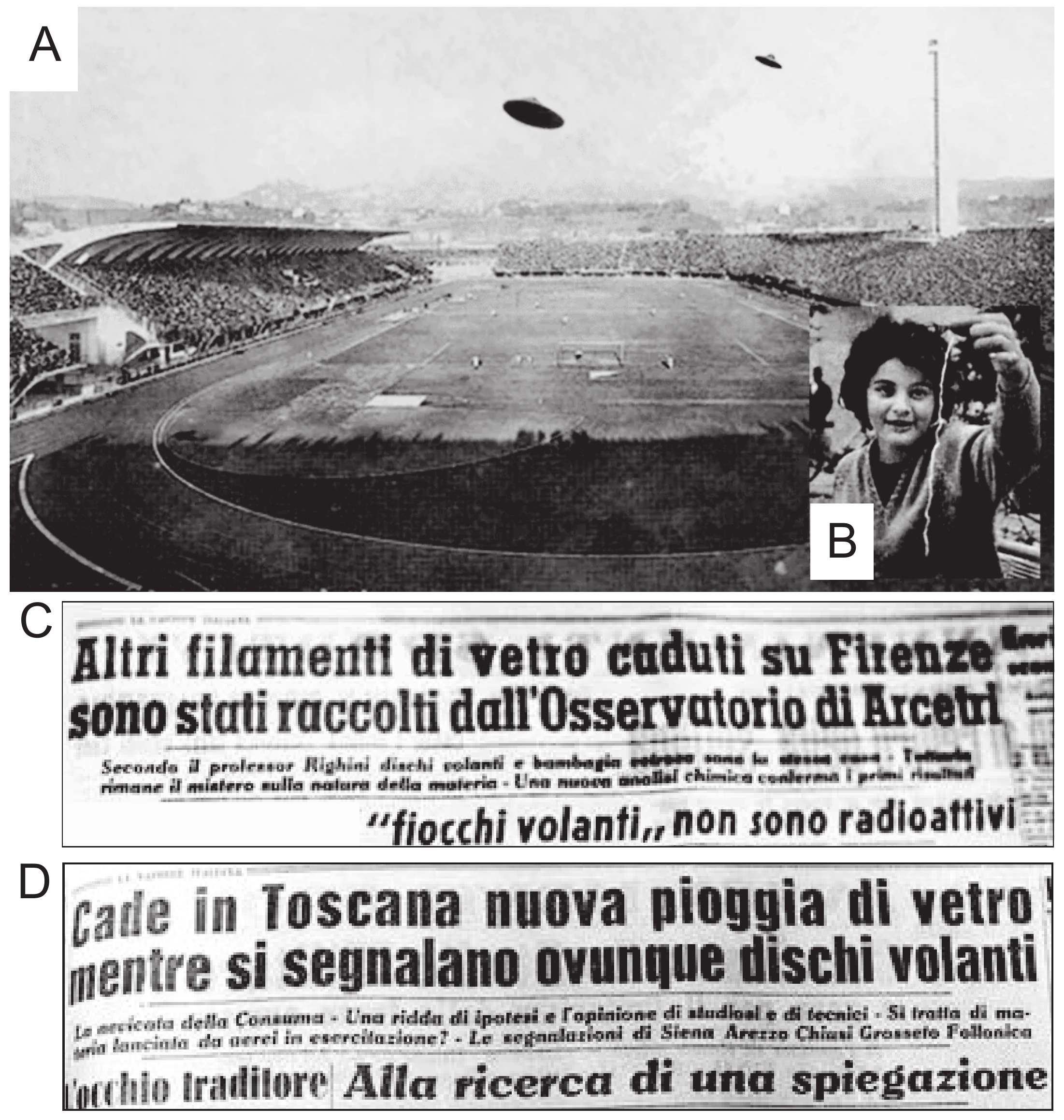}}
\caption{A. An image of the Stadio Artemi Franchi in Florence Italy over which the UFOs were seen (the two saucers were added by the source for illustrative purposes).  B. A young person holding up some of the angel hair that fell on the crowd.  (Source A and B: FUL Magazine [\url{https://firenzeurbanlifestyle.com/}] \cite{Caliani:2023}) C. Headlines of the Florentine paper \textit{La Nazione}, which reads ``Altri filamenti di vetro caduti su Firenze sono stati raccolti dall'Osservatorio di Arcetri'' (English Translation: Other glass filaments that fell on Florence were collected by the Arcetri Observatory.) D. Headlines of the Florentine paper \textit{La Nazione}, which reads ``Cade in Toscana nuova pioggia di vetro mentre si segnalano ovunque dischi volanti'' (English translation: New glass rain falls in Tuscany while flying saucers are reported everywhere.)} \label{fig:firenze}
\end{figure}

\addcontentsline{toc}{subsubsection}{\hspace{0.5cm}UFOs Interrupt a Soccer Match in Florence Italy}
\paragraph*{UFOs Interrupt a Soccer Match in Florence Italy}
    \mbox{}\\   
One of the most famous events involving UFOs and angel hair occurred in Florence Italy on 27 October 1954 at about 2:20 pm during a soccer match between the Fiorentina and Pistoiese teams at the Stadio Artemi Franchi (Fig. \ref{fig:firenze} A) \cite{Caliani:2023, Padula:2014, CentroRicerchePrato}.  Just after half-time, the crowd went silent and then roared as the ten thousand spectators no longer were watching the game but were instead pointing up at the sky watching as several cigar-shaped, or wing-shaped, objects flew fast towards the stadium and then stopped overhead.  Glittering glass-like filaments (angel hair) were observed to float down from the objects (Fig. \ref{fig:firenze} B, C, and D).  The game came to a stop as everyone, including the players, watched the spectacle.  The footballer Ardico Magnini was on the pitch and described the event as \cite{Padula:2014}
\begin{displayquote}
    It was something that looked like an egg that was moving slowly, slowly, slowly. Everyone was looking up and also there was some glitter coming down from the sky, silver glitter. ... We were astonished we had never seen anything like it before. We were absolutely shocked.
\end{displayquote}
A lifelong Fiorentina fan, Gigi Boni, described it as:  \cite{Padula:2014}
\begin{displayquote}
    I remember clearly seeing this incredible sight.  They were moving very fast and then they just stopped. It all lasted a couple of minutes. I would like to describe them as being like Cuban cigars. They just reminded me of Cuban cigars, in the way they looked.
\end{displayquote}

Roberto Pinotti, the president of Italy's National UFO Center, was 10-years old at the time, recounts the angel hair: \cite{Padula:2014}
\begin{displayquote}
    ... the same time the UFOs were seen over Florence there was a strange, sticky substance falling from above. In English we call this `angel hair'. \\
    ...\\
    The only problem is after a short period of time it disintegrates.\\
    ...\\
    I remember, in broad daylight, seeing the roofs of the houses in Florence covered in this white substance for one hour and, like snow, it just evaporated.\\
    ...\\
    No-one knows what this strange substance has to do with UFOs.
\end{displayquote}

Giorgio Batini, a journalist for the newspaper La Nazione, had received hundreds of calls about the sightings, which occurred in several towns across Tuscany that day and the days that followed.  Batini went to the top of the newspaper's building to see what was going on, and he recalled seeing ``shiny balls'' moving fast toward the dome of the Cathedral of Santa Maria del Fiore.  He then went to investigate and found that a wood outside the city was covered in white fluff.  Batini collected several samples by rolling the stringy material up on a matchstick.  He took the samples to the University of Florence, as did a number of other people. \cite{Padula:2014}

Prof. Giovanni Canneri, the Director of the Institute of Chemical Analysis at the University of Florence found that the material was not radioactive, and using elemental spectrographic analysis, he found that the material contained boron, silicon, calcium, and magnesium. \cite{Padula:2014}  He wrote that: \cite{CentroRicerchePrato}
\begin{displayquote}
    Substance with a fibrous structure, with notable mechanical resistance to traction and torsion. When heated, it darkens, leaving a fusible and transparent residue.
    The fusible residue is spectrographically shown to contain mainly: Boron, Silicon, Calcium and Magnesium.  Substance with a macromolecular structure, probably thread-like. Purely hypothetically, the substance examined in the microchemical scale could be: a borosilicate glass.
\end{displayquote}
Unfortunately, the hypothesis that the material was a glassy substance is problematic.  The lower limit for X-ray spectroscopy is in the neighborhood of the elements fluorine ($Z=9$) \cite{Jenkins:1984} and sodium ($Z=11$) \cite{Budinger:2025}.  This means that lighter elements, such as oxygen ($Z=8$), nitrogen ($Z=7$) and carbon ($Z=6$) are difficult to analyze using most elemental analysis techniques \cite{Jenkins:1984}.  Instead, if the material was spider or caterpillar silk, as in other cases that have been studied with more advanced equipment \cite{Budinger-UT017:2001, Budinger-UT096:2017, Budinger-UT008:2000}, the presence of the elements found by Canneri could be explained by dirt that contaminated the silk surface \cite{Budinger:2025}.

\section{Transmedium Travel and Water} \label{sec:trans-medium}
While the main historical and current focus has been on the aerial aspects of UAP, their transmedium capabilities and relationship with water are an extremely important aspect of their characteristics and performance.  

The three videos released by the US government were taken by Naval aircraft at sea, and in the 2004 Nimitz encounter case (FLIR 1 video: Sec. \ref{sec:AATIP}) the Tic-Tac-shaped UAP was first observed hovering over the roiling sea surface, which could have been an effect of the UAP on the water, but could have indicated the presence of a larger submerged UAP \cite{Powell+etal:2019, Gallaudet:2024}.  In addition, the US Homeland Security video of an object observed over the Rafael Hern\'{a}ndez Airport (BQN) in Aguadilla, Puerto Rico, in 2013 showed an ellipsoid-shaped object, which after flying over the airport, headed out to sea where it dipped into the ocean multiple times and traveled underwater without significantly slowing down or creating splashes, waves, or other disturbances \cite{Powell+etal:2015}.

To use a colloquialism, it does not take a rocket scientist (or an aerospace engineer, for that matter) to recognize and appreciate that craft with effective and capable transmedium abilities are quite beyond human competency.  For one example, a seaplane is both a very poor airplane and a very poor boat.  For this reason, the observed transmedium behaviors of UAP should perhaps give us the most pause.

\subsection{Historical Significance of USO (Unidentified Submerged Object) Encounters} \label{sec:HistoricalUSOs}

Observations of UAP at sea have been recorded in ship logs for several centuries \cite{Roberts:1967, Vallee+Aubeck:2010, Dolan:2025} making it clear that not all UAP can be explained away as being advanced secret, military, or commercial craft.  It is important to examine some of these old cases and appreciate both their diversity and similarities to modern encounters.  

When we explore the old encounters, we find many intriguing events that never receive the attention they appear to deserve.
The following event from 1887 occurred near Cape Race, off the coast of Newfoundland, Canada \cite{Parker:1887}\cite[pp. 442--443]{Mendenhall:1890}\cite[pp. 167--168]{Fort:1941}\cite{Dolan:2025}.  It reads as follows:
\begin{displayquote}
    Captain Moore, British steamship Siberian, reports, `Nov. 12, midnight, Cape Race bearing west by north, distant ten miles, wind strong south by east, a large ball of fire appeared to rise out of the sea to a height of about fifty feet and come right against the wind close up to the ship. It then altered its course and ran along with the ship to a distance of about one and one-half miles. In about two minutes it again altered its course and went away to the south-east against the wind. It lasted, in all, not over five minutes. Have noticed the same phenomenon before off Cape Race, and it seemed to indicate that an easterly or south-easterly gale was coming on.' 
\end{displayquote}
This event is interesting because the fireball emerged from the sea and appeared to maneuver intelligently by both moving against the wind and by changing course.
Following this report, in Fort \cite{Fort:1941}, there is a note of a similar event that occurred on the \nth{18} of June 1845 where from the brig \textit{Victoria} in the Mediterranean Sea about 900 miles east of Adalia in Asia Minor, ``three luminous bodies were seen to issue from the sea, at about a half a mile from the vessel.  They were visible about ten minutes.'' \cite[p. 168]{Fort:1941}.  The pheonomenon, which was reported from Syria and Malta, was described as two very large bodies ``nearly joined'' \cite[p.82--83]{Greg:1860}.

Such events would seem extraordinary if reported today. When one considers that this was recorded roughly a century and a half ago, it takes on even greater significance. As seems typical of such events, it was reported once and then forgotten.  Such events have sometimes been explained as ball lightning. However, such an explanation seems unlikely given that the objects emerged from the sea and had such long durations.

Another fascinating early encounter with an odd light phenomenon that reportedly engaged with the water occurred in West Africa in the late 1800s. This is a puzzling incident that is difficult to explain conventionally. 
In 1893, Mary Kingsley was on an exploration of Lake Ncovi between the Ogowe (modern spelling Ogooue) and Rembwe rivers \cite{Alexander:1990} where she had collected species of fish unknown to western science\footnote{The fish species \textit{Mormyrus kingsleyae} and \textit{Ctenopoma kingsleyae} are named after Kingsley. \textit{Ctenopoma kingsleyae} is a species of climbing gourami, which are named for their ability to climb out of water and walk short distances using their fins and tail for propulsion and gill plates for support.}. 

This area was in the Niger Protectorate and Gabon, about 250 miles inland from the West African coast. Kingsley wrote that, late at night, she went out alone to bathe and canoe on the lake waters. Then she saw something extraordinary: \cite[p. 254]{Kingsley:1897}
\begin{displayquote}
Down through the forest on the lake bank opposite came a violet ball the size of a small orange\footnote{Kingsley's encounter with a small luminous violet ball is strikingly similar to the close encounter with a small blinking violet light had by two patrolmen at Killeen Atomic Natonal Stockpile in 1949 (Sec. \ref{sec:nukes}).}. When it reached the sand beach, it hovered along it to and fro close to the ground. In a few minutes another ball of similarly colored light came towards it from behind one of the islets, and the two wavered to and fro over the beach, sometimes circling round each other. I made off toward them in the canoe, thinking — as I still do — they were some brand new kind of luminous insect. When I got onto their beach, one of them went off into the bushes and the other away over the water. I followed in the canoe, for the water here is very deep and, when I almost thought I had got it, it went down into the water and I could see it glowing as it sunk until it vanished in the depths.
\end{displayquote}
She later spoke to locals in the area who explained her sighting as an encounter with an ``Aku,'' or a devil bush. 

It is not surprising that Kingsley herself thought that she had seen a new type of glowing insect. Perhaps this was a reasonable assumption, except that we do not know of any such insect as she described. These objects might not be USOs in the sense that we tend to think of them, in other words, as a structured and advanced submersible craft.  However, we must keep in mind that UAP are not a single type of phenomenon.  What is important is that they were luminous (violet in color), unusual, and had the ability to enter bodies of water. That is, they appeared to be transmedium objects, whatever they were made of. Furthermore, they appeared to act intelligently and evasively. That is, when Kingsley attempted to approach them, they moved away from her.

\subsection{USOs: Unidentified Submerged Objects}
Sightings of unidentified flying objects have long been paralleled by encounters with unidentified submerged objects (USO).  A military forerunner were mystery submarines, which on several occasions have generated alarm of military navies and the interest of the media. The most famous series of sightings and sonar detections took place in the Norwegian fjords from November \nth{12} to \nth{22} in 1972 \cite{Holst:1974}. It was the Norwegian Navy who first used the term USO. The military and almost all commentators were convinced that it was actually a Soviet submarine, caught in the meshes of NATO defense, which suddenly disappeared, but UFO researchers noticed some parallels between the two phenomena.

The connection with the UFO phenomenon has been greatly reinforced by sightings of flying objects descending from the sky and submerging into seas (or lakes), as well as observations of
UFOs seen coming out of the water and taking flight.
One of the first authors to dedicate a full book to USOs was the British-American biologist Ivan T. Sanderson \cite{Sanderson:1970}.
However, Carl Feindt has compiled the most complete worldwide catalog of USO reports, both as an online database called WaterUFO \cite{Feindt:WaterUFO} and as a book \cite{Feindt:2016}.  Richard Dolan has published the most recent study \cite{Dolan:2025}.

So far, almost 2000 reports of USOs have been reported and compiled.  Marco Bianchini \cite{Bianchini:2003} has suggested a classification scheme based on the dynamic behavior observed of the objects:
\begin{itemize}
  \item underwater objects, which are observed and remain below the water surface;
  \item objects that are observed entering or falling into the water;
  \item objects that are observed leaving the water and rising into the air;
  \item objects that do not enter or leave the water but remain stationary or in movement (floating) on the surface (usually arriving or departing in flight).
\end{itemize}

Long remaining on the edge of the UFO topic, the issue of USOs took on an unexpected relevance in 2017 when some of the sightings by US military personnel were revealed, reporting ``transmedia objects''.
Particularly interesting was the confirmation by the Pentagon press office in May 2021 \cite{Schwartz+Stelloh:2021} that the filmed footage of a flying object sinking into the sea was a video taken by personnel in the command center of the battleship \textit{USS Omaha} on the \nth{15} July 2019, during a complex series of sightings of ``unidentified drones'' that hovered
at length around naval ships engaged in an exercise off the coast of California, with performances described as unexpected and superior to that of normal drones.

Precisely because of the involvement of the Navy, this series of events led to an unprecedented reversal.  For decades, UFO sightings were in the exclusive domain of the Air Force, but after these events, UFO sightings and related investigations were entrusted in 2019 to the US Office of Naval Intelligence (ONI). Furthermore, in a twist, the acronym UAP (which had replaced UFOs) changed in meaning from Unidentified Aerial Phenomena to Unidentified Anomalous Phenomena, and even Unidentified Aerospace-Undersea Phenomena, thus losing the historical connotation connected to air and flight. 

On July \nth{20} in 2022, the US Department of Defense announced the creation of the All-Domain Anomaly Resolution Office (AARO) to study UAP, which, in recognition of the transmedium nature of UAP, incorporated the phrase ``All-domain''.  This was followed by public statements by (current and former) military members, such as the retired US Navy Admiral, Tim Gallaudet, who had served as the head of the National Oceanic and Atmospheric Administration (NOAA), that UAPs could pose a threat to the national security of the United States of America \cite{Gallaudet:2024}.

There are several interesting texts that focus on the underwater aspects of UAP: \cite{Sanderson:1970, Cochrane:1980, Rullan:2002, Feindt:2016, Dennett:2018, Dolan:2025}.  Oceanographer and retired Rear Admiral Gallaudet recently published a paper that emphasizes the importance of the underwater aspects of UAP and the difficulty in studying them \cite{Gallaudet:2024}.

\section{Scientific Field Studies}
There are several scientific studies currently being conducted by academics, scientists, and private citizens.  In this section, we briefly describe several prominent and notable studies.  These studies vary dramatically in focus and scope, as well as in structure and relationship to the academic and scientific communities.

To put these current studies into context, it is important to review prominent past efforts and individuals.  Although a comprehensive review is beyond the scope of this article, a useful summary is provided in the Appendix (Sec. \ref{sec:Appendix}).

\subsection{Academic/University Programs}

\subsubsection{Project Hessdalen, Hessdalen Norway. 1984 -- Present} \label{sec:Hessdalen}
Research had shown that anomalous phenomena sometimes tend to occur in specific areas with reasonable recurrence, and this inspired the creation and maintenance of Project Hessdalen as one of the longest-running instrumented ``targeted search'' studies of UAP research.  As such, Project Hessdalen, which has been led by Erling Strand and supported, in part, by {\O}stfold University College, has been operating in the Hessdalen area of Norway since 1984 \cite{Strand:1984}.

The results of the research carried out there show prominent recurrent anomalies in the behavior manifested by the observed phenomenon \cite{Teodorani:2004}. This project has demonstrated that similar areas around the world can be used as a laboratory to study the phenomenon systematically using multimode and multiwavelength instrumentation \cite{Teodorani:2024}. 

The “Hessdalen lights” \cite{Hessdalen} represent the prototype of anomalous light phenomena in the atmosphere. They generally consist of multi-shaped and multicolored balls of light, characterized by a long duration, and sometimes by high-energy emission. The lights range in size from half a meter to 30 meters. Very short-lasting ``flashes'' in the sky most often precede Hessdalen Light phenomena. Sometimes plasma-like orbs overlap with the transient appearance of apparently structured phenomena: The reason for this connection is not known yet, but it should be investigated in depth. The established characteristics of recurrence make this type of physical event very suitable for systematic measurement campaigns. The project demonstrated for the first time that the Hessdalen phenomenon was measurable and that, in particular, the luminous events were clearly correlated with magnetic disturbances and produced a strong but intermittent radar signature and sometimes sudden and short duration VHF radio waves. In one occasion, the light phenomenon showed a photoreactive behavior when a laser beam was aimed at it. 

The topic was discussed in 1994 at an important international workshop on ball lightning, where the unanimous conclusion was reached that the phenomenon deserves careful analysis according to the protocol of physical sciences. In fact, the observational developments of this research have rapidly evolved. In time, two automatic sensing stations have been activated (the `Blue Box' and the `White Box'), which allowed the acquisition of data around the clock, including videos of the phenomenon. Over the course of about 30 years, observers have made it possible to build reliable statistics on the phenomenon, showing that it tends to appear with peaks in the winter period, occurs mostly in the time slot from 10pm to 1am, but appears everywhere in the sky and on the ground, and does not follow preferential airways \cite{Teodorani:2004}. Now, the Hessdalen Project is being relaunched with new and more advanced sensors and research strategies \cite{Hessdalen}.

The main open problems, which make the Hessdalen phenomenon an interesting subject for physics investigation, can be summarized as follows:

\begin{enumerate}
\item The mechanism of confinement of the plasma in a small volume and without energy losses.
\item The high energy produced both optically and in the radio band and, sometimes, the long duration of the light phenomena.
\item The constancy of temperature, the self-containment characteristics of the light balls, and the coexistence of white and red spheres of equal size.
\item The sudden way in which the phenomenon turns on and off, and the formation of clusters of light balls, also with characteristics of the expulsion of secondary orbs.
\item The apparent characteristics of “solidity” shown by some of the phenomena encountered.
\end{enumerate}

Hessdalen is not the only world area of interest regarding recurring anomalous phenomena. Similar phenomena have been reported in locations such as Brown Mountain, Cape Fear River, Joplin, Piedmont, Uintah Basin, the Hudson Valley, and the Yakima Indian reservation, in the United States, for instance, as well as other areas of the world.

The main goal of this research is to try to understand the physics of the observed phenomena, whatever they may be. The physics of what is reported and measured in Hessdalen and similar areas of the world could deal with a natural phenomenon of possible geophysical origin \cite{Freund:2003,Pettigrew:2003,Vargemezis+etal:2024} as well as a possible propulsion mechanism that has nothing in common with those used in conventional aircraft and rockets. These phenomena can be investigated using exactly the same procedures as used in astrophysics and standard methodologies of science \cite{Teodorani:2024}. 


\subsubsection{UAlbany-UAPx Collaboration}
The Physics Department at the University at Albany, State University of New York (SUNY), collaborates with UAPx to study UAP through the collection of field data.  UAPx is a federal 501(c)(3) non-profit organization, cofounded by US Navy veterans Gary Voorhis and Kevin Day, who were involved in the 2004 \textit{Nimitz} carrier strike group tic-tac encounter \cite{Powell+etal:2019}. 

UAPx specifically recruited Ph.D. faculty in physics to work with engineers and technicians to collect and analyze their own data on UAP. The Collaboration's strategy is designed to maintain the quality of raw data, their chain of custody, their provenance, and their quantitative analysis. 

The physicists and engineers in the UAlbany-UAPx Collaboration represent a broad range of unique sets of skills, including knowledge of radiation detection, image processing, exoplanets, quantum foundations, and systems engineering. The members also differ in their opinion on the nature of UAP, while striving for internal skepticism, introspection, and balance.

Like other efforts, the UAlbany-UAPx Collaboration uses sensors and techniques similar to those of both the Galileo (Sec. \ref{sec:GP}) and the Eye on the Sky -- Nightcrawler (Sec. \ref{sec:Nightcrawler}) Projects. The UAlbany-UAPx Collaboration deploys existing equipment in novel ways while working to develop new detectors as necessary. Just as in the Galileo project, the UAlbany-UAPx Collaboration seeks diversity in sensors, going far beyond just having multiple cameras, using an agnostic, statistical-based approach inspired by particle physics when looking for anomalies, which UAPx defines separately from ambiguities in its paper.  Specifically, the UAlbany-UAPx Collaboration paper, published in this issue \cite{Szydagis+etal:2024}, suggests that (scientific) UAP researchers adopt the following conventions: A \textit{Coincidence} is defined as a ``simultaneous occurrence'' within the temporal resolution, and spatial resolution when germane; whereas an \textit{Ambiguity} requiring further study is a coincidence between two or more detectors or data sets at the statistical level of $3~\sigma$ or more, with a declaration of a genuine \textit{Anomaly} requiring, the High Energy Physics (HEP)-inspired, $5~\sigma$ coincidence. The statistical significance must be defined relative to a null hypothesis, in our case accidental coincidence, combined with causally linked hypotheses, like cosmic rays striking a camera pixel.  In this way, one can rigorously quantify the meaning of extraordinary evidence in the same way as has been done historically by particle physicists, who have established a very high bar to clear.

In general, the UAlbany-UAPx Collaboration strives for openness, planning for data release after publication, and scientific publications in well-established scientific and engineering journals. 

\subsubsection{IFEX} \label{sec:IFEX}
The Interdisciplinary Research Center for Extraterrestrial Studies (Interdisziplin\"{a}res Forschungszentrum f\"{u}r Extraterrestrik, IFEX) is an institution of the Faculty of Mathematics and Computer Science at the Julius Maximilian University of W\"{u}rzburg. It was founded in 2016 by Prof. Hakan Kayal, who also heads the center today. The scientific research priorities of IFEX include space exploration, objects in our solar system, the search for signs of life, the search for extraterrestrial intelligence (SETI), and the study of Unidentified Anomalous Phenomena (UAP). In doing so, technologies are being developed to support the aforementioned research fields on Earth and in space, studies are being conducted on a variety of topics, and case investigations are being carried out as needed. IFEX also engages in public relations work to further expand science communication and organizes UAP workshops. As a result, there are many media reports on the work of IFEX.

The center maintains and expands national and international relationships to advance scientific work in the aforementioned research fields. In addition to the voting members from the faculty, a growing number of associated members from the national and international interdisciplinary environment support the work of IFEX; there are currently 21 members.

Among the projects currently funded by IFEX are SATEX, a study on the use of small satellites in extraterrestrial research, NEAlight, a study on small satellite concepts for the investigation of the near-Earth asteroid Apophis and VaMEx3-MarsSymphony, a demonstration of communication technologies and sensor systems in realistic exploration scenarios with a view to a future Mars mission in the Valles Marineris region.

An aspect of VaMEx3-MarsSymphony that is highly relevant to this paper is that it includes a UAP system element. In this context, VaMEx3-MarsSymphony is developing an UAP detection system designed for Mars, which will be demonstrated on Earth in 2025. An important feature is that VaMEx3-MarsSymphony is the first time in Germany that a UAP detection system is being developed as part of a government-funded project. The system can also be used to detect other transient events such as lightning or meteors in the Martian atmosphere.

In addition, IFEX is developing and operating detection systems such as SKY-CAM on Earth in close cooperation with the Professorship of Space Technologies. Based on its own earlier work in this field since 2009, SKY-CAM5 has been in permanent operation since December 2021 and SkyCAM-6 since March 2024. Both systems support the ongoing development of sensor systems for the detection and observation of UAPs and enable the accumulation of experience that will be used in the development of a more complex system in the future.

In space, the satellite SONATE-2 was developed and built at the professorship.  It is currently actively operated from its own mission control center and ground stations. SONATE-2 demonstrates the application of AI technologies in space for anomaly detection. The possibility of training the system on board is unique worldwide. This will make it possible, in the future, to search autonomously for anomalies, artifacts, or possibly even artificial signatures in new, unknown environments, such as new asteroids.

As part of the KI-SENS project, terrestrial and space-based detection and observation systems for objects in space are being developed and are now in operation with the support of students association W\"{u}Space. These currently include an active telescope system that is being further developed for detection.

\subsubsection{SETI Kingsland\\ Lough Key, County Roscommon, Ireland. 2000 - Present}
\label{sec:SETI-Kingsland}
SETI Kingsland, situated at Kingsland Observatory and directed by Eamonn Ansbro, was founded in 2000 and funded by Space Exploration Ltd. to establish permanent instrumentation for UAP astrophysical research.  The instrumentation was specifically located near Lough Key, County Roscommon in Ireland, where there is a known high frequency of manifestation of UAP.   The instrumentation was designed based on the experience and knowledge about the characteristics of UAP gained from 150 case studies over a 10-year period.

The instrumentation involved the design and development of 11 cameras using high quantum efficiency (QE) sensors in the visible range.  These cameras monitor the sky and are used to control instrumentation mounted on two optical platforms.  These optical platforms, which are separated by 150 yards, are activated when triggered by one of the 11 cameras.  The two optical platforms, which are equipped with long-range cameras, slew to the UAP, allowing the distance to the UAP to be determined by triangulation \cite{Ansbro:2001a}.
Additional sensors were added in 2002 to widen the range for gamma ray and infrared detection.  However, the key instrument was a unique spectrograph that has a scanning slit that covers a wide field of view when the UAP is moving \cite{Ansbro:2003}.

The operation has detected UAP of different shapes.  Photometric analysis showed significant plasma characteristics both within the UAP and also in the outer layers of these constructs \cite{Butler+Nally:2006}.
In addition to the high frequency of UAP occurrences at Lough Key, a pattern was discovered to the occurrences of the phenomena.  Analysis and results, using data from a number of databases, suggest that this pattern is part of a global phenomenon involving an average orbital period of 66 minutes with an orbital inclination between 0 and 18 degrees.  This raises the possibility that UAPs are technosignatures 
\cite{Ansbro:2001a, Ansbro+Overhauser:2001b, Ansbro+Ro:2024}.

\subsubsection{The Galileo Project} \label{sec:GP}
The Galileo Project (GP) is led by Avi Loeb of Harvard University and was established in July 2021 with co-founder Frank Laukien \cite{Loeb+Laukien:2023}.  The goal of GP is to search for extraterrestrial artifacts in the solar system.  One branch of the project is concerned with the study of interstellar objects (ISOs) \cite{Siraj+etal:2023}.  Early results from this work include the analysis of exotic ocean floor materials recovered near the atmospheric entry site of meteor CNEOS 2014-01-08, the first meteor reportedly of interstellar origin \cite{Loeb+etal:2024}; a follow-up expedition is planned for summer 2025.

The GP's UAP investigation is dedicated to ground- and satellite-based detection and characterization of UAP. The project is constructing multimodal ground-based observatories with the goal of developing instrumentation, data processing methods, and software to characterize the physical environment (the background) and to search for ``scientific anomalies'' in the atmosphere: ie, statistically unusual objects and phenomena that resist explanation in terms of prevailing scientific beliefs \cite{Watters+etal:2023}. As of October 2024, one of the Galileo Project observatories serves as a development platform and is in operation, while a second is under construction; a third is in the initial planning stages; the project is striving to make all three operational in 2025. A parallel effort has focused on building and testing inexpensive, portable, instrumented stations.  The UAP investigation is described in
\cite{Watters+etal:2023}, which contains a detailed description of (i)
motivations for the study; (ii) methods to determine instrument requirements; (iii) project-level requirements; (iv) a science traceability matrix (STM) connecting observations with physical parameters and instrument requirements; (v) a procedure to identify field sites. Additional papers have described the design of other subsystems \cite{Mead+etal:2023,Randall+etal:2023,Cloete+etal:2023}. The project aims to share methods and data products and has framed success in terms of laying a solid foundation for future work using modern methods.

So far, instrument development has focused on (i) microphones sensitive to the infrasonic and audible bands; (ii) radio spectrum analyzers sensitive to 100 MHz - 3.3 GHz; (iii) visible light and long-wave infrared (LWIR, 7.5 μm - 13.5 μm) camera arrays; and (iv) a nT-precision magnetometer. During the commissioning phase of each instrument, the project has used aircraft ADS-B transponder telemetry to calibrate, characterize, and thoroughly understand the in situ performance, including detection volumes and efficiency, allowing the evaluation and refinement of measurements.

To characterize and understand the signatures of ordinary phenomena and ordinary aerial objects, an aerial and environmental census will be conducted both during and after the commissioning phase. The GP team relies on a mixture of traditional methods (including manual review) and machine learning methods for novel class discovery (previously unknown aerial objects or phenomena) and outlier detection. These methods are being developed using instrument-acquired data (labeled and unlabeled) as well as synthetic data. Confirmed outliers and novel classes will be used to motivate refinements of instrumentation and will be targeted in future observations as part of hypothesis-driven investigations.

Aerial object detection in camera imagery has been accomplished using a combination of traditional methods and machine learning models. In the commissioning phase of its all-sky infrared camera array, the GP has amassed a database of $\sim 5\times 10^5$ aerial objects and their reconstructed 2-D trajectories in angular coordinates (mostly aircraft, birds, insects, and leaves) \cite{Domine:dalek}. These tracks have been the focus of efforts for novel class discovery and outlier analysis. The team has also developed and deployed lightweight, portable, multimodal, instrumented field stations with GPS-synchronized clocks, and has used these to acquire videos for deriving triangulated trajectories of nocturnal lights (e.g., aircraft, meteors, satellites) \cite{Szenher+etal:2023}. Acoustic instrumentation efforts have focused on (i) collecting a labeled database of sounds associated with specific aircraft, trajectories, and environmental conditions; (ii) developing machine learning models to recognize signatures of known craft; (iii) a direction-of-arrival (DOA) instrument using an array of five audible band microphones \cite{Mead+etal:2023}. To date, magnetometry efforts have focused almost exclusively on calibration and understanding signals that deviate from the published data of a nearby United States Geological Survey (USGS) magnetic observatory, a member of the International Real-Time Magnetic Observatory Network (INTERMAGNET). Notable events to date include video capture of a Falcon 9 rocket 2nd stage reentry and magnetometer recordings of the 10 May 2024 geomagnetic storm.

The GP is also developing methods to detect unusual aerial objects in satellite images of the Earth's surface and atmosphere. So far, these efforts have focused on detecting unusual objects in $\sim$ 4~m resolution photographs acquired by Planet Labs \cite{Keto+Watters:2023}. The GP has demonstrated successful detection of objects in motion by locating features that occur along a track in multiple color-band images that were acquired asynchronously. The measured apparent displacements that occur as a consequence of object speed and parallax can be used to identify a set of degenerate solutions for velocity and elevation. Aerial objects for which this set of solutions reside entirely outside the expected performance envelope can be classified as unusual. A shape analysis will also be used to provide further information.  Additional details are described in Section \ref{sec:sats}.

\subsubsection{VASCO} \label{sec:VASCO}
The Vanishing and Appearing Sources during a Century of Observations (VASCO) project \cite{Villarroel+etal:2016,Villarroel+etal:2019}, led by Beatriz Villarroel, is focused on studying and comparing astronomical photographic plates taken over the last century to identify transient objects.  This project has the unique advantage of involving a century's worth of trusted professional astronomical data, much of which is free of contamination by artificial satellites, which did not exist prior to the launch of Sputnik-I in October of 1957.  In addition to providing an opportunity to observe the variability of astronomical objects on scales ranging from decades down to hours, this project has the capability of identifying photographic evidence of unidentified aerospace phenomena in the near-Earth space environment.

In June 2021, the VASCO project published a paper featuring a serendipitous discovery made while searching for vanishing stars. The team had been comparing images from the early 1950s with modern images of the sky and found a peculiar image from April 1950, where nine stars appear in and out within an exposure time, ~50 minutes \cite{Villarroel+etal:2021}. In a previous image taken half an hour earlier, the nine stars were not there and the nine stars were never seen again. The team used the world’s largest optical telescope, the Gran Telescopio Canarias (GTC) 10.4 meter telescope, and did deep photometry, without finding any counterparts to the nine mysterious transients. No astrophysical phenomena can cause the nine transients, and no instrumental effect was identified as the root cause. The team wondered whether plate contamination (plate defects) could be responsible, but found no support for the explanation as the transients have star-like brightness profiles. 

Furthermore, the team proposed that solar reflections from artificial objects could produce subsecond transients. The team followed up with a prediction paper in \textit{Acta Astronautica} \cite{Villarroel+etal:2022:Glint} on how to identify potential nonhuman artifacts more clearly in geosynchronous orbits around the Earth by searching for several transients aligned in a row. Aligned transients help to secure the separation between potential plate defects and nonhuman artifacts, and, indeed, two such statistically significant examples were identified \cite{Villarroel+etal:2022:Geosynchronous}. More recently, the team identified a bright triple transient that also appears and vanishes in 50 minutes, also without explanation \cite{Solano+etal:2024}.  Ironically, the triple transient and the most improbable candidate alignment occurred on 19 July 1952 and 27 July, respectively \cite{Villarroel:2024}, two dates associated with the Washington UFO flyovers when multiple UFOs were observed by highly credible witnesses, both visually and over multiple airport radars. 

In order to verify the existence of the multiple transient phenomenon, the VASCO scientists have now started the new ExoProbe project \cite{Villarroel+Marcy:2023:ExoProbe}, where the goal is to search for multiple transients and signatures of extraterrestrial probes with the help of multiple wide-field telescopes equipped with fast cameras, aiming to detect, validate, characterize the spectrum and localize any eventual probe.

\subsection{Private and Citizen Science Research Efforts}

\subsubsection{The Hudson Valley - Pine Bush Studies. 1980---2003}
\label{sec:Hudson}

Hynek et al. \cite{Hynek+Imbrogno+Pratt:1998} provided detailed accounts and witness descriptions for more than 62 individual sightings and some group sightings, selected from more than 7,000 reports from six Hudson valley counties.
The wave of Hudson valley sightings was also described in \cite{Dennett:2008} and \cite{Zimmerman:2013}.

UFO researcher Ellen Crystall conducted an extensive ten-year study of UAP phenomena in the vicinity of Pine Bush, NY, starting in 1980 \cite{Crystall:1991}. 
Whereas the Hudson Valley sightings tapered off in 1986 after spreading north to Kingston, New York, and south to southern Long Island, New York (see Sec. \ref{sec:Nightcrawler}), and northern New Jersey, Crystall along with other witnesses continued to observe anomalous objects near Pine Bush into late June 1988. The observed objects were described as triangular, rectangular, boomerang and dome-shaped, as well as orb-like light phenomena. Geologist Bruce Cornet joined the research effort in the early 1990s.  Cornet hypothesized that UAP activity in the area was related to the regional geology.  This led him to perform a five-year, ground-based, geological study in the Wallkill River Valley, which runs through the Pine Bush region, with a focus on the mapping and monitoring of magnetic anomalies.

UAP activity in and around Pine Bush rapidly dropped off in 1997, and moved elsewhere, reappearing and surging in Suffolk County (eastern Long Island) and Northern New Jersey around 2012, but never completely disappearing from Pine Bush and east of the Hudson Valley \cite{Costa+Costa:2017}.

\subsubsection{Eye on the Sky} \label{sec:Nightcrawler}
The Eye on the Sky platform is a fully instrumented mobile sensor developed by John and Gerald Tedesco, enabling the observation, detection, tracking, data collection, and identification of unidentified objects in our skies and along our coastline. This state-of-the-art experimental expeditionary vehicle, a fully instrumented high-tech mobile surveillance system, offered a centralized location and system integration for environmental surveys and safety assurance. As engineers in this study, the Tedescos deployed two iterations of a cutting-edge proto-type mobile laboratory-observatory vehicle. Nightcrawler-1, a robust `bottom-up construction design, was used for the first three months of the study. Nightcrawler-2, an RV with a robust 'top-down' modification design and a retrofitted instrument payload, was used for the subsequent seven months of scientific field study due to its superior speed and mobility. It remains the current mobile system and is in continuous field use.

 John Tedesco, Gerald Tedesco, and Donna Nardo conducted a year-long investigation field research study at Robert Moses State Park (RMSP), which is located on a barrier island (Fire Island National Seashore) off the south shore of Long Island, NY \cite{Tedesco+Tedesco+Nardo:2024,Tedesco+Tedesco:2024}.  The location is known for its unique environmental conditions and potential for UAP. The research involved ten months of data collection and analysis at the field site, with 12 hour overnight observation shifts, including two months of research team meetings for careful review, analysis, and collation of the data. The objective was to determine whether aerial phenomena of an unknown nature exist in a coastal location and to characterize their properties and behaviors. In this data-centric study, a combination of primary (qualitative) and secondary (quantitative) field observation methods was utilized. The principles and methodologies of forensic engineering guided the study. The challenges set forward were object detection, observation typology, and characterization, where cutting-edge multispectral electro-optical devices and radar were employed due to limited visual acuity and intermittent presentation of the phenomena. The primary means of detection used a 3 cm X-band radar operating in two scan geometries, the x- and y-axis. Multispectral and hyperspectral electro-optical devices were used as a secondary detection and identification method. In addition, emphasis was placed on the use of high-frequency (HF) and low-frequency (LF) detectors and spectrum analyzers that incorporate electromagnetic (EM) field transducers (ultrasonic, magnetic, and RF) to record spectral data in these domains. Data collection concentrated on recording a wide or broad bandwidth of the electromagnetic spectrum, including visible, near-infrared (NIR), short-wave infrared (SWIR), long-wave infrared (LWIR), ultraviolet (UVA, UVB, and UVC), and the higher-energy spectral range of ionizing radiation (alpha, beta, gamma, and X-ray) recorded by Geiger–Müller counters as well as special-purpose semiconductor diode sensors.

A complete telemetry monitoring system, with meteorological, atmospheric, and environmental displays, is inside the Nightcrawler. What makes this system unique is the use of detection systems such as active radar and hyperspectral optical systems. Active radar required the team to obtain an experimental license from the FCC to deploy this technology on a land-based mobile vehicle for UAP research. Furthermore, the radar system is innovative and proprietary. By using two Radome antennas scanning a target in two geometries, horizontal and vertical, an object size defined by its physical cross-sectional profile produces much more detail scanned in this way. This helps in better understanding the target's distance, speed, and size.

\subsubsection{The UAP Tracker Project} \label{sec:UAPTracker}
The UAP Tracker Project \cite{UAPTracker:2024} was founded in and has been operational since October 2021 \cite{UAPTracker:Highlights:2023}.  It is a citizen science project designed to monitor the skies for possible UAP events.  The project uses five cameras, one with a fisheye lens and four directional cameras, a HIKVISION DS-2TD2637T-15/P therma/optical camera, as well as a passive forward scatter radar system \cite{Morgan:2023}, a MADAR-III-B electromagnetic sensor, and a Automatic Dependent Surveillance–Broadcast (ADS-B) tracking system to identify known aircraft.  The project relies on off-the-shelf commercial grade equipment whose data output is processed by software, the majority of which is free and open source. The data and video are live streamed over YouTube. \\(\url{https://www.youtube.com/channel/UCKO4qkQJvaGqsWLtqllbqxA}),

\subsubsection{UFO Data Acquisition Project (UFODAP)}
The UFO Data Acquisition Project (UFODAP) system was developed as a collaboration between Ron Olch \cite{Olch:2021} and the previous UFO Camera Project team: Wayne Hollenbeck and Christopher O'Brien, to develop a low-cost optical tracking system for UAP studies.  The purpose of UFODAP is to provide low-cost technology to recognize, track and record anomalous objects while simultaneously collecting data from multiple sensors.  The UFODAP system is focused on visible / near-IR imagery as it performs moving target detection and tracking using a Pan-Tilt-Zoom (PTZ) camera and the Optical Tracking Data Acquisition Unit (OTDAU) software, which utilizes a commercial machine vision algorithm to identify moving targets and track them by rotation of the PTZ camera as needed. The UFODAP system was used by the UAlbany-UAPx Collaboration in their 2021 study of the Catalina Channel in southern California, published in this issue \cite{Szydagis+etal:2024}.

\subsubsection{Sky360}
Sky360 is an open source international sky observation project.  Their observation stations consist of an AllSkyCam with a wide angle fish-eye lens and a Pan-Tilt-Zoom (PTZ) camera, for which the wide-angle camera registers all movement and enables the PTZ camera for tracking.  Their software, which is based on a computer vision ``background subtraction'' algorithm that monitors differences in successive video frames to detect any motion, performs a rough analysis of events in real time and decides whether to activate and deploy other sensors, such as the PTZ camera.  In addition to providing observation stations, Sky360 envisions facilitating ``a citizen science project to observe the skies and all their phenomena around the globe, 24/7 and provide harmonized high-quality results and analysis --- available to everybody'' \cite{Sky360}.

\section{Organizations}
\subsection{American Institute of Aeronautics and Astronautics (AIAA)}
The American Institute of Aeronautics and Astronautics (AIAA) established the Unidentified Anomalous Phenomena Integration and Outreach Committee (UAPIOC) in 2021 to advance the scientific understanding of Unidentified Anomalous Phenomena (UAP) and improve aerospace safety. Serving as a neutral, scientifically focused group, the UAPIOC brings together aerospace professionals, academics, industry experts, and policymakers to collaboratively address the challenges associated with UAP research and integration into aerospace systems.

The UAPIOC is dedicated to mitigating barriers to the scientific investigation of UAP and fostering an environment conducive to open, data-driven inquiry. The committee's efforts are organized around two primary subcommittees:
\begin{description}[style=nextline,leftmargin=2em,labelsep=2em]
   \item[Hardware Factors Subcommittee]
   The Hardware Factors Subcommittee focuses on the detection, characterization, and evaluation of UAP using advanced sensor systems. This group works to build an understanding of which multi-modal sensor packages best enable a deeper insight into UAP phenomena. Key activities include:
   \begin{description}[style=nextline,leftmargin=2em,labelsep=2em]
       \item[Sensor Evaluation and Selection:] Evaluation of a variety of sensor systems, including radar, optical, infrared, and acoustic sensors, to determine the most effective combinations for UAP detection and characterization.
       \item[Collaborative Research:] Partnering with private research groups on sensor selection, placement, and performance evaluation. These collaborations improve the quality and scope of the collected data.
       \item[Technology Integration:] Develop methodologies for integrating sensor data into comprehensive analytical frameworks to improve the detection and analysis of UAP.
   \end{description}
   \item[Human Factors Subcommittee]
   The Human Factors Subcommittee addresses the human element in UAP incidents. Comprising of human factors engineers, cognitive scientists, and aviation safety experts, this group focuses on:
   \begin{description}[style=nextline,leftmargin=2em,labelsep=2em]
       \item[Pilot Engagement and Interviews:] Conduct direct and structured interviews with pilots who have encountered UAP, using a two-phase approach to gather detailed experiential data.
       \item[Collaboration with Americans for Safe Aerospace:]
       Working closely with Americans for Safe Aerospace (ASA), a nonprofit organization that receives UAP reports from commercial and US veteran aviators. With pilots' permission, the subcommittee assesses incident cases for analysis.
       \item[Safety Risk Evaluation:] 
       Assessing aviation safety risks associated with UAP based on collected data and pilot testimonies.
       \item[Recommendations for Reporting Mechanisms:]
       Formulating recommendations to enhance existing reporting systems for aviators, aiming to improve data collection and safety protocols.
   \end{description}
   \item[Scientific Contributions and Outreach]
   The UAPIOC is involved in both advancing and promoting the scientific study of UAP through the following activities:
   \begin{description}[style=nextline,leftmargin=2em,labelsep=2em]
       \item[Annual UAP-Focused Paper Sessions:] 
       The UAPIOC holds yearly paper sessions at the AIAA Aviation and ASCEND conferences. Throughout the year, the committee solicits papers from academia and industry that align with the missions of the Hardware Factors and Human Factors Subcommittees. These sessions aim to expand the understanding of UAP through rigorous scientific inquiry. Since its inception, the UAPIOC has facilitated the introduction of 15 UAP-related papers.
       \item[Panel Discussions and Community Engagement:]
       The committee organizes panel sessions at conferences to engage with the broader scientific and engineering community within AIAA. These panels provide platforms for experts to discuss current challenges, share insights, and foster collaborative solutions related to UAP research.
       \item[Collaborations with Academic Institutions:]
       By partnering with leading universities and research organizations, the UAPIOC promotes interdisciplinary research efforts that address both technological and human aspects of UAP incidents.
   \end{description}
   \item[Impact and Future Directions]
   The efforts of the UAPIOC have significantly advanced the scientific study of UAP and contributed to improving aerospace safety. By focusing on both technological advancements and human factors, the committee addresses the multifaceted nature of UAP incidents. Looking ahead, the UAPIOC plans to:
   \begin{description}[style=nextline,leftmargin=2em,labelsep=2em]
       \item[Expand Collaborative Research Initiatives:]
       Continue partnerships with private research groups and academic institutions to advance sensor technologies, data analysis methods, and understanding of UAP phenomena.
       \item[Enhance Pilot Reporting Mechanisms:]
       Utilize findings from the Human Factors Subcommittee to advocate for improvements in US and international reporting systems, making it easier for aviators to report UAP encounters without fear of reprisal.
       \item[Increase Educational Outreach:]
       Develop training materials and workshops to educate pilots and aviation professionals about UAP detection, reporting procedures, and safety protocols related to UAP.
       \item[Influence Policy and Standards Development:]
       Work with government agencies and international organizations to influence policy changes and establish global standards for UAP reporting and investigation.
   \end{description}
\end{description}
Through its dedicated subcommittees and collaborative efforts, the AIAA UAPIOC plays a pivotal role in the new science of monitoring UAP. By integrating technological advances with human factors research, the committee improves the collective understanding of UAP and contributes to the safety and integrity of aerospace operations. The UAPIOC remains committed to fostering an open and scientific approach to UAP investigation, promoting transparency, and facilitating the advancement of aerospace knowledge.

\subsection{National Aviation Reporting Center on Anomalous Phenomena (NARCAP)}
The National Aviation Reporting Center on Anomalous Phenomena (NARCAP)
was established late in 1999 by Richard F. Haines (Chief Scientist), (see \ref{sec:Haines}) and Ted Roe (Executive Director) to deal with the fact that commercial pilots were not reporting the aerial phenomena they had seen, of highly unusual and unexplainable characteristics, to their management. At the time, it was likely that military pilots were doing much the same. 

The aim was to offer a website (\url{https://www.narcap.org/}) on which all American pilots could submit the details of their encounters no matter how strange and unbelievable they were without discrimination or ridicule. NARCAP presented a safe place for them while evaluating their reports to: 
\begin{enumerate}
\item see if they might contain valuable scientific and/or engineering data that could be relevant to UAP \footnote{Richard Haines (NARCAP) first used the UAP acronym.} studies that might improve our understanding.
\item publish any findings on the NARCAP website for the benefit of any and all interested parties.
\item help motivate other pilots to report their sightings as well.
\end{enumerate}
All pilot and air traffic controller names were kept confidential to protect them from unwarranted and often biased recrimination. 
It was decided to confine the studies only to US pilots and, if there was any possibility that the sighting might contain restricted data, NARCAP would not get involved.

In order to help NARCAP gain and maintain an image of objectivity and neutrality in this controversial area of study, one that had long-collected a distorted image smeared by skeptics for decades, and be accepted by aviation authorities (FAA, commercial airlines, government agencies and its many departments, and others) NARCAP deliberately avoided all contact with UFO groups. NARCAP operated under the premise that if the work of private UAP organizations was ever going to be accepted by the government and the aviation community, they would have to prove themselves by providing some kind(s) of useful, practical data of value.  

The ``useful product'' generated by NARCAP was safety as related to aviation.  Today, there are 715 licensed commercial airlines in the world. Of these, US airlines carried 853 million passengers in 2022 and 980 million the following year. Worldwide, airlines carried 342 billion passengers in 2022 and 468 billion in 2023. This represents a very large number of potential eye witnesses of UAP not even counting the flight crews.  Unfortunately, eyewitness reports from airplane passengers provide very little reliable information on UAP. However, in 2022, a total of 366 UAP aviation reports were received by US government officials and about 171 of them were evaluated as unexplained. In 2023 there were 270 reports reported to the Department of Defense with a few from commercial pilots. In fact, most of them came from military pilots. During NARCAP's twenty-four years of operation, NARCAP received many valuable reports from private pilots as well.

Why are pilot and related aircrew reports valuable?
\begin{enumerate}
    \item Pilots are highly trained, educated, and experienced witnesses. They have seen a wide variety of weather and related visual phenomena in their careers and so, may not be so easily fooled to misinterpret something. This is not to say they can’t be fooled by some highly unusual opto-visual atmospheric effects as
    skeptics often suggest.
    \item If pilots do make a report they usually do so only after eliminating a long list of ``more acceptable'' explanations and are finally left with an unknown---a UAP. When they do report, they often end up facing the negative scrutiny of their employers. This is why NARCAP guarantees their anonymity.
    \item They are flying airplanes carrying a greal deal of sensitive instrumentation that might sense, and even record, electro-magnetic radiation related to the UAP, local air temperature and/or pressure changes, and more.
    \item They are in radio contact with ground support personnel such as radar operators who can check if there may be some other sufficiently reflecting and/or emitting object(s) nearby or even scramble other airplanes to assist.
    \item They can maneuver to avoid a collision or achieve a different perspective on the UAP, for example, to get above it with the Earth as background and thereby establish its maximum visual slant range.
\end{enumerate}
Pilots make good witnesses.

NARCAP called together an advisory group made up of two notable physicists, several engineers and pilots, a meteorologist, an air traffic controller, a human factors specialist, and several others. They provided invaluable advice on technical and policy issues. See Section \ref{sec:Vallee} in this report regarding Jacques Vall\'{e}e who also served on their Advisory Board.

How could UAP flying near an airplane possibly be dangerous?  First, if the UAP came close enough, the pilot might make a rapid avoidance maneuver (in addition to its various consequences to passengers and subsequent responses by air traffic controllers on the ground). The flight crews could then become so fascinated by their presence that they interfere with their continuous monitoring of their flight instruments.

Disturbances of various cockpit instruments by electro-magnetic energy (seemingly) emanating from the UAP could impact flight safety \cite{Haines:1992:EM}. This has also definitely happened.  Having reliable E-M data may contribute to a fuller understanding of the physics related to UAP.

The NARCAP model has helped to foster similar organizations in other nations. Since aviation safety is so important to so many people, it is not surprising that NARCAP’s mission-objectives and reporting procedures have led others in Chile and Germany to form their own national organizations. There may well be others as well.  NARCAP has been in communication with pilots and officials in many other nations.

NARCAP has produced reports, which have been subdivided into the
following categories (all are available at \url{https://www.narcap.org/}): Technical Reports: (In-depth reports of
specific cases); Topical Reviews: (topical subjects possibly related to operations and existence of UAP); Investigator Support Papers: (a variety of useful information for interested researchers
prepared by American and foreign authors); International Air Safety Reports: (Reports of foreign UAP encounters and related phenomena of interest to US investigators); and International Technical
Specialist Reports: (collection of broad statistics and topics of general interest).

\subsection{The Scientific Coalition for UAP Studies (SCU)} \label{sec:SCU}
The Scientific Coalition for UAP Studies (SCU), a nonprofit organization
dedicated to the scientific study of the phenomena, was organized in 2017 and
incorporated as a 501(c)(3) non-profit in 2018. The organization is managed by a Board of Directors and eight Board Advisors that meet monthly.

The SCU mission is the scientific exploration of unidentified aerospace phenomena, and it achieves its mission with the following goals and objectives:
\begin{itemize}
\item to establish a foundation and resource for credible, objective, scientific peer reviewed content on the subject of UAP;
\item to create partnerships in scientific and technological endeavors to explore UAP with other scientific organizations, universities, and governments bodies to help improve the data collection, data dissemination and research of the UAP subject;
\item and to sponsor public and networking environments and events to augment
communication within the serious research community.
\end{itemize}

The organization is international in scope and is currently composed of over 350 members from a wide diversity of scientific disciplines and the majority with advanced degrees. SCU is highly regarded for its transparency and credibility. In addition to numerous published scientific studies \cite{Schulze+Powell:2010, Powell+etal:2015, Powell+etal:2019, Powell+etal:2022, Reali:2022, Powell+Little:2023, Hancock+etal:2023:pattern, Hancock+etal:2023:indications, Porritt:2024, Hancock+etal:2024:activities}, it holds annual conferences, publishes a quarterly newsletter, supports UAP project teams, hosts a podcast, and provides periodic press releases.

\subsection{The Sentinel Center}
The Sentinel Center \cite{Sentinel:website} is an association created in 2024 to promote scientific research and archiving of documentation concerning maritime, land, air and space surveillance.

The Sentinel Library \cite{Sentinel:library} facilitates access to documents through its Internet portal, allowing the referencing of texts, analyses, and references listed in chronological order to facilitate the work of scientists. The Library also has its own physical library, hosted in France, which can be consulted by a residency in the Champagne region, alongside archives that are held under its care.

The Sentinel Lab \cite{Sentinel:lab} designs and tests surveillance platforms aiming to record fast moving targets that would otherwise evade astronomical instruments. It consists of a mobile observatory able to be deployed during high-intensity events and a fixed observatory located in a low light pollution environment, hosted at the same site as the Library. Finally, Sentinel News \cite{Sentinel:news} acts as an information platform covering scientific and military news related to UAP in English, French, Spanish, and Italian.

\subsection{3AF-SIGMA2}
SIGMA2 is a Technical Commission of 3AF (Association A\'{e}ronautique et Astronautique de France) comprised of 25 experts in aeronautics and astronautics on physical observables, as well as the effects of unknown aerospace phenomena. 3AF (\url{http://www.3af.fr}) is a technical society registered under the French legislation (France, Loi 1901, non-profit organization), which gathers about 1500 individual members as well as aerospace industrial, academic and institutional partners including CNES (French Space National Agency), DGAC (French Civil Aviation Directorate), DGA (French MoD Directorate for armament programs and exports), ISAE (French aerospace school of engineers), Airbus, Dassault Aviation, MBDA, SAFRAN, THALES and many other groups.

The 3AF UAP technical commission was created in 2008 bringing together aeronautics experts. It was renamed SIGMA, then SIGMA2 in 2013. SIGMA2 was tasked with a new and specific mandate. Its mission is not to conduct investigations in the field on UAP cases (GEIPAN mission for cases in France), but to study unexplained UAP cases (foreign or D type in GEIPAN typology, which amount to unexplained cases based on reliable testimony) using a scientific methodology. SIGMA2 consists of 25 high-level experts from the civil or military domains (research or industry laboratories). They are skilled in aeronautics and space, air, space and missile defense, electromagnetic radiation physics and weapons, high-energy plasma, medicine, biology, missile propulsion, radar, IR, acoustical detection. They are mainly engineers, Ph.D.s., astronauts, fighter and civil pilots, air traffic controllers, and some experts in political science, psychologists, and physics. Their objective is to conduct analysis on the UAP cases based on physical observables, on physical or psychological effects, which can involve multiple domains from medicine to electronics warfare and directed energy weapons.

As a result, SIGMA2 studies of UAP cases are conducted scientifically by relying on physical data like radar, IR/electro-optical or electromagnetic
observables, with the goal of understanding the physics related to the UAP observables, including the effects induced on human beings. For this reason, the efforts of SIGMA2 are interdisciplinary in nature, simultaneously involving physicists employing optical, infrared, and electromagnetic sensors to study the observables of UAP, but also involving psychologists studying the effects induced on humans, including the psychological effects of UAP on people.

Many reports and documents are published in the 3AF-SIGMA2 media repository \cite{Sigma2:2024:Media} among them the 3AF-SIGMA2 Progress Report 2021, which includes a list of cases analysis and a tour of the physical features of UAP \cite{Sigma2TechnicalCommitte:2021}.  3AF-SIGMA2 also organizes webinars on physical observables of UAP \cite{Sigma2:2024:webinar}.

\subsection{The Society for UAP Studies (SUAPS)}
The Society for UAP Studies (\url{https://www.societyforuapstudies.org/}) is a 501(c)(3) nonprofit organization, which was organized in September 2022 to play a vital role in creating the necessary academic foundations for a stable and sustainable transgenerational scholarly field focused on the many aspects of UAP.  The Society takes as its primary mission to promote and actively aid in the development of UAP studies as a proper area of scholarly research.

As a professional scholarly association, the Society for UAP Studies exists both for the kind of professional development expected of such an organization, such as organizing conferences and colloquia; conducting public roundtable discussions on key issues; developing academic curricula and pedagogy in the field,  among many other activities and functions, and for broader community outreach. With its outreach, the Society aims not only to raise awareness about the subject, but --- and most importantly --- to help educate the general public on how it is that scholars approach their subject, even one as complex and challenging as UAP.

SUAPS manages and publishes the only English-language, peer-reviewed scholarly journal devoted entirely to UAP Studies: \textit{Limina - The Journal of UAP Studies}. It is hosted by Scholastica at: https://limina.scholasticahq.com/.

The Society plays a vital role in demonstrating to other academics that not only does UAP Studies have as valid a place within mainstream academia as any other established field (even though it is still a young, developing area of study), it also plays a vital role in demonstrating that the empirical study of UAP (i.e. as an object of investigation for the natural sciences) can be done by using the accepted methods and practices of the sciences not to dismiss the subject (or explain it away), but rather to sincerely pursue UAP as a genuine problem in the physical sciences --- one that potentially requires explanatory options \textit{outside} of what is accepted as conventional. Mainstream scholars are right to worry about stepping outside of conventional explanations when it comes to UAP; but equally, scholars and scientists should be able to (and indeed should be \textit{expected} to) engage in the more challenging meta-theoretical and philosophical considerations that must be tackled when attempting to challenge or exit altogether conventional thinking - whether for UAP or for any other object of careful study. This, finally, is also what the Society exists for: organized, systematic inquiry at the meta-theoretic and more philosophical levels, as an aid to navigating the terrain of unconventional - or what we might call ``liminal'' --- epistemologies.

\subsection{The Sol Foundation}
The Sol Foundation addresses the complex political, scientific, and social challenges raised by UAP by combining an academic research institute and a policy think tank into one organization. This allows the foundation to provide policymakers with scientific insights into the puzzling data they face while guiding academics to listen to the US government as it releases UAP information. Sol’s approach has led to requests for advice from several national governments, including the US Senate and House, the Parliament of Canada, the EU Parliament and Commission, and the Parliament of Japan. The institute's non-applied research focuses on three problems: the provenance of potential UAP materials, human narratives and experiences about so-called non-human intelligence, and how multi-agent AI can be used to reduce UAP misinformation.

\subsection{UAP Check}
An international network called UAP Check was launched from Europe in 2023 \cite{Vaillant:2024}. Among its goals are to:
\begin{enumerate}
    \item Survey ongoing research activities about UAP worldwide, update a list of who is doing what, and connect those who are working on similar projects.
    \item Build a vetted bibliography on the UAP topic.  The first two parts of the bibliography were published in 2024:
    \begin{itemize}
      \item The first part consists of a collection of 460 university theses and dissertations on UFO / UAP, which is now available as both a catalog \cite{Toselli:2024} and a search tool \cite{UAPCheck:2024:theses}.
      \item The second part is a selection of 690 articles published on the specific topic in peer-reviewed scientific journals \cite{UAPCheck:2024:library}.
    \end{itemize}
\end{enumerate}
These efforts aim to improve outreach efforts, reduce stigma, and provide reliable information about UAP.

\section{Social Sciences}
Although this paper has focused mainly on the physical sciences, it is important to recognize that UAP have also been the subject of considerable research within the social sciences. While a thorough analysis of UAP research in this field would require an exploration as detailed as the current one on the physical sciences --- a task beyond the scope of this paper --- we can nonetheless identify key limitations that the social sciences have encountered and suggest alternative directions they might pursue if these limitations were overcome.

The central methodological approach in social science studies of UAP has been the suspension of judgment regarding their existence. This principle allows researchers to evaluate controversial claims more objectively and shields them from the stigma that still often surrounds the topic of UAP. By avoiding a stance on the actual existence of UAP, social scientists focus on people's beliefs and perceptions. They explore how these beliefs should be best understood—as myths \cite{Peebles:1994, Bullard:2016, Halperin:2020} religious phenomena \cite{Ellis:2000,Denzler:2001}, forms of rationality \cite{Denzler:2001} or irrationality \cite{Ellis:2000}, or alternative lay sociological constructs \cite{Lepselter:2016}—--and analyze the social consequences that arise, such as the development of new religious movements \cite{Palmer:2004}, forms of social estrangement \cite{Bader+etal:2011} or shifts in political orientations \cite{Dean:1998}.

This approach means that social scientists primarily focus on the individuals who claim that UAP are real, rather than investigating anything about UAP themselves. By suspending judgment or presuming nonexistence, these studies tend to offer purely sociological or psychological explanations for why certain individuals or communities hold beliefs considered fringe. Although the assumption that one can analyze a phenomenon socially without considering its actual existence has been a powerful methodological tool, it also risks a form of social reductionism: treating UAP solely as social constructs without acknowledging any independent social consequences their existence might entail.

An analogy with climate change helps to clarify this point. Beliefs about climate change have significant social (and existential) consequences that merit study, but climate change itself has tangible societal effects that extend beyond individual or collective beliefs. If researchers indefinitely suspend their judgment on the reality of climate change, they miss the opportunity to study its direct impact on social structures and practices. Similarly, by suspending judgment on the existence of UAP, social scientists limit the range of questions they can ask witnesses and may overlook the potential significance of patterns in testimonies. For example, when witnesses across decades and regions report objects that show intelligent control, incredible speeds, non-aerodynamic shapes and maneuvers, and near-complete silence, a social theorist might focus on how these accounts spread as sociocultural memes and express long-standing cultural anxieties \cite{Eghigian:2015}. This focus risks ignoring what these reports might reveal about the underlying phenomenology of whatever is controlling the UAP. As a result, this approach can distort our understanding of witness testimonies by reducing them solely to historical and sociocultural phenomena.

Despite various studies on UAP within the social sciences, researchers have seldom taken the possibility of their actual existence seriously enough to challenge the foundational assumptions of their disciplines. They have not explored how their methodologies and theories would need to evolve if it were the case that a higher non-human intelligence has been interacting with humans over time. Critical questions remain unaddressed: How might such interactions have shaped societal structures? How can we accurately discern and characterize their social influences? How would this shift our interpretations of witness testimonies? What recurring patterns should we prioritize in reports and what significance should be attributed to them? These questions have largely been neglected because the social sciences have not seriously considered the prospect of technologically advanced non-human intelligences interacting with humanity. The tendency to suspend judgment—a stance that has been understandable given the stigma around UAP studies—has led to reductionist explanations that risk overlooking critical dimensions of the phenomenon. However, researchers risk professional censure if they deviate from this norm.

There have been notable exceptions to this general trend. A prominent example is the argument of Alexander Wendt and Raymond Duvall \cite{Wendt+Duvall:2006} that if UAP were real, governments would be compelled to deny their existence. They contend that acknowledging UAP would threaten the anthropocentric foundations of state sovereignty and, by extension, sovereignty itself. Official recognition of UAP would imply a lack of control over national security, thus undermining the legitimacy of the government. Their argument provides a compelling theoretical framework for understanding why governments might resist disclosing information about UAP. However, this line of thought has not been further developed or explored by other social scientists, leaving critical questions unaddressed.

For example, governments are complex entities composed of interconnected and overlapping bureaucracies that rarely operate in perfect unison. This complexity suggests that gaps within these structures could be exploited, potentially leading to some level of official recognition of UAP. Wendt and Duvall’s argument also opens the possibility of exploring how internal factions within governments—some pushing for greater disclosure, others for secrecy—could create public confusion about the existence of UAP. This dynamic, evident today in the conflicting statements from various government sources, may also help explain the social and political confusion surrounding UAP in recent years \cite{David:2022}\cite{Graff:2023}. These aspects of Wendt and Duvall’s work highlight promising avenues of inquiry that could shed light on what we are seeing today, yet have largely been left unexplored by the broader field.

Another potential extension of Wendt and Duvall’s argument could explore the idea that if UAP are indeed advanced nonhuman intelligences, governments might actively suppress related scientific research to maintain national security, a fundamental aspect of state sovereignty. This suppression could include efforts to reverse engineer nonhuman technology, with nations competing against each other to gain technological advantages ---- a claim that some highly placed government officials have recently made \cite{Kean+Blumenthal:2023}. The impact of such secrecy on the development of a legitimate academic science of UAP remains largely unexamined, as are the broader implications for international relations. These unexplored areas highlight the limitations of suspending judgment on the existence of UAP and challenge the notion that the social consequences of UAP can be fully understood without confronting the reality of the phenomenon itself.

Moving beyond these questions of governmental dynamics, scholars like Diana Pasulka \cite{Pasulka:2019,Pasulka:2023} and Jeffrey Kripal \cite{Kripal:2010,Kripal:2024} have offered fresh perspectives on the UAP phenomenon by examining its intersections with religious and mystical experiences. They argue that UAP encounters bear remarkable resemblances to experiences recounted in religious and mystical texts, blurring the lines between the sacred and the secular, the technological and the spiritual. Without claiming that historical religious encounters were actually UAP sightings or that modern UAP encounters are simply contemporary religious or mystical experiences, Pasulka and Kripal highlight how these similarities challenge fundamental categories. Their work suggests that a purely physical scientific approach may be insufficient to fully comprehend UAP. By emphasizing these phenomenological parallels, they call attention to the need for new concepts, categories, and methodologies that transcend traditional disciplinary boundaries, enabling a more nuanced apprehension of UAP.

This has significant implications not only for religious studies but for the social sciences and the scientific community as a whole. By exploring these dimensions, researchers could begin to address critical questions previously neglected, such as how UAP might interact with human consciousness or influence cultural narratives. Incorporating such perspectives encourages a more comprehensive understanding of UAP phenomena, moving beyond reductionist explanations and acknowledging that existing scientific paradigms may not fully capture their complexity. This approach underscores the necessity of interdisciplinary collaboration and innovative methodological frameworks to advance our understanding of both human experiences and the nature of UAP themselves.

In summary, the social sciences have typically approached the UAP topic by suspending belief in their existence and treating believers as subjects of a peculiar social phenomenon. This analytical stance separates the study of social impacts from the question of the actual existence of UAP, assuming that the social analysis would remain the same regardless. This approach contrasts sharply with how social scientists engage with more conventional, legitimate fields of science and technology. In those areas, researchers explore how scientific communities establish evidentiary standards \cite{Latour+etal:2013}, develop reasoning frameworks \cite{Knorr-Cetina:1999}, and form epistemic objects \cite{Pickering:1999, Rheinberger+Muller-Wille:2018} --- all while acknowledging the existence of the phenomena they study. For example, social scientists examining virology would never question the existence of viruses, as doing so would fundamentally change the interpretation of scientific practices and findings.

However, in the case of UAP, the focus tends to remain on why people hold these beliefs, without seriously considering that such beliefs might be rooted in actual events. This oversight often leads to the dismissal of sociological evidence that might point toward the existence of UAP. To deepen our understanding, social scientists may need to move beyond the safe harbor of suspended judgment and consider more openly the potential existence of UAP and the profound implications this could have on society, research methodologies, and the foundational assumptions of their disciplines.

\section{The Scientific Methodology and Best Practices for Collecting UAP Data}
In this section we summarize the scientific methodology that has been found to be useful, as well as best practices, to detect, monitor and study UAP.  Several papers have been published that present useful summaries.  Most notable are the papers by Ailleris \cite{Ailleris:2011, Ailleris:2024}, the Galileo Project \cite{Cloete+etal:2023, Keto+Watters:2023, Mead+etal:2023, Szenher+etal:2023, Watters+etal:2023}, the Hessdalen Project \cite{Hessdalen, Strand:1984, Teodorani+Strand:2001, Teodorani:2004, Teodorani:2024}, the UAlbany-UAPx Collaboration in this volume \cite{Szydagis+etal:2024}, the UAP Tracker Project \cite{UAPTracker:Highlights:2023, UAPTracker:2024}, the Tedescos' Eye On The Sky Project \cite{Tedesco+Tedesco:2024,Tedesco+Tedesco+Nardo:2024}, and the numerous IFEX projects, such as SKY-CAM, SONATE-2, KI-SENS, and VaMEx3-MarsSymphony, which are summarized in Sec. \ref{sec:IFEX}.

It is generally agreed that the optimal methodology to study UAP relies on many different types of instruments, spatially separated, to dramatically reduce the possibility of error.  This is the only way in which the scientific community will recognize truly anomalous data.  The modern equivalent is that of multi-messenger astronomy, which uses instruments of multiple modalities \cite{Watters+etal:2023, Szydagis+etal:2024}.
The UAlbany-UAPx Collaboration makes the following recommendation, ``We recommend at least two of each type of sensor and 2+ distinct sensor types.'', along with a statistical procedure to define terms, such as \textit{coincidence}, \textit{ambiguity} and \textit{anomaly}: \cite{Szydagis+etal:2024}
\begin{displayquote}
    We suggest (scientific) UAP researchers adopt the following conventions: An ambiguity requiring further study is a coincidence between two or more detectors or data sets at the level of 3$\sigma$ or more, with a declaration of genuine anomaly requiring (the HEP-inspired) 5$\sigma$... Coincidence here is defined as ``simultaneity'' within the temporal resolution, and spatial when germane. This way, one rigorously quantifies the meaning of extraordinary evidence, in the same way it has been done historically by particle physicists, who have established a very high bar to clear. The statistical significance must be defined relative to a null hypothesis, in our case accidental coincidence, combined with causally-linked hypotheses, like cosmic rays striking camera pixels.
\end{displayquote}
Additional discussion of these concepts and the criteria for recognizing outliers, anomalies, and novel classes in the formal analysis of multimodal instrument data can be found in \cite{Watters+etal:2023, Domine:dalek}.

To achieve useful results, precision is required.  The experimenter must know both the position and pose (orientation) of the instrumentation and acquire data with precise timing.

\subsection{Time Synchronization}
Time synchronization of the equipment is important so that recorded events can be precisely correlated.  The precision required should be a function of the rate at which the data is collected, which is typically different for different instruments.  For example, typical video frame rates are on the order of $24\unit{fps}$ (frames per second), which corresponds to a characteristic time of $41.6\unit{ms}$.  Since some video cameras can handle frame rates of $300\unit{fps}$, this would suggest that better time synchronization than $3\unit{ms}$ would be desirable.

\subsection{Instrument Positioning}
One way to control the positioning of the instrumentation is to mount the equipment on a rack so that they have known relative positions.  The positions and orientations of multiple racks would then have to be measured, most likely with precisions appropriate to the sizes of the specific pieces of equipment, which will be of the order $1-10\unit{cm}$.  The positions of multiple pieces of instrumentation can be more easily measured and monitored using a Local Positioning System (LPS) \cite{Aitenbichler:Muhlhauser:2003, Gulden+etal:2009, Wu+Casciati:2014, Hasan+etal:2018}.

\subsection{Diverse Types of Instrumentation}
It is desirable to have many different types of instrumentation to measure different aspects of the phenomenon.  The papers published by the different currently operating groups, mentioned above at the beginning of this section, provide a great deal of information on the methodologies, measurements, and instrumentation employed.

\subsubsection{Light-Based Imagery}
In this section, we discuss UAP imagery based on light acquisition.  This includes optical imagery ranging from infrared (IR) through visual (VIS) to ultraviolet (UV) imagery.  Given both the large number of UAP images recorded using visible-light cameras, as well as the US Navy and Homeland Security imagery recorded in the near-infrared range using Forward Looking InfraRed (FLIR) Cameras, imagery in these modalities is already rather familiar.

Acquiring imagery in the near- and far-infrared is important, as in addition to imagery at wavelengths inaccessible to the human eye, IR imagery has the capability to provide information that enables one to estimate the temperatures of the object.  This is especially important with UAP, as it has been observed that many UAP are extremely cold (sometimes at temperatures down to -50 C or $-60^{\circ}$ F).  Such imagery can provide important information about the lift and propulsion mechanisms employed by the UAP.  In addition, obtaining ultraviolet (UV) imagery extends the imaging window in the other direction of the spectrum, which can provide additional important information.

It is extremely important to obtain information about the light beyond the images.  Spectra can provide information about the elements or molecules involved in the light production, which, in addition to informing us about the kinds of lights being used, can also provide information about the UAP's lift and propulsion mechanisms.  Obtaining imagery using a polarizer can provide information about the polarized light environment surrounding the UAP.  The axis of polarized light can be rotated by strong magnetic fields via a phenomenon known as the Faraday effect \cite{Schatz+McCaffery:1969},  Given that the scattered light in the sky is generally polarized, one can use the rotation of the axis of polarization to estimate magnetic field strengths around UAP.

Most UAP groups employ multiple cameras.  This is especially important because it is helpful to record imagery with different kinds of lenses, ranging from wide-angle fisheye lenses to high-quality telephoto lenses, the latter of which will require some kind of computer-controlled tracking.  In addition, different cameras could employ different polarizing lenses and be set up with different exposure times to be recorded at the expense of detailed imagery.  This can help to provide valuable information about the speed and trajectory of a UAP in three dimensions, especially if multiple spatially-separated cameras are used for triangulation.

\subsection{Remote Sensing Satellites for Scientific UAP Research}\label{sec:sats}
 UAP researchers are now considering the air and space domains as open-air laboratories, utilizing these vast environments for systematic scientific inquiry. This chapter examines the role of remote sensing satellites in this scientific endeavor, highlighting the context, advantages, current efforts, and future possibilities in UAP research.

Since the launch of Sputnik 1, Earth observation (EO) technology has seen tremendous developments, leading to a significant increase in the launch of EO satellites in recent years \cite{EROS:2024}.
These satellites serve a multitude of applications, revolutionizing industries ranging from agriculture, infrastructure monitoring, disaster management, security, and defense to environmental monitoring. Modern EO science is characterized by frequent and systematic image acquisitions of the Earth's surface and atmosphere, providing high-resolution data that democratize the use of satellite imagery and extend its benefits to various levels of society. Continuous advances in science, informatics, communication, and instrumentation have facilitated ongoing improvements in Earth monitoring across various parts of the electromagnetic spectrum.  

Currently, we benefit from higher spatial, temporal, and spectral resolution, along with an expanding volume of imagery and accessible data, which ensures continuity for decades. In the past, access to satellites and their data was predominantly restricted to major governments and military and defense organizations. However, thanks to significant technological advancements and the proliferation of commercial satellite services, access to satellite data has expanded dramatically.  In addition, rapid advances in information and communication technologies have opened new avenues for many more actors. Tools such as cloud computing, artificial intelligence (AI), and machine learning (ML) now enable scientists to gather, store, transmit, and analyze data more efficiently than ever before. Today, a wide range of entities, including private companies, academic institutions, and individual researchers, can leverage satellite technology for various purposes. This democratization of satellite data opens new possibilities for scientific research, including the study of UAP, enabling a more broader and inclusive exploration of this field \cite{Ailleris+Knuth:2020,Knuth+Ailleris:2020}.

Compared to other instruments, satellites offer distinct advantages for UAP research. They provide a wide coverage and continuous monitoring of large areas of the sky, increasing the chances of detecting UAP that ground-based observations might miss.  In addition, satellite imagery provides the unique opportunity to verify, inform, and supplement ground observations of UAP, especially large UAP, after the fact, especially in daytime over land, where satellite imagery is autonomously collected and archived.
Unfortunately, satellite coverage over the oceans is still sporadic, and night-side imaging is generally limited as the activity during the night-side portions of the orbit are typically reserved for data download.

Satellites are equipped with reliable and accurate measurement sensors capable of collecting scientifically useful data. Their high-resolution imaging capabilities allow for the capture of detailed images of UAP, facilitating identification and analysis. In addition, satellites can access remote locations such as polar regions, deserts, and oceans, which are difficult for humans to reach. Furthermore, utilizing remote sensing satellites represents a cost-effective approach for UAP research, as many relevant space-borne systems are already financed and deployed into orbit. The costs of acquiring and processing satellite images are decreasing, and some data, such as those from the European Copernicus satellites (EU/ESA), are made freely available daily. The global nature of both UAP phenomena and satellite observation underscores the importance of this approach. Satellites have the potential to enable the collection of validated independent data on a global scale, which is essential for a comprehensive understanding of UAP.

Historically, UAP researchers were already envisaging the potential of satellite technology as early as the 1970s, long before public access to satellite data and the proliferation of Earth's satellite coverage. Despite this early interest, scientific discussion on the satellite detection of UAP has been almost nonexistent until the very recent years, with most information stemming from books or anecdotal reports rather than confirmed data from high-tech sensors.  Occasionally, UFO literature discusses infrared satellite sensors, particularly the Space-Based Infrared System (SBIRS). Originally designed to detect missile launches from the former Soviet Union and China, this network of satellites in geosynchronous and highly elliptical orbits, supported by ground-based data processing, detects a wide range of heat sources, including nuclear explosions, aircraft afterburners, forest fires, volcanoes, meteors, and re-entering satellites. Preceded by the Defense Support Program (DSP) satellites, the SBIRS has been rumored to record unidentified objects entering Earth's atmosphere and maneuvering (unlike meteors). However, today’s context is quite different, as recent official declarations have suggested that UAP-like phenomena might indeed have been detected by defense satellites. For example, in 2021, the former US Director of National Intelligence John Ratcliffe highlighted the potential of satellites in UAP research, noting that some apparently anomalous phenomena have been detected by satellites \cite{McCarthy:2021}. Moreover, the United States Office of the Director of National Intelligence and Department of Defense Fiscal Year 2023 Consolidated Annual Report on UAP emphasized the need to fully integrate the space domain into the processes of the All-Domain Anomaly Resolution Office (AARO). The collaboration of several agencies related to space awareness with the AARO, including the National Aeronautics and Space Administration (NASA), the National Reconnaissance Office (NRO), the National Geospatial Intelligence Agency (NGA), and the United States Space Force (USSF), underscores the significant role that space remote sensing satellites and technologies can play in UAP detection and analysis.

Building on advances in satellite technology and data accessibility, recent developments promise potential scientific progress. The 2023 NASA UAP Independent Study Team Report \cite{Spergel+etal:2023:NASA-REPORT} emphasizes the use of existing and planned space observing assets, alongside archived historic and current datasets \cite{McCarthy:2021}. Although current NASA satellites may not have direct UAP detection capability, their sensors can investigate local environmental conditions associated with UAP reports. As stressed in this report, the commercial remote sensing industry offers a potent mix of Earth-observing sensors capable of directly resolving UAP events. Commercial satellite constellations provide imagery at sub to several meters spatial resolution, suitable for typical spatial scales of known UAP. Moreover, the high temporal cadence of commercial remote-sensing networks enhances the likelihood of retroactively covering UAP events initially observed through other means. However, it is important to note that high-resolution coverage of Earth's surface by commercial satellites is limited at any given time, requiring fortunate timing to obtain observations of specific UAP events from space. Although Earth observation satellites have shown great potential in sensing technical capabilities, current satellites lack the necessary algorithms for UAP detection. Artificial intelligence will thus play a vital role in identifying rare occurrences such as UAP within large datasets, leveraging NASA's robust data handling capabilities. 

The Society for UAP Studies (SUAPS) has recently proposed a promising research angle in UAP scientific research, highlighting the importance of citizen science as noted in the 2023 NASA report \cite{Spergel+etal:2023:NASA-REPORT}. The report suggests several citizen science projects including a citizen reporting center, smartphone capture and reporting, ground instrumentation, and the use of commercial satellite remote sensing Earth observation imagery. Citizen science projects offer a compelling avenue for UAP research, taking advantage of the collective power of the public to gather and analyze data. 
The success of citizen science projects such as Galaxy Zoo, which led to the discovery of Boyajian's star, demonstrates the potential to develop a citizen science project focused on extracting valuable insights from satellite data for UAP research, aligned with rigorous evidence-based methodologies.

Following this, it is important to highlight the contributions of the Galileo Project \cite{Keto+Watters:2023} (see Sec. \ref{sec:GP}), which has already begun to assess the feasibility of using satellite data for UAP detection and characterization. The project aims to develop software that employs pattern recognition techniques for the automatic identification of moving objects in commercial satellite images provided by the commercial company Planet Labs. The primary objective is to identify objects that exhibit velocities, accelerations, sizes, or shapes that deviate from those expected from natural phenomena, common vehicles, or projectiles. This includes satellite data that capture objects entering Earth's atmosphere that do not follow ballistic orbits, such as meteors or rockets, although the task poses significant complexities and challenges.  Since 2023, the Galileo Project's research has made significant progress in analyzing the potential of satellite data for UAP detection. The project has demonstrated at least two distinctive and easily recognizable characteristic motion signatures in Earth observation images made with push-broom scanning, indicating that the proposed method is successful and enables the measurement of the apparent velocity of moving objects using archived information \cite{Keto+Watters:2023}. This novel method uses multispectral images from push-broom scanning satellites to detect and analyze the velocities of moving objects. By estimating the relative acquisition times between different spectral bands, the method accurately determines velocities even without precise timestamp information, as demonstrated by comparing aircraft velocities derived from satellite images with those reported by onboard ADS-B transponders. Keto and Watters \cite{Keto+Watters:2023} applied this image analysis technique to one of the most notorious UAP incidents of recent times: the flight of a Chinese spy balloon over the US, which was eventually shot down by an Air Force fighter jet. They also analyzed imagery of a different spy balloon that passed over Colombia at about the same time. Using Planet’s SuperDove satellites, they created a baseline to interpret spectral-band images and provided estimates of the altitudes of the balloons.  These techniques could be used to detect phenomena that are even more exotic than spy balloons, with the aim of collecting high-quality data that could be useful in the search for objects of non-human origin.  In future work, the Galileo Project aims to generalize their method to analyze accelerating objects and to define any limitations. This includes recognizing objects with uncharacteristic flight patterns, such as sudden acceleration, which could be of particular interest \cite{Keto+Watters:2024:PushBroom}.  

It is not only in the US that the interest in satellite-based instruments for UAP detection has increased, especially given the inherently limited scope of Earth-based observations. In 2022, the Julius-Maximilians-Universit\"{a}t of W\"{u}rzburg (JMU) made history by becoming the first prominent western university to formally recognize UAP as a legitimate subject for academic inquiry. The University's Interdisciplinary Research Center for Extraterrestrial Science (IFEX) expanded its scope to include UAP research alongside its primary focus on space exploration and the search for extraterrestrial life. Under the supervision of Hakan Kayal, IFEX issued a research paper titled ``Detection of UAP with a Nano Satellite'' in early 2022 \cite{Reitemeyer+Weinmann:2022}. The paper evaluates the feasibility of using two proposed satellites for full-time observation, employing onboard preprocessing of data to facilitate orbital searches for UAP. Visual detection was deemed to be the most achievable and two satellite versions with appropriately sized buses and solar panels were modeled, which met most of the user requirements. The study presented a tactic and structure for onboard data processing specialized for UAP detection, reducing unnecessary data transmission to the ground. Such a platform would provide global data on UAP occurrences, forming a basis for more extensive research. The document states that a nanosatellite of 12 to 30U\footnote{In the context of satellites, ``U'' stands for ``Unit'' which is a standard measurement used to describe the size of CubeSats. One ``U'' is defined as a cube with dimensions 10 cm x 10 cm x 10 cm.  Retrieved from [\url{https://www.asc-csa.gc.ca/eng/satellites/cubesat/what-is-a-cubesat.asp}. (Accessed: 14 Jan 2024)]} could fulfill this role at a significantly lower cost than larger systems.

However, it is recognized that a more extensive project with appropriate funding would be needed to realize the satellite concept. This concept could also be expanded to larger satellites for more varied and higher quality measurements or multiple satellites for more frequent coverage, depending on available funding. As emphasized by military and scientific experts, a greater understanding of the UAP will be very valuable. With high enough reliability, the satellite could also be linked to other systems, either on the ground or in orbit, notifying them of a detection to collect more data when needed.

In a significant expansion of UAP research, IFEX (see Sec. \ref{sec:IFEX}) has also extended its efforts beyond traditional Earth-based observations to include the near-space environment.  IFEX's initiatives exemplify a strategic leap forward in integrating terrestrial and space-based observations to advance our understanding of unexplained aerospace phenomena.

In conclusion, the integration of satellite technology, artificial intelligence, and citizen science initiatives represents a pivotal advancement in UAP research, offering new avenues for rigorous scientific inquiry and data-driven exploration. Future missions and technological innovations hold promise for overcoming current limitations and expanding our observational capabilities. In the coming years, collaborations between academia, commercial entities, and governmental space organizations, including NASA, ESA, CNES, and DLR, will be crucial in advancing UAP research. Using the global network of remote sensing satellites and harnessing the power of AI and citizen science, researchers can approach UAP research with unprecedented breadth and depth. This interdisciplinary approach underscores the transformative potential of integrating cutting-edge technologies into the study of unexplained aerospace phenomena.



\section{Conclusion}
It is important to keep in mind that UAP are a class of unknown phenomena, and not a single thing \cite{Knuth:2023}. For this reason, the instruments used to study them need to be sufficiently diverse to be able to provide useful information about a wide class of phenomena.

We have seen that, as a class, UAP describe a wide range of at least initially unidentifiable aerial and sometimes undersea phenomena with characteristics that present to the sciences a number of challenges --- both in terms of their physical properties and also in terms of the manner in which they manifest. The latter is perhaps the key difficulty in studying the phenomena strictly scientifically, as their seeming randomness or ephemerality (or ``elusiveness'', as the French philosopher Bertrand M\'{e}heust recently described it) requires constant monitoring of wide swaths of terrestrial parameter space, which in turn requires significant resources (technological and personnel) to be devoted to the research for long periods of time. But because of their tantalizing elusiveness, the default mode of empirical study has been forensic in nature.  As Eghigian \cite{Eghigian:2024} recognized in his recent history of the UFO phenomenon, attempts to empirically engage UAP have been largely confined to mere forensic cold case chases (as if dealing with a crime). But this is in truth inadequate to the very nature of the phenomena: if UAP are elusive or ephemeral, then no mere forensic study (which is what we might say ``classical ufology'' was all about) can possibly be definitive; only long-term, transgenerational research programs, such as enjoyed by many research programs well established and stabilized within academic science now for many decades, can possibly yield the proper data on which a potential resolution to UAP can be founded. Yet, when a definite resolution \textit{was} attempted, scientists and governments reached the default mode of forensic investigations and --- unsurprisingly --- often came to the conclusion that there is not much scientific value to be obtained from studying UAP.

What we have also seen in this essay is that, starting decisively in the year 2021, and with the founding of academic projects such as Harvard's \textit{Galileo Project} or W\"{u}rzburg University's IFEX, or the UAlbany-UAPx Collaboration, there is a concerted effort by serious academic researchers to transition away from classical ``ufology'' --- a forensic or forensic-like science which relies principally on the UAP case report (rather than on reliable data about UAP themselves) --- to a stricter, non-forensic observational research paradigm that is not primarily reliant on the case report but is, rather, in pursuit of data produced on and from reliably calibrated, well-characterized and accurately synchronized suites of instruments. Thus, the decisive historical change we have attempted to convey and document in this paper is the move away from the forensic cold case chase such as typifies classical ufology --- often conducted by the ``citizen scientist'' --- to the stricter forms of observational (\textit{Galileo Project}, IFEX, UAlbany-UAPx) and experimental / observational science (i.e. VASCO and ExoProbe) conducted by university-trained professional scientists. Only long-term, transgenerational, sustained university research programs devoted to gathering --- and then analyzing --- data on UAP, we argue, can break the paradoxical loop of dismissal in mainstream science we have seen throughout the pre--2021 history of attempts to study UAP empirically: UAPs are not taken seriously because they are not studied seriously by trained academic scientists; but UAPs are not studied seriously by academic scientists because they are not considered a serious topic for study! Forensic cold case chases will not break this loop. Only well-funded, transgenerational university research can. That is a very long game for UAP research, like any other academic research program, but so far this has not happened. This paper shows that this, we hope, is changing.

Reflecting further on the history of the conflict between UFOlogy and the scientific community, we are inspired by the Pocantico Meeting organized by Sturrock in 1997.  Although Sturrock's account details some of the familiar frustrations experienced by both the investigators and the academics, the recommendations made by the panel were both reasonable and productive.  Perhaps holding more meetings with the same planning, organization and care might be equally productive.

\section*{Acknowledgements}
The authors express their gratitude to Peter Sturrock, who passed away while this paper was being written, and whose influence on UAP studies cannot be overstated.  
There are a number of UFO researchers who could have been mentioned along with their colleagues in the appendix.  In the end, both space and time became constraining factors, but we wish to emphasize that this in no way lessens our appreciation of their contributions.  In this spirit, the authors express their gratitude to Jan Aldrich and Richard Hall.

We are deeply grateful to the volunteers who made the ``Flying Saucer Review'' a quality publication for so many years, ensuring that the work of dedicated scientists and investigators would be recorded for the evolving research community. We especially recognize the five long-term editors who succeeded each other in this thankless and difficult task: Derek Demster, Brinsley Le Poer Trench, Waveney Girvan, Charles Bowen, and Gordon Creighton.  

Avi Loeb is warmly thanked for his contributions to the description of the Galileo Project and, more broadly, for his efforts to draw rigorous scientific attention to the UAP topic.

The authors thank Larry J. Hancock, Ian M. Porritt, Sean Grosvenor, Larry Cates, and Ike Okafor of SCU for providing the histograms of UAP sightings near nuclear weapon complexes (Fig. \ref{fig:atomic-sites}).

KHK thanks Michael Way for assisting with the Swedish references from the 1930s.

JT and GT thank their friend and colleague Donna Nardo for her efforts as an Investigator/Senior Field Technician for the Nightcrawler - Eye on the Sky Project. 

Last, while it is perhaps awkward to thank one's coauthors, KHK expresses his gratitude to them for their efforts in studying UAP (some of whom have been doing this their whole lives), for inspiring him to take on this exciting, albeit daunting, and at times, intimidating, challenge, for assisting in writing this altogether enormous and comprehensive paper, and for several years of friendship and support through what have been the most challenging and exciting years of his life.  Most importantly, KHK thanks his family and especially his wife, Emily, for their support.


\appendix
\section{Prominent Past Efforts and Individuals} \label{sec:Appendix}
Four years ago, the director of Harvard College Observatory, the Administrator of NASA, the Director of the CIA and the Archbishop of Washington stood at the National Cathedral to announce that the enigma represented by UAP (UFO) reports remained an unsolved mystery.  Following such statements by government and military authorities about the reality of unexplained sightings of unexplained objects in the Earth’s environment, the public and the scientific community have begun to revise their negative impressions of the subject. Naturally, they now seek reliable information on its nature.   For this reason, reconstructing and presenting the history of research on the topic of UAP has become critical.

The range of phenomena known as UAP (formerly referred to as UFOs) was never accepted as a subject of interest in the Hard Sciences.  Even quality studies into the material nature and behavior of reported objects have been consistently discouraged by major scientific magazines---journals such as \textit{Science}, \textit{Nature} or even \textit{Scientific American}.  Their pages continue to be denied for such studies, even when submitted by established scientists with credentials and a history of publication in other areas.

Much of the UAP literature of various quality is now found in the form of student theses, conference transcripts, and papers about witness psychology, reporting statistics, medical symptoms, and sociological interpretations of sightings, but not in formal papers about physics, engineering, or site investigation of actual cases. As for computer compilations of the many thousands of reported cases, few sources are publicly available and their reliability is debatable.
Such documentation does exist, however, in the form of research studies dating back to the 1960s, but it is easy to miss because it had to be published in rare amateur magazines that have only survived in obscure Internet archives or in the confusion of online depositories.  An important exception is the professionally curated and edited magazine called \textit{Flying Saucer Review}, which was published in England from 1955 to 2000.  The researchers who submitted their results to such an obscure publication were particularly dedicated and passionate about the subject (and no, they did not always agree).

In this section, we summarize the research efforts and works from a number of prominent PhD scientists and academics who studied UFOs relatively early on.  We explicitly mentioned and described Flying Saucer Review because it is one of the only publications to which many of these individuals were able to submit their work.

\subsection{David Akers, Yakima Valley, Washington USA}
David Akers coordinated with local residents to study anomalous light phenomena on the Yakima Indian Reservation in the vicinity of Toppenish, WA, from 1972 to the early 2000s.  This activity started in 1971 and spiked in the late summer of 1972, consisting largely of nocturnal lights very near the ground or at low altitude.  The initial phase of investigation lasted from August 20 to August 31, 1972, during an intense wave of sightings; many of the most interesting results were derived from this period.  In this interval, nocturnal lights were reported daily in the towns of Union Gap, Harrah, Zillah, Toppenish, Wapato and Mabton.  Akers used optical SLR cameras (VIS-NIR), a compass spin detector, infrared photometer, microphone and casette recorder, and energetic particle detector (Geiger-Muller counter) (see \cite{Akers:2007} Appendix B).  Relying on community reports, Akers deployed his instruments where sightings were most recently reported (a ``hit-and-run'' approach) and was successful in witnessing and recording events in this way.  Activity decreased dramatically after the first two days of the study.

Akers acquired several photographs of what he described as ``large, flickering orange objects'' \cite{Akers:2007}, although no quantitative size determination was possible.  An object viewed with binoculars was described as a ``sharply defined, brownish orange disc `` with a mottled surface''.  This object appeared to be flat and illuminated the ground with an orange glow.  The appearance and motion of the light phenomena were highly distinctive and easily distinguished from those of conventional aircraft.  On multiple occasions, attempts to acquire photographs were foiled when illumination ceased abruptly just as a camera was trained on the object.  The most striking example of this occurred at a relatively close range, while an object hovered directly overhead \cite{Akers:2007}.

On at least one occasion, a light was recorded in a background exposure, captured when no lights were visible to the naked eye, suggesting near-infrared emission or visible band emission with a pulse duration and repetition rate below the perception threshold of human sight.  Akers also noted the possibility that the objects may have been moving in a discrete or noncontinuous fashion, as suggested by the observed flickering (when observed by eye) and discontinuities in long-exposure photographs; this was not definitively distinguished from ordinary flashing and displacement while the objects are nonluminous.  During the initial 11-day field campaign, none of the other instruments succeeded in recording the anomalous events.  On one occasion, Akers and a colleague experienced over a minute of extremely bright and disorienting light flashes in their peripheral vision, later tentatively ascribed to strong transient magnetic field fluctuations later observed at the same location \cite{Akers:2007}.

\subsection{Phyllis A. Budinger, Frontier Analysis, Ltd.}
\label{sec:Budinger}
Phyllis Budinger is a chemist who is one of the few scientists who has worked to study the physical traces produced during UFO encounters.  She received her B.S. in chemistry in 1961 from Baldwin-Wallace University in Berea, Ohio, and her M.S. in chemistry in 1964 from Miami University in Oxford, Ohio.  She began working for Sohio, which was the original Standard Oil Company that later became BP/Amoco.  She accumulated 35 years of analytical chemistry experience, during which she headed a group that specialized in spectroscopy and magnetic resonance.  She retired as a research scientist from BP/Amoco in 1999 \cite{Budinger-CUFOS:2025,Budinger-BlackVault:2025}.

In 2000, Budinger founded Frontier Analysis, Ltd. as a laboratory, which has been involved in analyzing physical traces related to UFO cases, such as UFO debris \cite{Budinger-UT084:2014, Budinger-UT086:2014} and deposits \cite{Budinger-UT009:2001, Budinger-UT011:2001, Budinger-UT061:2009}, as in the 1971 Delphos, Kansas USA encounter \cite{Budinger-UT001:1999, Budinger-CUFOS:2025,Budinger-BlackVault:2025}, and angel hair \cite{Budinger-UT008:2000}, as well as chemical traces related to claims of abductions \cite{Budinger-UT025:2003, Budinger-UT041:2005, Budinger-UT047:2006, Greenewald:2020}, unexplained cattle mutilations \cite{Budinger-UT016:2001}, and crop circles.

\subsection{Dr. Erol A. Faruk}
\label{sec:Faruk}
Dr. Erol A. Faruk was born in London, UK, in 1951. While undertaking postdoctoral organic chemical research at Nottingham University in England, he requested and received, through CUFOS, soil material that had reportedly been subjected to physical and chemical changes by an illuminated UFO observed hovering just above the ground. This event \cite{Phillips:1972a, Phillips:1972b} occurred in November 1971 on a farm in Delphos, Kansas, USA and left an eight-foot diameter ring of soil whose surface was found to glow brightly once the illuminated UFO had left the area. Subsequent field investigations showed that the ring soil had now become impervious to water down to a depth of 14 inches, such that when snow had fallen later there was no absorption by the ring, leaving only a snow-covered shape while all around was dark muddy soil.

After receiving the material Faruk approached his academic supervisor (Dr. B.~W. Bycroft) who, after being described the event and associated glow, became curious enough to refer him to the resident chemiluminescence expert (Dr. Frank Palmer) who became similarly intrigued by the case. Together they enabled Faruk to analyze the material while Palmer undertook recording of fluorescence spectra on samples taken as the investigation proceeded. Faruk discovered that the ring soil was impregnated by an air-sensitive organic compound that on chromatographic purification revealed it to have all the attributes necessary to generate light, a process known as oxidative chemiluminescence. He was able to use the soil data to propose a viable theory as to how the ring had actually formed, essentially confirming the witnesses’ description of the nearby presence of a hovering aerial device of unknown technology and origin \cite{Faruk:2014, Faruk:2021}.

\subsection{Dr. Pierre Gu\'{e}rin, Institut d'Astrophysique}
Dr. Pierre Gu\'{e}rin (1926 - 2000) was a French astrophysicist who specialized in planetary studies. Working between his office in Paris and observatories in the South of France (primarily at the Saint Michel Observatory in Haute Provence and at the Pyr\'{e}n\'{e}es observatory at Pic du Midi), he produced some of the finest photographs of Mars, Jupiter, and Venus in pre-Sputnik days. He was also the discoverer of a new ring of Saturn and initiated improvements in optical and photographic planetary studies for space-based exploration.

A former Ph.D. student of G\'{e}rard de Vaucouleurs, (the man who brought his expertise in planetary and galactic astrophysics to the University of Texas in Austin), Pierre Gu\'{e}rin was among a cohort of French scientists who had submitted articles about UFOs to the 'official' academic press and had been rebuked, with the implication that such efforts would be detrimental to their career in the rigid mindset of French science. Decades before, Director LeVerrier of Paris Observatory had fired Camille Flammarion, a fine astronomer, because he tried to popularize science and dared to speculate about life on other worlds.

As director of research at the Astrophysical Institute in Paris, Dr. Gu\'{e}rin was among the senior scientists who objected. Finding no professional journal willing to touch the subject, he turned to the British \textit{Flying Saucer Review} (FSR) because---far from being the work of amateurs--- it was professionally edited by a group of able specialists. His first two communications in FSR appeared in March 1970 and were followed by many others. Here we give a selection of the most significant ones:
\cite{Lemaitre:1969:plan,Lemaitre:1969:parallel,Guerin:1970:letter,Guerin:1970:warminster,Guerin:1973,Guerin:1976,Guerin:1976:Letter,Guerin:1976:review,Guerin:1979,Guerin:1983:coverup,Guerin:1983:impatient,Guerin:2000}.

The above selection of papers shows the mind of a scientist anticipating many of the issues that plague research to this day: the diversity of testimony, the tendency to pick one set of observed characteristics over another, and the complexity of technology when used in field conditions.

\subsection{Dr. Richard F. Haines, NASA-Ames Research Center} \label{sec:Haines}
Dr. Richard F. Haines, Ph.D., (born 1937) long served as a research scientist for NASA beginning in 1967.  While working at NASA Ames Research Center, Haines investigated numerous aviation accidents and incidents for the United States Federal Aviation Administration (FAA) and the United States National Transportation Safety Board (NTSB). This professional work includes over two decades of ``Human Factors in Space'' research studies. Dr. Haines has also worked with the FAA on several projects including "Head Up Display" (Flight Standard Hqs.). A member of the International Society of Air and Safety airline investigators, Dr. Haines has interviewed many military and commercial airline pilots, and spoken with many air traffic controllers to amass reports of more than 3,000 UFO sightings.  Dr. Haines has been researching this subject now for over 37 years. He has written three books on the topic of UFOs \cite{Haines:1987:Missing,Haines:1994:ProjectDelta,Haines:1999}, over 50 scientific and technical articles and more than 20 serious articles on unidentified aerial phenomena. We present here a selection of some of his more significant publications:
\cite{Haines:1976,Haines:1980:B36,Haines:1980:NZ,Haines:1982:Part1,Haines:1982:Part2,Haines:1982:Supplementary,Haines:1984,Haines:1984,Haines:1987:Missing,Haines:1987:Review,Haines:1992:Joint,Haines:1993:Reverse}.

\begin{figure}
\centering
\makebox{\includegraphics[width=0.75\columnwidth]{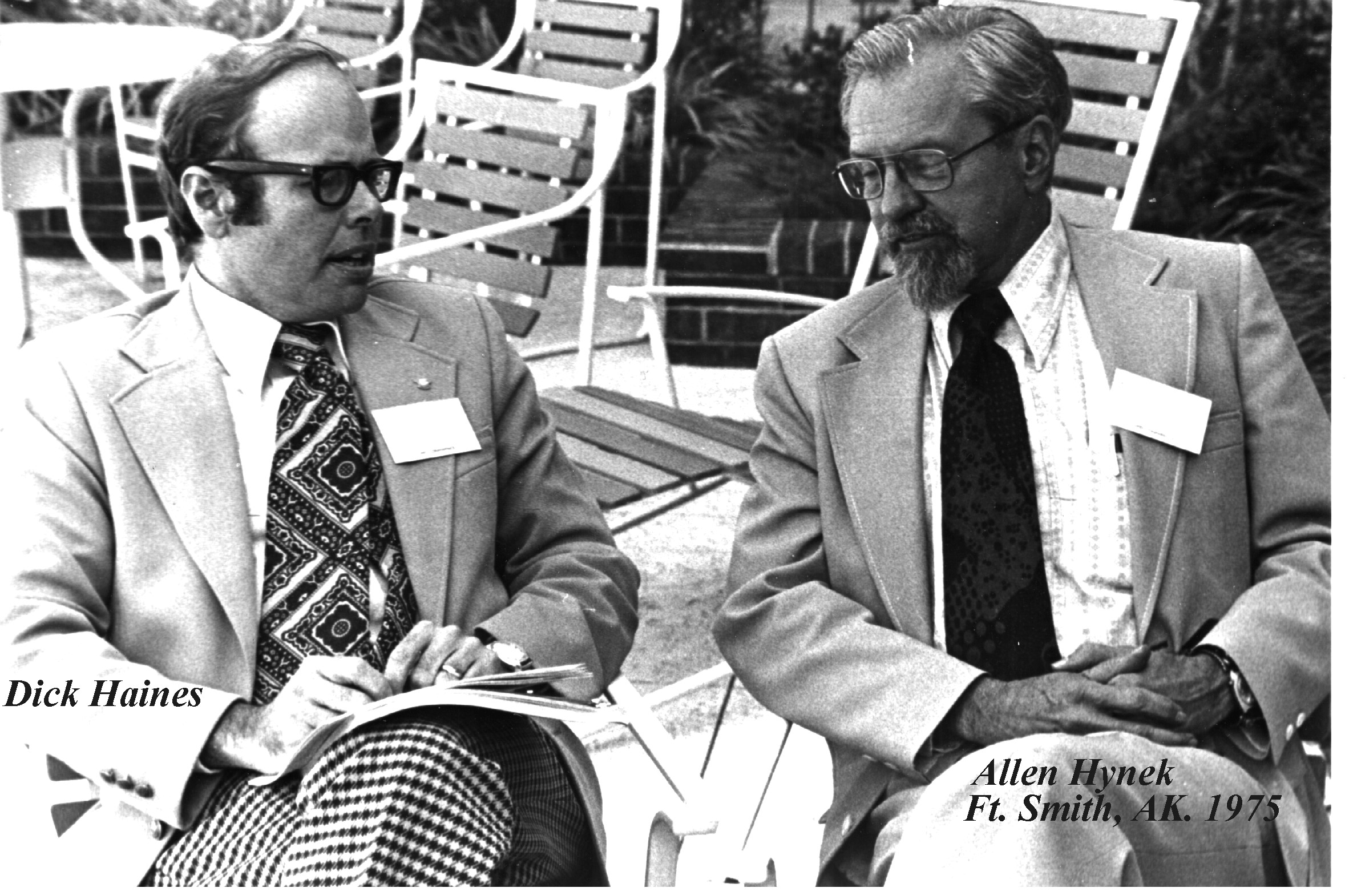}}
\caption{Richard F. Haines (left)and J. Allen Hynek (right).  Image taken in 1975 at a conference in Ft. Smith, Arkansas, USA.} \label{fig:Haines+Hynek}
\end{figure}

\subsection{Paul R. Hill, NASA Langley Research Center}
Paul R. Hill (1909-1990) was a distinguished American aerodynamicist who made significant contributions to mid-20th-century aerospace research. He obtained in his BS in mechanical engineering from the University of
California, Berkeley in 1936, and joined the National Advisory Committee of Aeronautics (NACA), the predecessor to NASA, in 1939. He worked on various NASA
aerospace projects, including aircraft design, space missions, hypersonic propulsion, and the Apollo moon missions. He was one of NASA's chief scientists and
was awarded NASA's Exceptional Service Medal in 1969. The historian James Hansen describes Hill as one of two ``key members of Langley's early space station research.'' \cite[pp. 275--277]{Hansen:1995}, and as the specialist on configuration and propulsion on the HYWARDS (HYpersonic Weapon And R and D System) Program \cite[p. 368]{Hansen:1987}.

Hill is is widely known for his book \emph{Unconventional Flying Objects: A Scientific Analysis}, which was rejected by many publishers and published after his death.  After his considerable analysis of reported cases and noting their advanced performance characteristics, he accepted that they were most likely not from Earth. As Hill applied his expertise in aerodynamics to the study of the phenomena, his book focused on the scientific analysis of the types of craft and their possible propulsion systems.  He concluded that UFOs, ``obey, not defy, the laws of physics.'' \cite{Hill:1995}

In the early 1950's, Paul had his first UFO sighting in the Chesapeake area of Virginia which he reported to Project Blue Book. He witnessed four objects flying in a V-formation over Chesapeake Bay, near Langley Air Force Base; that, after being investigated, were classified as being aircraft.   He had yet another encounter a decade later which he did not report but it is mentioned in his book \cite{Hill:1995}.

\subsection{Richard Lee Hoffman}
Rich Hoffman was born in Jasper, Indiana, in 1951 and at age 4 his parents moved to Dayton, Ohio. He graduated from West Carrollton High School in 1969 and went on to college at Ohio University in Athens, Ohio. He transferred to Wright State University and graduated with a BA degree in Organization Communications in 1995. He worked at a variety of companies as a Training and Development Manager for a variety of companies (Emery Worldwide, Miami Valley Hospital, and Ponderosa Steakhouses) and later transitioned into Information Technology and for the past three decades has been a defense contractor serving the US Army Materiel Command HQ. he is presently with Chugach Government Solutions LLC as a Senior Systems Engineering Analyst.

It was in 1964, timed with the Socorro, New Mexico sighting at age 13, that Rich began investigating the UFO/UAP phenomena. By age 15, he was lecturing to churches, service clubs, and more and found himself on the Phil Donahue Show. He became the ``go-to guy'' for UFOs in Dayton. He got connected with Project Blue Book at Wright Patterson AFB and was provided a hot line number to RAPCON (Radar Approach Control) at the Base able to confirm traffic in and around Dayton when conducting investigations. 

In 1969, Rich joined the Midwest UFO Network (MUFON) and was the State Section Director. He served as the Director of Investigations for an Ohio-based UFO organization called the Ohio UFO Investigators League. It was this organization that held the 1978 MUFON Symposium in Dayton, attended by ~3,000, the largest in UFO Conference in UFO History. 

During the 60 years, Rich conducted more than 1000 investigations, conducted field investigator training sessions, and held numerous additional positions with MUFON. He was the State Director for Alabama and Mississippi, the Deputy Director of Investigations and Star Team Manager from 2009 to 2015 and later the Director of Strategic Projects, a role that sought to bring technology to the forefront. In 2017, Rich left his MUFON leadership positions and cofounded the Scientific Coalition for UAP Studies (SCU). The organization has grown rapidly, now close to 400 members worldwide. The vast majority are scientists, engineers, researchers, military, and many other disciplines. Rich is also an advisor and supporter for UAPx and the Society for UAP Studies (SUAPS).

Rich has been on numerous TV, radio, podcasts, and news media and still lectures at universities and clubs. He is now in his fifth season as an expert on the History Channel show \textit{The Proof is Out There}. 

In 2022, Rich represented the US along with Ryan Graves for a NATO workshop in Bologna, Italy. The workshop was focused on the Space Domain and the UAP as multi-domain phenomena and potential threat that NATO needs to consider as it develops its policies.

\subsection{Dr. J. Allen Hynek, Northwestern University}
\label{sec:Hynek}
\begin{figure}
\centering
\makebox{\includegraphics[width=0.75\columnwidth]{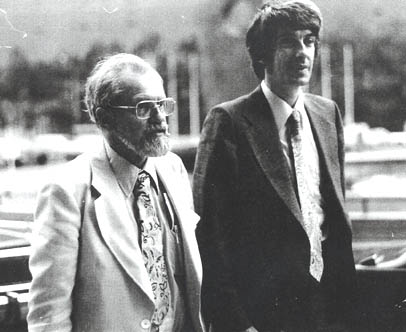}}
\caption{J. Allen Hynek (left)and Jacques Valll\'{e}e (right).  The image is in the Public Domain.} \label{fig:Hynek+Vallee}
\end{figure}

Dr. Josef Allen Hynek (1910-1986) was born in Chicago (see Figure \ref{fig:Hynek+Vallee}). He graduated from the University of Chicago with a Bachelor of Science degree in 1931 and then spent four years at the Yerkes Observatory of the Department of Astronomy and Astrophysics at the University of Chicago, where he received his PhD in astrophysics in 1935. In 1936 Ohio State University hired him as an instructor in the Department of Physics and Astronomy and promoted him to assistant professor in 1939. 

During World War II, Hynek worked on developing a proximity fuse for the US Navy at Johns Hopkins University’s Applied Physics laboratory, returning to Ohio State in 1946 as associate professor and director of the McMillin Observatory. In 1950, he was promoted to full professor and became assistant dean of the Graduate School.

While at Ohio State University, in 1948, Hynek had begun consulting for Project Sign at Wright-Patterson Air Force Base in Dayton, Ohio to investigate reports of UFOs. Hynek was initially skeptical of reports on UFOs, and stated that ``the whole subject seems utterly ridiculous.''  Hynek continued working for the Air Force as Project Sign became Project Grudge, and later remained a scientific consultant for Project Blue Book while he was at Northwestern University.

Like most people, Hynek was a complicated individual, as he was skeptical and enjoyed his role as a debunker for the Air Force.  In an interview in 1985, Hynek said: \cite{Stacy:1985}
\begin{displayquote}
    My own investigations for Project Sign added to that [negative attitude surrounding UFOs], too, I think, because I was quite negative in most of my evaluations.  I stretched far to give something a natural explanation, sometimes when it may not have really had it. I remember one case from Snake River Canyon, I think it was, where a man and his two sons saw a metallic object come swirling down the canyon which caused the top of the trees to sway. In my attempt to find a natural explanation for it, I said that it was some sort of atmospheric eddy. Of course, I had never seen an eddy like that and had no real reason to believe that one even existed. But I was so anxious to find a natural explanation because I was convinced that it had to have one that, naturally, I did in fact, it wasn't until quite some time had passed that I began to change my mind. 
\end{displayquote}
Come to change his mind, he did.  Hynek slowly came to resent the ``negative and unyielding attitude of the Air Force'', which never gave the ``UFOs the chance of existing, even if they were flying up and down the street in broad daylight. Everything had to have an explanation.'' Furthermore, ``the caliber of the witnesses began to trouble'' him. \cite{Stacy:1985}.

Hynek described his complicated relationship with the Air Force and scientists interested in the UFO phenomena in his 1985 interview \cite{Stacy:1985}:
\begin{displayquote}
    Dr. James E. McDonald, a good friend of mine who was then an atmospheric meteorologist at the University of Arizona, and I had some fairly sharp words about it. He used to accuse me very much, saying you're the scientific consultant to the Air Force, you should be pounding on generals' doors and insisting on getting a better job done. I said, Jim, I was there, you weren't you don't know the mindset.  They were under instruction from the Pentagon, following the Robertson Panel of 1953, that the whole subject had to be debunked, period, no question about it.  That was the prevailing attitude. The panel was convened by the CIA, and I sat in on it, but I was not asked to sign the resolution.  Had I been asked, I would not have signed it, because they took a completely negative attitude about everything. So when Jim McDonald used to accuse me of a sort of miscarriage of scientific justice, I had to tell him that had I done what he wanted, the generals would not have listened to me. They were already listening to Dr. Donald Menzel and the other boys over at the Harvard Astronomy Department as it was. 
\end{displayquote}

Hynek believed that UFOs should be properly studied by the scientific community.
He intensified his personal message when the Air Force closed Project Blue Book after a 1969 report by Dr. Edward Condon of the University of Colorado concluded that UFOs did not merit further inquiry.  In his 1985 interview, Hynek stated, 
\begin{displayquote}
   the Condon Report said that a group of scientists had looked at UFOs and that the subject was dead. The UFOs, of course, didn't bother to read the report and during the Flap of 1973, they came back in force.
\end{displayquote}
After 1969, Hynek was virtually alone in the scientific community in supporting the continued study of UFOs.  Largely in response to the Condon report, Hynek wrote The UFO Experience \cite{Hynek:1972}, laying out the three classes of ``Close Encounters.'' Northwestern administrators were embarrassed by the publicity, demanding that Hynek’s Center for UFO Studies be kept separate from the University. Hynek retired from Northwestern in June 1978, and moved the Center to his home in 1981, disconnecting the toll-free hotline in 1982. In search of funds and better research opportunities, Hynek moved the Center for UFO Studies to Arizona in 1984.

It is a sad commentary on American science during the second half of the \nth{20} century that Dr. Hynek and his team were never allowed to publish their research in the scientific press, despite multiple efforts. Instead, much of the fundamental work done during this period found a home in \textit{FSR}. A list of selected articles by Dr. Hynek follows:
\cite{Hynek:1966,Hynek:1970,Hynek:1970:AAAS,Hynek:1972,Hynek:1984,Hynek:1984:myopia,Hynek:1985}
Dr. Hynek contributed many other items such as Letters to Editors and comments on current events in the field throughout his life. He was also a popular guest on TV programs and on local talk shows.

\subsection{Dr. Bruce Maccabee, US Naval Surface Warfare Center}
Bruce S. Maccabee (1942--2024) was a career physicist in optics for the US Navy. He received a B.S. in physics from Worcester Polytechnic Institute in Worcester, Massachusetts, and his M.S. and Ph.D. from American University, Washington, DC. He worked at the Naval Ordnance Laboratory, White Oak, Silver Spring, Maryland; which later became the Naval Surface Warfare Center Dahlgren Division.  He worked with optical systems and various aspects of the Strategic Defense Initiative (SDI) and Ballistic Missile Defense (BMD) using high-power lasers.

Dr. Maccabee's interest in UFOs began in the 1960s. He authored several books on the subject \cite{Walters+Maccabee:1997}\cite{Maccabee:2000}\cite{Maccabee:2018} and has written dozens of articles on specific UFO sighting reports. As an optical scientist, his analysis of the luminosity of UFOs \cite{Maccabee:1999}, as well as his analysis of two UFO photos taken in McMinnville, Oregon \cite{Maccabee:McMinnville:1995} were invaluable . His UFO research and investigations also included the Kenneth Arnold sighting of 1947 \cite{Maccabee:Arnold:2017}, the F4 engagements of a UFO over Tehran in 1976 \cite{Maccabee:Iran:2006}, the Kaik\={o}ura New Zealand videos and sightings of December 1978 \cite{NZHerald:2018, Maccabee:NZ:1987}, the Japan Airlines (JAL1628) sighting and radar data of 1986 \cite[pp. 17--18]{Powell:2024}, and the Phoenix Lights sightings of March 1997 \cite{Kitei:2010}.

\subsection{Dr. James McDonald, University of Arizona}
Dr. James McDonald (1920--1971), a nationally known University of Arizona physicist, is credited with advances in cloud physics and weather modification research \cite{Battan:1971}.  One of his first influential papers was on the shape of raindrops \cite{McDonald:1954}.  His meteorological work focused on radiation, atmospheric optics, cloud physics, nucleation, and cloud/weather modification \cite{McDonald:1958}.

Born and raised in Duluth, Minnesota, he served as a cryptographer in the United States Navy during World War II. He received a B.A. in chemistry from the University of Omaha in 1942, an M.S. in meteorology from the Massachusetts Institute of Technology in 1945, and his Ph.D. in physics at Iowa State University in 1951. In 1953, Dr. McDonald, as a professor of meteorology at the University of Arizona in Tucson, helped to establish a meteorology and atmospherics program at the University of Arizona.  He later became the head of the Institute of Atmospheric Physics.  McDonald had a major impact on atmospheric physics through his service on the Panel on Weather and Climate Modification of the National Academy of Sciences \cite{Battan:1971}.

McDonald gave a lecture on ``The Problem of UFOs'' to the American Meteorological Society in Washington, D.C. in October 1966.  McDonald estimated that only about 1\% of UFOs were unknowns, and that the UFOs of interest were described by the ``credible and trained observer as machine-like `craft' which remained unidentified in spite of careful investigation.''

McDonald found support to research UFOS from the Office of Naval Research in 1967, where he was to officially study the idea that some UFOs were misidentified clouds.  During his studies, McDonald was able to examine the files of Project Blue Book at Wright Patterson Air Force Base.  He concluded that the Air Force was mishandling UFO evidence. 

The Secretary General of the United Nations, U Thant, also supported McDonald's studies, and he arranged for McDonald to speak to the United Nations' Outer Space Affairs Group on June 7, 1967. During the address, McDonald declared that ``There is no sensible alternative to the utterly shocking hypothesis that UFOs are extraterrestrial probes'' \cite{Randles:1993}.

On 29 July 1968, McDonald addressed the House Committee on Science and Astronautics of the United States Congress on the topic of UFOs.  In his statement, McDonald notes \cite{McDonald:1968:UFOs}
\begin{displayquote}
    the scientific world at large is in for a shock when it becomes aware of the astonishing nature of the UFO phenomenon and its bewildering complexity. I make that terse comment well aware that it invites easy ridicule; but intellectual honesty demands that I make clear that my two years' study convinces me that in the UFO problem lie scientific and technological questions that will challenge the ability of the world's outstanding scientists to explain - as soon as they start examining the facts. [...] the scientific community [...] has been casually ignoring as nonsense a matter of extraordinary scientific importance.
\end{displayquote}

McDonald wrote detailed criticisms of the Condon report \cite{McDonald:1969}, and was especially disturbed by the fact that Condon, in his introduction, stated that all UFOs could be explained as hoaxes or misidentified prosaic objects, despite the fact that the report has classified 30\% of the cases investigated as ``unknown''.  In a letter to Condon, dated June 6, 1969, McDonald wrote: \cite[Appendix 16-B, p.574]{Druffel:2003}
\begin{displayquote}
    Enclosed are copies of a number of summaries of recent talks in which I criticized your Report. Your conclusions do not at all seem to be supported by the Reports content. The contents of the Report argue need for much more careful examination of the UFO problem. And this in spite of what I must view as many gross inadequacies of scientific aspects of the investigations you headed. I am unable to understand how you approached this task as you did. Your Philadelphia and Irvine talks indicate you must have no awareness of the weakness of the position you have developed. I am in process of preparing discussions of many features of the Report which seem to me to attest to that weakness. 
    
    In giving the Academy such a Report, I believe you did science a direct disservice. That the Academy processes could lead to endorsement is disturbing.
\end{displayquote}

In 1969 McDonald presented his influential lecture ``Science in default: Twenty-two years of inadequate UFO investigations'' to a UFO Symposium of the American Association for the Advancement of Science \cite{McDonald:1972}.  An audio recording of McDonald's presentation can be heard here \cite{McDonald:1969:audio}.

McDonald was deeply concerned about society and spent hundreds of hours researching the effects of the proposed supersonic transport (SST) emissions on the upper atmosphere.  He concluded that SST posed a great danger in that it would adversely affect the ozone layer of the Earth, increasing the amount of ultraviolet light reaching the surface and increasing the occurrence of skin cancer \cite{Battan:1971}.  In 1970, at a US congressional hearing on the development of supersonic transport (SST), McDonald testified that the SST could potentially harm the Earth's ozone layer.  Congressman Silvio O. Conte of Massachusetts, who supported the SST because of the participation of businesses in his district, discredited McDonald by referring to his UFO research stating that anyone who ``believes in little green men'' was, in his opinion, not a credible witness''.


A representative list of his articles on UFOs follows: \cite{McDonald:1968:UFOs,McDonald:1970:Kirtland,McDonald:1970:Lakenheath,McDonald:1970:RB-47,McDonald:1972}

\subsection{Dr. Donald Menzel, Harvard College Observatory} \label{sec:Menzel}
Dr. Donald Menzel (1901-1976) was born in Colorado. At the age of 16, Menzel enrolled in the University of Denver to study chemistry and stayed on for a master's degree in chemistry and mathematics, which he received in 1921. The solar eclipse of 8 June 1918 and the eruption of Nova Aquilae in 1918 stimulated his interest in astronomy. He found a position as a summer research assistant to Harlow Shapley at the Harvard College Observatory.  Menzel then received a Ph.D. in astronomy from Princeton University in 1924, and in 1926 he was appointed assistant professor at Lick Observatory of the University of California in San Jose, California; later moving to Harvard University.

His colleague at Harvard, Dr. Dorrit Hoffleit, recalls that one of his first actions was asking his secretary to destroy a third of the astronomical photographic plates (sight unseen) \cite{Hoffleit:2002}.  This resulted in their permanent loss from the record, which was referred to as the ``Menzel Gap''.  It is important to note that the Menzel Gap adversely affects the modern efforts by Dr. Beatriz Villaroel and her VASCO Project (see \ref{sec:VASCO}) to identify astronomical transients and possibly connect some of them with UAP \cite{Villarroel:2024}.

Menzel was a prominent skeptic concerning UFOs. He authored or co-authored three popular books debunking UFOs: Flying Saucers - Myth - Truth - History \cite{Menzel:1953}, The World of Flying Saucers \cite{Menzel+Boyd+Gifford:1963}, and The UFO Enigma \cite{Menzel+Taves:1977}, all of which argued that UFOs are misidentifications of prosaic phenomena, along with several relevant papers \cite{Menzel:1964:lines,Menzel:1964:cut-and-thrust,Menzel:1964:part1,Menzel:1964:part2}.

\subsection{Dr. Richard Niemtzow, Brooks Air Force Base}
Richard Charles Niemtzow was born on 18 April 1942 in Philadelphia, Pennsylvania. He received his medical degree, specializing in radiation oncology, in 1976 from the Universit\'{e} de Montpellier Facult\'{e} de M\'{e}decine, Montpellier, France, his Ph.D. in biological sciences in 1985 from Pacific Western University, and his Master's Degree in Public Health in 1992 from the Medical College of Wisconsin. He studied acupuncture at the University of California at Los Angeles Medical School.

Dr. Niemtzow attended the United States Air Force School of Aerospace Medicine, and became a Flight Surgeon, class of 891003, Brooks Air Force Base, Texas. He introduced innovative techniques of pain control using acupuncture, and is best known as the developer of battlefield acupuncture for pain relief \cite{Moore:2010}.

Apart from his work with the Air Force, Dr. Niemtzow investigated numerous cases of witness injuries after UFO encounters, working with John Schuessler of MUFON and civilian teams.  He established Project UFOMD, which consisted of a network of medical doctors to intensively study UFO related injury cases.  His publications include: \cite{Niemtzow:1980, Niemtzow:1982, Niemtzow:1983, Niemtzow:1984, Niemtzow:1991}

\subsection{Dr. Claude Poher, CNES} \label{sec:Poher}
Dr. Claude Poher, who holds a PhD in Astrophysics, began his career as an electrical engineer, and was an instructor with Air France, and astronomical researcher with the Centre National de la Recherche Scientifique (CNRS). Poher served for 30 years as a scientist in Space Research and Aeronautics with the Centre National d'Etudes Spatiales (CNES), contributing to scientific experiments in space at the time of the NASA lunar explorations, using the various USSR and US space stations and various planetary exploration missions. 

Poher has organized numerous international congresses of astronautics and advanced technical studies concerning the future feasibility of interstellar travel.
In 1977, Poher created and directed (1977-79) the Groupe d'Etudes des Ph\'{e}nom\`{e}nes A\'{e}rospatiaux Non-identifi\'{e}s (GEPAN) in Toulouse as a government-funded research group that examines unidentified aerospace phenomena (UFOs or ``OVNIs'' in French). Now known as ``GEIPAN'', the group remains the only civilian, unclassified research organization devoted to the compilation, investigation and analysis of UFO reports. 

In 1996, Poher retired from CNES to pursue personal research into the physics of gravitation. Experiments based on his theory has recently allowed him to develop a series of patented devices with potential applications in various fields. Dr. Poher has been awarded the French National Order of Merit Medal, the AAAF Astronautical Prize, and the CNES Medal. 

Among Dr. Poher's influential works are these two papers investigating patterns in UFO sightings \cite{Poher:1972, Poher:1974}, and this paper detailing the UFO encounters at Minot Air Force Base in North Dakota USA in 1968 \cite{Poher:2005}, which was critical in assisting to establish the high accelerations and speeds of UFOs/UAP noted earlier in 1954 by German rocketry pioneer Hermann Oberth \cite{Oberth:1954}, and by Coumbe \cite{Coumbe:2022}, and Knuth, Powell, and Reali \cite{Knuth+etal:2019}, as summarized by Knuth \cite{Knuth:2023}.

\subsection{Dr. Harley Rutledge, Southeast Missouri State Univ.}
Physicist Dr. Harley D. Rutledge of Southeast Missouri State University, together with his PhD students, carried out a monitoring campaign called Project Identification in Piedmont, Missouri USA from 1973-1981 \cite{Rutledge:1982}, where a recurrent light phenomenon occurred. Rutledge's team used magnetometers, telescopes, and optical and radio spectrometers to obtain measurements.  These measurements were collected together with detailed reports from numerous witnesses.  This was the first scientific-instrumented attempt to study these phenomena, and while the researchers were driven and determined, they could not develop the type of systematic or automated approach that one would aspire to today due to the lack of computer technology.  Despite this, Project Identification was  somewhat successful, especially since it both inspired and informed subsequent scientific studies of the phenomena.

\subsection{Dr. Peter Sturrock, Stanford University} \label{sec:Sturrock}
Dr. Peter A. Sturrock (1924-2024) studied at Cambridge University, England, 
and was awarded the University Rayleigh Prize for mathematics. His work on
electron physics at Cambridge, the National Bureau of Standards in D.C., and
the \'{E}cole Normale Sup\'{e}rieure in Paris, earned him a PhD in 1951. Much of his career was spent at Stanford University as the Professor of Space Science and Astrophysics in the Applied Physics Department. He was a consultant for NASA, Brookhaven National Laboratories, CERN, Boeing, and the the British Atomic Energy Research Establishment.

In 1977, Dr. Sturrock organized a survey of members of the American Astronomical Society \cite{Sturrock:1994:part_1} in which he found that out of the 1356 respondents, 23\% stated that UFOs should certainly be studied and 30\% said that UFOs should probably be studied as opposed to the 17\% that said UFOs should probably not be studied and 3\% that said that UFOs should certainly not be studied.  Moreover, Sturrock found that 62 of the respondents had either witnessed or obtained an instrumental record of something that they could not explain and that of those responding witnesses, 63\% of them were night sky observers.

Dr. Sturrock helped establish the Society for Scientific Exploration in 1982 to provide a scientific forum to subjects such as UFOs that are neglected by the mainstream. Their publication, the Journal of Scientific Exploration, has been published since 1987.  Sturrock organized a scientific panel in 1997 to review various types of physical evidence associated with UFOs. The panel's conclusions were that the cause of UFOs was complex, it needed more physical evidence, and more attention from the scientific community was required. Sturrock provided the details of the work by this scientific panel in The UFO Enigma \cite{Sturrock:1999} as well as his view of the need for a new paradigm in the study of UFOs in A Tale of Two Sciences \cite{Sturrock:2009}.

\subsection{Dr. Michael Swords, Western Michigan University} \label{sec:Swords}
Dr. Michael D. Swords, Ph.D., (born 1940) obtained his doctorate in the  History of Science and Technology from Case Western Reserve University and his Masters of Science in Biotechnology from Iowa State University.

Professor Swords is a long term board member of the J. Allen Hynek Center for UFO Studies, and he edited the Journal of UFO Studies for six years. With over 40 years of experience studying the UFO subject, Professor Swords has written extensively on the two most famous examinations of the subject: The Robertson Panel and the Condon Committee. His writings are referenced in \cite{Swords:1995/96}, \cite{Swords+etal:2012}, and \cite{Swords:1995}.

\subsection{Dr. Jacques F. Vall\'{e}e, Documatica Research} \label{sec:Vallee}
Dr. Jacques Vall\'{e}e (born 1939) holds a bachelor’s degree in mathematics from the Sorbonne, and a master’s in astrophysics from Lille University (see Figure \ref{fig:Hynek+Vallee}).  He co-developed the first computer-based map of Mars at the University of Texas before joining Northwestern University, where he implemented the first interactive database of UAP cases and completed his PhD in computer science/AI. Joining Stanford University, he did research between the computation center and the Plasma Research Institute in studies of pulsar and quasar energy and the solar corona. 

As one of the Principal Investigators for the Advanced Research Projects Administration of the United States Department of Defense in 1972, Jacques led the team that developed the first network conferencing system on the Advanced Research Projects Agency Network (ARPANET), which was the foundation of the internet.

After 1987, as co-founder and General Partner in the Euro-America Ventures group of funds, he invested in over 60 high-technology startups (including four future ``unicorns'' on NASDAQ) with partners in the US and Europe, targeting advanced software, medicine and space technology as major themes for the portfolio. 

In 2006, Dr. Vall\'{e}e and his two partners were selected by NASA to form Red Planet Capital, a specialized space fund.

In 2010, Jacques Vallée and Chris Aubeck published Wonders In the Sky \cite{Vallee+Aubeck:2010}, a chronological list of aerial phenomena dating back several millennia, questioning the prosaic hypothesis of explaining witness sightings by tests of advanced technology.

His current interest focuses on medical and industrial investments in AI and space imagery.  We present here a selection of some of his more significant publications:
\cite{Vallee:1963,Vallee:1964:EntitiesFacts,Vallee:1964:EntitiesI,Vallee:1964:EntitiesII,Vallee:1964:GhostRockets,Vallee:1965,Vallee:1965:I,Vallee:1966,Vallee:1966:II,Vallee:1968,Vallee:1969:landings,Vallee:1969:witness,Vallee:1971,Vallee:1973:Symbolism,Vallee:1973:UFOWave,Vallee:1987,Vallee:1998,Vallee+Aubeck:2010}

\bibliographystyle{elsarticle-num}
\bibliography{references}

\end{document}